\documentclass[useAMS,usenatbib]{mn2e}

\newcommand{\be}{\begin{equation}}
\newcommand{\ee}{\end{equation}}
\newcommand{\bea}{\begin{eqnarray}}
\newcommand{\eea}{\end{eqnarray}}
\newcommand{\bfig}{\begin{figure}}
\newcommand{\efig}{\end{figure}}
\newcommand{\bt}{\begin{table}}
\newcommand{\et}{\end{table}}
\renewcommand{\vec}[1]{ {\bf #1} }
\newcommand{\nablavec}{\vec{\nabla}}

\usepackage[dvips]{graphicx}

\title[Radiative transfer in GADGET]{An implementation of radiative transfer in the
cosmological simulation code GADGET}
\author[M. Petkova and V. Springel]{Margarita Petkova$^1$\thanks{E-mail: mpetkova@mpa-garching.mpg.de} and
Volker Springel$^1$\thanks{E-mail: volker@mpa-garching.mpg.de}\\
$^1$Max-Planck-Institut f\"ur Astrophysik, Karl-Schwarzschild-Strasse 1, 
85748 Garching, Germany}

\begin{document}

\date{Accepted 2009 March 30. Received 2009 March 27; in original from 2008 December 9}

\pagerange{\pageref{firstpage}--\pageref{lastpage}} \pubyear{2009}

\maketitle

\label{firstpage}


\begin{abstract}
  We present a novel numerical implementation of radiative transfer in
  the cosmological smoothed particle hydrodynamics (SPH) simulation
  code {\small GADGET}. It is based on a fast, robust and
  photon-conserving integration scheme where the radiation transport
  problem is approximated in terms of moments of the transfer equation
  and by using a variable Eddington tensor as a closure relation,
  following the `OTVET'-suggestion of Gnedin \& Abel.  We derive a
  suitable anisotropic diffusion operator for use in the SPH
  discretization of the local photon transport, and we combine this
  with an implicit solver that guarantees robustness and photon
  conservation.  This entails a matrix inversion problem of a huge,
  sparsely populated matrix that is distributed in memory in our
  parallel code. We solve this task iteratively with a conjugate
  gradient scheme.  Finally, to model photon sink processes we
  consider ionisation and recombination processes of hydrogen, which
  is represented with a chemical network that is evolved with an
  implicit time integration scheme.  We present several tests of our
  implementation, including single and multiple sources in static
  uniform density fields with and without temperature evolution,
  shadowing by a dense clump, and multiple sources in a static
  cosmological density field. All tests agree quite well with
  analytical computations or with predictions from other radiative
  transfer codes, except for shadowing. However, unlike most other
  radiative transfer codes presently in use for studying reionisation,
  our new method can be used on-the-fly during dynamical cosmological
  simulation, allowing simultaneous treatments of galaxy formation and
  the reionisation process of the Universe.
\end{abstract}


\section{Introduction}

The absence of Gunn-Peterson troughs in the spectra of high redshift quasars
up to $z \leq 6$ \citep{Fan2006, White2003} suggests that hydrogen is highly
ionised at low redshift, with a volume averaged neutral hydrogen fraction of $x_{\rm HI}
\leq 10^{-4}$ \citep{Fan2007}. The current generation of simulation codes for cosmological
structure formation calculates the self-gravity of dark matter and cosmic gas,
and the fluid dynamics of the cosmic gas, but radiation processes are
typically not taken into account, or only at the level of a spatially uniform,
externally imposed background field. However, we know that the radiation field
has been highly inhomogeneous during certain phases of the growth of
structure, and may have in fact provided important feedback effects for galaxy
formation \citep[e.g.][]{Iliev2005, Yoshida2007, Croft2008}. Therefore, it would be ideal to be able
to calculate self-consistent simulations that {\em simultaneously} follow
cosmic structure growth and the ionisation history of the Universe. 

However, the high calculation cost and complexity of radiative transfer has so
far prevented such simulations, apart from very few exceptions
\citep[e.g.][]{GO1997, Kohler2007, Shin2008, Wise2008}. Instead, the cosmic reionisation process has
been most often treated separately in postprocessing, based on static
simulated density fields \citep[e.g.][]{Ciardi2003, Sokasian2004, Iliev2006, Zahn2007, Croft2008, Li2008}.  
It is the goal of this study to
develop a new numerical scheme that makes self-consistent
radiation-hydrodynamic simulations possible based on the high-resolution
Lagrangian cosmological code {\small GADGET-3} \citep{gadget1, gadget2}.

A large number of different approximate methods have been developed to make
problems of radiative transfer numerically tractable, which are in their full
generality only solvable analytically for very few special cases.  The known
numerical methods include long characteristics ray-tracing schemes \citep{MM,
  ANM1999, Sokasian2001, Cen2002, Abel2002b, Razoumov2005}, short characteristic ray-tracing
schemes \citep{KA1988, NUS2001, MIAS2006, Whalen2006, Alvarez2006a,
  Susa2006, Altay2008, CFMR2001, MFC2003}, hybrid
ray-tracing schemes \citep{Rijkhorst2006, Trac2007},  moment methods and direct solvers \citep{Mellema1998,
  GA2001, Shapiro2004, 
  Whitehouse2005, AT2007, Finlator2008} and other particle-based transport methods
\citep{Ritzerveld2003, Pawlik2008}. 

In the long characteristics method each test cell in the computational volume
is connected to all other relevant cells. Then the equation of radiative
transfer is integrated individually from that test cell to each of the
selected cells. While this method is relatively simple and straight-forward,
it is also very time consuming, since it requires $N^2$ interactions between
the cells. Moreover, parallelization of this approach is cumbersome and
requires large amounts of data exchange between the different processors.
Short characteristic methods integrate the equation of radiative transfer
along lines that connect nearby cells. Here a cell is connected only to its
neighboring cells, and not to all other cells in the computational
domain. This reduces the redundancy of the computations and makes the scheme
easier to parallelize.

Stochastic integration methods, specifically Monte Carlo methods, employ a
ray-tracing strategy where the rays are discretized into photon packets. For
each photon packet, the emission, its frequency and its direction of
propagation are determined by sampling the appropriate distribution functions
that have been assigned in the initial conditions. A particular advantage of
this approach is that comparatively few approximations to the radiative
transfer equations need to be made, so that the quality of the results is
primarily a function of the number of photon packets employed, which can be
made larger in proportion to the CPU time spent. A disadvantage of these
scheme is the comparatively high computational cost and the sizable level of
noise in the discretized radiation field.

Using moments of the radiative transfer equations instead of the full
radiative transfer equation can lead to very substantial simplifications that
can drastically speed up the calculations. In this approach, the radiation is
represented by its mean intensity field throughout the computational
domain. Instead of following rays, the moment equations are solved directly on
the grid or for the particles. Due to its local nature, the moment approach is
comparatively easy to parallelize, but it of course depends on the nature of
the investigated problem whether the simplifications made still provide
sufficient accuracy.

In this paper we develop a moments method that is closely related in spirit to
the Optically Thin Variable Eddington Tensor (OTVET) scheme proposed by
\citet{GA2001}. However, we try to implement it directly on top of the
irregular set of positions sampled by the particles of Smoothed Particle
Hydrodynamics (SPH) simulations, and we use quite different numerical
techniques to solve the resulting transport equations.  In OTVET, the system
of moment equations is closed by estimating the local Eddington tensor with a
simple optical thin approximation, i.e.~one pretends that all sources of light
are `visible' at a given location. Once the Eddington tensors are found, the
local radiation transfer reduces to an anisotropic diffusion problem. The
particular attraction of this moment-based formulation is that it is
potentially very fast, allowing a direct coupling with cosmological
hydrodynamic simulations. In particular, if a rapid method for calculating the
Eddington tensors can be found, the scheme should be able to easily deal with
an arbitrary number of sources. Also, the radiation intensity field does not
suffer from the Poisson shot noise inherent in Monte Carlo
approaches. Together with the local nature of the diffusion problem, this
makes this approach particularly attractive for trying to address the
cosmological reionisation problem with self-consistent simulations of galaxy
formation, since it is likely that low-mass star-forming galaxies of high
number density play an important role for the reionisation process.  We
therefore adopt in this paper the suggestion of \citet{GA2001} and work out an
implementation of the OTVET scheme in SPH. As we shall see, this entails a
number of numerical challenges in practice. We will describe our solutions for
these problems, and carry out a number of tests to evaluate the accuracy of the
resulting implementation.

The near complete lack of analytical results for non-trivial radiative
transfer problems makes it actually hard to validate different numerical
techniques and to compare their performance with each other. A very useful
help in this respect is provided by the cosmological radiative transfer code
comparison project, carried out by \citet{Iliev2006b}. In their paper, they
present a comparison of 11 independent cosmological radiative transfer codes
when applied to a variety of different test problems. A number of our tests
are based on this study, which hence allows a comparison with results of these
other codes.

We start this paper with a brief introduction to the radiative transfer
equations in Section~\ref{SecIntro}. We then describe in
Section~\ref{SecMoments} the moment-based method that is the basis for our
approximate treatment of the radiation transfer problem.  In
Section~\ref{SecNumerics} we elaborate in detail the numerical implementation
of this scheme in SPH. This is followed by a presentation of results for
various test problems in Section~\ref{SecResults}.  Finally, we conclude with
a summary and an outlook in Section~\ref{SecConclusions}. In an appendix, we
give for reference further details on our rate equations, and on the conjugate
gradient approach.


\section{The Radiative Transfer Problem} \label{SecIntro}

\subsection{Equation of radiative transfer} 

Let us briefly derive the radiative transfer (RT) equation in comoving
coordinates, which is also useful for introducing our notation. Let
$f_\gamma(t, \vec{x}, \vec{p})$ be the photon distribution function for
comoving coordinates $\vec{x}$ and comoving photon momentum \be \vec{p} =
a\frac{h\nu}{c}\vec{\hat{n}} \, , \ee where $a \equiv a(t)$ is the
cosmological scale factor, $h$ is the Planck constant, $\nu$ is the frequency
of the photons, and $\vec{\hat{n}}$ is the unit vector in the direction of
photon propagation. Then the number of photons in some part of the Universe is
\be N_{\gamma} = \int {\rm d}\vec{x}\, {\rm d}\vec{p} \, f_\gamma(t, \vec{x},
\vec{p}) \, .  \ee

We can further define the phase-space continuity equation for the distribution
function $f_\gamma \equiv f_\gamma(t, \vec{x}, \vec{p})$ of
photons as
\be
\frac{\partial f_\gamma}{\partial t} + \frac{\partial}{\partial
\vec{x}}(\dot{\vec{x}} f_\gamma) + \frac{\partial}{\partial
\vec{p}}(\dot{\vec{p}} f_\gamma) = \left . \frac{\partial f_\gamma}{\partial t} \right |
_{\rm sources} - \left . \frac{\partial f_\gamma}{\partial t} \right |
_{\rm sinks} \, .
\label{eqn2.3}
\ee

Here the source and sink terms on the right hand side of the equation
represent photon emission and absorption processes, respectively. We define
the specific radiation intensity $I_\nu$ as the energy of photons in a
frequency bin $\Delta \nu$ that pass through an area $\Delta A$ and solid
angle $\Delta \Omega$ for a time $\Delta t$. The specific intensity
$I_\nu $ is then related to the photon distribution $f_\gamma$ as follows
\be
I_\nu = h\nu f_\gamma \frac{{\rm d}^3 x \,{\rm d}^3 p}
{{\rm d}\nu \, {\rm d} \Omega \, {\rm d}A\, {\rm d}t} = \frac{h^4
  \nu^3}{c^2}f_\gamma \, .
\ee

Substituting into equation (\ref{eqn2.3}), rearranging and adding the
proper absorption and emission terms, one obtains the following radiative
transfer equation in comoving variables \citep{GO1997}:
\be
\frac{1}{c}\frac{\partial I_\nu}{\partial t} +
\frac{\vec{n}}{a}\frac{\partial I_\nu}{\partial \vec{x}} -
\frac{H}{c}\left(\nu\frac{\partial I_\nu}{\partial \nu} - 3I_\nu \right) =
-\kappa_\nu I_\nu + j_\nu \, ,
\label{theeqn}
\ee where $H \equiv \dot{a}/a$ is the Hubble expansion rate, $\kappa_\nu$ is
the absorption coefficient and $j_\nu$ is the emission
coefficient. 

Unfortunately, this full radiative transfer (RT) equation is in practice very
difficult to solve in full generality. In particular, the high dimensionality
(comprised of 3 spatial variables, 2 directional angles, 1 frequency variable,
and time) of this partial differential equation makes a direct discretization
on a mesh highly problematic.  We will hence apply in Section~\ref{SecMoments}
simplifications to the RT equation that yield an approximation that can be
more easily calculated in cosmological codes.

\subsection{Basic physics of hydrogen photoionisation}

If we consider pure hydrogen gas, the rate equations describing
photoionisation and photoheating processes become comparatively simple. For
the most part we will restrict ourselves to this chemical composition in this
paper, but we note that an extension of our formalism to include other
elements is readily possible. In fact, we have already implemented helium as
well, but we omit an explicit discussion of it in the following for the sake
of simplicity.

The photoionisation rate $k_{\rm ion}$ of hydrogen ($H + h\nu \to H^+ + e^-$)
is given by
\be
k_{\rm ion} = \int {\rm d}\Omega \int_{\nu_o} ^ \infty {\rm d}\nu\,
\frac{I_\nu \sigma_\nu}{h\nu} \, ,
\ee
where $h\nu_o = 13.6\,{\rm  eV}$ is the hydrogen ionisation potential and
$\sigma_\nu$ is the photoionisation cross-section:
\be
\sigma_\nu = \sigma_o \left( \frac{\nu}{\nu_o} \right) ^4 \frac{\rm
exp\{4 - [(4\,\rm tan^{-1}\epsilon)/\epsilon]\}}{1-\rm exp(-2\pi /
\epsilon)} \mbox{   for $\nu \geq \nu_o$},
\ee
where $\sigma_o = 6.30 \times 10^{-18} \, \rm cm^2$ and $\epsilon =
\sqrt{(\nu/\nu_o) -1}$.

The corresponding
photoheating rate of hydrogen is given by
\be
\Gamma = n_{\rm HI} \int {\rm d}\Omega \int_{\nu_o} ^ \infty {\rm d}\nu\, \frac{I_\nu
\sigma_\nu}{h\nu} (h\nu - h\nu_o) ,
\label{eqn:realgamma}
\ee
where $n_{\rm HI}$ is the number density of neutral hydrogen. Furthermore,
the change in the neutral gas density due to recombinations is given by
\be
\frac{\partial  n_{{\rm HI}}}{\partial t} = \alpha \, n_{\rm e}  n_{\rm
p} ,
\ee 
where $\alpha$ is the temperature-dependent recombination coefficient, $n_{\rm
  e}$ is the electron number density and $n_{\rm p} \equiv n_{\rm HII}$ is the
proton number density, which is in turn equal to the ionised hydrogen number
density (for a pure hydrogen gas).  

The change of the density of the neutral
gas due to ionisations is given by
\be
\frac{\partial  n_{{\rm HI}}}{\partial t} = - c\, \sigma_o n_{\rm HI}
n_\gamma ,
\ee  
where $c$ is the speed of light and $n_\gamma$ is the number density of
ionising photons. Thus, the total change in the neutral gas density,
due to ionisations or recombinations, is
given by
\be
\frac{\partial  n_{{\rm HI}}}{\partial t} =  \alpha \, n_{\rm e}  n_{{\rm
HII}} - c\, \sigma_o n_{\rm HI}  n_\gamma .
\ee 

For all our calculations described in this paper we use the
  on-the-spot approximation \citep{OF2006}, i.e.~photons emitted due to
  recombinations to excited levels are re-absorbed immediately by neutral
  hydrogen atoms in the vicinity. This behavior is described by the so called
  case-B recombination coefficient $\alpha_{\rm B}$.

\section{The variable Eddington tensor formalism} \label{SecMoments}

We now turn to a description of the moment-based approximation to the
radiation transfer problem that we use in this study.  The first three moments
of the specific intensity, the mean intensity $J_\nu$, the radiation flux
vector $F^i_\nu$, and the radiation pressure tensor $P^{ij}_\nu$, are defined
as follows: \bea
J_\nu &=& \frac{1}{4\pi}\int {\rm d}\Omega\, I_\nu ,\\
F^i_\nu &=&  \frac{1}{4\pi}\int {\rm d}\Omega\, n^i I_\nu , \\
P^{ij}_\nu &=& \frac{1}{4\pi}\int {\rm d}\Omega\, n^in^j I_\nu ,  \eea
where $n$ is a direction vector and the indices $i \, {\rm and} \, j$ run
through the three elements of the vector in Cartesian space. We can
further define $h^{ij}$, the so-called Eddington tensor, based on $P^{ij}_\nu
= J_\nu h^{ij}$.

We can for the moment ignore the frequency derivative in the RT equation if we
can assume that the Universe does not expand significantly before a photon is
absorbed. With this simplification, the first moments of the RT equation take
the form: \bea
\label{m1}
\frac{1}{c}\frac{\partial J_\nu}{\partial t} +
\frac{1}{a}\frac{\partial F^i_\nu}{\partial x^i}
 &=&
-\hat{\kappa}_\nu J_\nu + j_\nu , \\
\frac{1}{c}\frac{\partial F^j_\nu}{\partial t} +
\frac{1}{a}\frac{\partial J_\nu h^{ij}}{\partial x^i}
 &=&
-\hat{\kappa}_\nu F^j_\nu \label{m2} ,
\eea
where 
\bea
\hat{\kappa}_\nu &=& \kappa_\nu + \frac{3H}{c} .
\eea
In the second moment equation (\ref{m2}), we can ignore the 
term of the order
$c^{-1}$ and solve for the flux
\be F^j_\nu = - \frac{1}{\hat{\kappa}_\nu}  
\frac{1}{a}\frac{\partial J_\nu h^{ij}}{\partial x^i} , \ee
which we then insert back into 
equation (\ref{m1}).
This leads to the following approximation to the RT
equation:
\be
\frac{\partial J_\nu}{\partial t}=\frac{c}{a^2}\frac{\partial}{\partial
x_j}\left( \frac{1}{\hat{\kappa}_\nu}\frac{\partial J_\nu h^{ij}}{\partial
x_i} \right) - c\hat{\kappa}_\nu J_\nu +  c j_\nu .
\label{eqnRTMom}
\ee

This form of the RT equation is already much simpler than the fully general
form of equation~(\ref{theeqn}). In particular, each of the terms in equation
(\ref{eqnRTMom}) has a simple physical interpretation. The time evolution of
the local mean radiation intensity is given by a transport term, described by
the anisotropic diffusion term on the right hand side, a sink term describing
absorptions, and an emission term that accounts for sources.  However, in
order to be able to solve this equation an expression for the Eddington tensor
$h^{ij}$ is needed, which is left undefined by these moment equations. We
therefore need to assume a certain form for the Eddington tensor, or in other
words, a {\em closure relation}.

For the closure relation, we follow \citet{GA2001} and estimate the local
Eddington tensor with an optically thin approximation. This means that we
assume that a reasonable approximation to the Eddington tensor can be obtained
by approximating all lines-of-sight to the sources as being optically thin.  The
radiation intensity pressure tensor $P^{ij}$ in this optically thin regime can
then be computed as \be P^{ij} \propto \int {\rm d}^3 x' \rho_{\ast}(\vec{x}')
\frac{(\vec{x}-\vec{x}')_i(\vec{x}-\vec{x}')_j} {(\vec{x}-\vec{x}')^4} ,
\label{edd1}
\ee
and thus the Eddington tensor is given by
\be
h^{ij} = \frac{P^{ij}}{{\rm Tr}(P)} .
\label{edd2}
\ee Note that the Eddington tensor only determines in which {\em direction}
the local radiation propagates, but the magnitude of the radiation intensity
tensor is unimportant as far as the Eddington tensor is concerned. This means
that even in situations where the lines-of-sight to the sources are not
optically thin at all, one will often end up with fairly accurate estimates of
the Eddington tensor based on equations (\ref{edd1}) and (\ref{edd2}), simply
because the radiation will typically mainly propagate away from the sources,
even in optically thick cases. In particular, note that the above
approximation is always correct for a single source. When there are multiple
sources of equal strength, the optically thin approximation will weight the
sources that are closest most strongly, in accordance with the $1/r^2$ decay
of the intensity. While this can be expected to result in reasonably accurate
estimates of the Eddington tensor in many situations (especially in the
vicinity of a dominating source), errors can certainly arise in particular
situations, for example at locations that are equidistant from two sources of
equal strength. How serious these errors are in problems of interest needs to
be analyzed with appropriate test problems.

As we describe later in more detail in Section \ref{ETcalc}, we note that
equation~(\ref{edd1}) can be accurately calculated with a hierarchical
multipole approach similar to the one applied in gravitational tree
algorithms. This allows a fairly efficient treatment of an arbitrarily large
number of sources, which is a distinctive advantage of the moments based
approach compared with other methods.

\subsection{Choice of convenient Lagrangian variables}

We will now rewrite the RT equations into a form that is more convenient for
use with a Lagrangian method such as SPH. In particular, it is advantageous to
pick variables in the numerical scheme that are normalized to unit mass, not
unit volume. For example, if we express the ionisation state of the gas as the
number density of ionised hydrogen per unit volume, then we have to readjust
this number somehow any time we reestimate the local gas density (which may
change if the gas moves around), otherwise the ionised fraction would
change. However, if we use convenient variables that are normalized to unit
mass, we do not need to worry about such corrections.

For chemical networks of hydrogen, it is convenient (and often done in
practice) to express abundances relative to the total abundance of hydrogen
nuclei:

\be n_{\rm H} = \frac{X_{\rm H}\,\rho}{m_{\rm p}} . \ee 
Here
$X_{\rm H}=0.76$ is the cosmological mass fraction of hydrogen
and ${m_{\rm p}}$ is the proton mass. In the following, we use the notation $n_{\rm
  HI}$ for neutral hydrogen, and $n_{\rm HII}$ for ionised hydrogen, such
that
\be
 n_{\rm H} = n_{\rm  HI} + n_{\rm HII} .
\ee

In our actual numerical code, we will use a variable $\tilde n_{\rm HII}$ to
express the abundance of ionised hydrogen, defined as \be \tilde n_{\rm HII} =
\frac{n_{\rm HII}}{n_{\rm H}}, \ee where $n_{\rm HII}$ is the ordinary number
density of HII atoms (i.e. number of protons per unit volume). Note that this
quantity is now normalized to unit mass, as desired. In addition, it is
dimensionless, which avoids numerical problems due to large numbers if we use
astronomical length units.

A similar reasoning also applies to the radiation intensity itself. In
principle, the fundamental quantity we work with is the frequency dependent,
angle-averaged mean intensity. However, we cannot afford to carry around a
full spectrum with each fluid element in a hydrodynamical code.  This would be
too cumbersome and also is not really necessary of we are interested only in
the reionisation problem. Instead, it is sufficient to store the intensity
integrated over a narrow frequency interval around the ionisation potential of
hydrogen. Or in other words, a more convenient quantity to work with would be
something like the number density of photons capable of ionising hydrogen. We
now formulate the relevant equations using this concept.

In general, the photon number density is \be n_{\gamma} = \frac{1}{c} \int
\frac{4 \pi J_{\nu}}{h\nu}\,{\rm d}\nu .  \ee However, we will only consider
the spectrum in a small band around the frequencies of interest. For
simplicity, we assume that the spectrum has the form \be
J_\nu = J_{0}\, \delta(\nu-\nu_{0}) \ee around the ionisation
frequency $\nu_0$, where $h\nu_0=13.6\,{\rm eV}$ is the hydrogen ionisation
potential. This form of the radiation intensity limits the spectrum to
effectively just the hydrogen ionisation frequency. We therefore
obtain this simple form \be
n_{\gamma} = \frac{1}{c} \frac{4 \pi J_{0}}{h\nu_0} \ee for the number density
of ionising photons.

We also note that the absorption coefficient $\kappa_{\nu}$ for
ionisation in the equation for
radiation transport is  
\be 
\kappa_\nu = n_{\rm HI}\, \sigma_\nu,
\ee
where $\sigma_\nu$ is the 
cross-section for hydrogen ionisation. If we multiply the loss term
$\kappa_\nu J_{\nu}$ in the RT equation
by $4\pi/(c\, h\nu)$ and integrate over $\nu$, we get the so-called
ionisation rate $k_{\rm ion}$, given by
\be
k_{\rm ion} = \int \frac{4\pi J_\nu}{h\nu} \sigma_\nu {\rm d}\nu .
\ee
For our narrow spectrum, this leads  to the simple expression
\be
\int {\rm d}\nu \frac{4\pi}{c h\nu}\, \kappa_\nu J_\nu =  \sigma_0
n_{\rm HI}\,n_\gamma \, ,
\ee
where $\sigma_0$ is the cross section at the resonance.
Another consequence of these
definitions is that
we can
write the number density evolution of ionised hydrogen due to new
reionisations as 
\be
\frac{{\rm d}n_{\rm HII}}{{\rm d}t} = c \, \sigma_0 \,n_{\gamma}\, n_{\rm HI}.
\ee
The photon field loses energy at the same rate, i.e.~the loss
term for the radiation field should be of the form
\be
\frac{{\rm d}n_{\gamma}}{{\rm d}t} = - c \, \sigma_0 \,n_{\gamma}\,
n_{\rm HI},
\ee
which is also what the loss term in the RT equation
gives in this notation.

The above suggests that we can cast the moment-based RT
equation into a more convenient form if we multiply it through with
$4\pi/(c h\nu)$ and integrate over $\nu$. This gives:
\be
\frac{\partial n_\gamma}{\partial t}= c \frac{\partial}{\partial
x_j}\left( \frac{1}{\kappa}\frac{\partial n_\gamma h^{ij}}{\partial
x_i} \right) - c\,\kappa\, n_\gamma +   s_\gamma ,
\ee
where $\kappa = \sigma_0 n_{\rm HI}$ and the cosmological scale
factor has been dropped for simplicity. The source function  $s_\gamma$ gives the
rate per unit volume at which new ionising photons are produced.

Finally, we also want to express the photon number density in terms of
the local density of hydrogen atoms, $\tilde n_\gamma  = {n_\gamma}/{n_{\rm H}}$.
Then we get:
\be
\frac{\partial \tilde n_\gamma}{\partial t}= c \frac{\partial}{\partial
x_j}\left( \frac{1}{\kappa}\frac{\partial \tilde n_\gamma h^{ij}}{\partial
x_i} \right) - c\,\kappa\, \tilde n_\gamma +   \tilde s_\gamma .
\label{eqA}
\ee This is the formulation of the RT equation that we  implemented in
this work in the simulation code {\small GADGET-3}.

We have to augment equation (\ref{eqA}) with the changes of the different
chemical species as a result of interactions with the radiation field. If we
consider only hydrogen, this is just: 
\be \frac{\partial \tilde n_{\rm HII}}{\partial t} = c \,\sigma_0 n_{\rm H}\, \tilde n_{\rm HI} \tilde
n_\gamma - \alpha \, n_{\rm H} \, \tilde n_{e} \tilde n_{\rm HII} \, ,
\label{eqB}\ee 
where $\alpha$ is the temperature-dependent recombination
coefficient. If we have only hydrogen, we can  set $\tilde n_{e} =
\tilde n_{\rm HII}$ and $\tilde n_{\rm HI} = 1 - \tilde n_{\rm
HII}$.

For hydrogen reionisation problems, we want to solve the two basic equations
(\ref{eqA}) and (\ref{eqB}) as efficiently, accurately and robustly as
possible.  We recall that the three terms of (\ref{eqA}) have a
straightforward interpretation. The first term on the right hand side is a
diffusion like equation, which is conservative, i.e.~it leaves the total
number of photons unchanged. The second term describes photon losses, and each
photon lost will cause one hydrogen atom to be ionised. Finally, the third
term is the source term, and describes the injection of new photons.  This
suggests that the total number of ionisations must always be equal to to the
total number of photons lost. If we can maintain numerically accurate photon
conservation, then this property should ensure a proper speed of the
ionisation front even for relatively inaccurate time stepping, as the
propagation of the front should
largely be determined by the injection rate of photons at the source.

The above suggests a simple possibility for treating the time evolution of the
photon number density in each timestep in terms of three parts, corresponding
to an {\em operator-splitting}, or {\em fractional-step} approach: One may
first inject new photons according to the source function, then transport
photons conservatively by treating the diffusion part, and finally, advance
the ``chemical network'' (eqn. \ref{eqB}) by treating only ionisations and
recombinations, making sure again that we do not lose any photons.  The
chemical equations can be easily ``subcycled'' or treated with an integrator
for stiff differential equations, if needed, because they are completely
local. On the other hand, the most expensive part of the time advance is given by
the diffusion part. This not only involves a coupling with neighboring fluid
elements but also cannot easily be integrated with an explicit time
integration scheme, because the diffusion equation becomes easily unstable in
this case. We will therefore treat this part with an implicit method. While
involving an expensive iteration scheme, this provides good stability and
allows for comparatively large timesteps.

\section{Numerical Implementation} \label{SecNumerics}

In this section we describe the numerical formalism we have implemented in
order to solve the moment-based RT equations coupled to the parallel Tree/SPH
code {\small GADGET-3}, which is a significantly evolved and extended version
of the public {\small GADGET-2} code \citep[][]{gadget2}.  We first give a
very brief overview of the basic concepts of smoothed particle hydrodynamics
(SPH), and then present a derivation of a new anisotropic diffusion operator
in SPH, which is needed for the radiation transfer in moment form when a
spatially varying Eddington tensor is used. We also explain how the diffusion
equation can be integrated robustly in time based on an implicit scheme with
an iterative sparse matrix solver.  The time integration of the rate equations
for ionisation and recombination also requires special methods because they
involve stiff differential equations. Finally, we describe the calculation of
the Eddington tensors, and how this can be best combined existing algorithms
in the {\small GADGET-3} code.

\subsection{The basics of smoothed particle hydrodynamics}

SPH is a widely used Lagrangian scheme that follows the evolution of gas
properties based on discrete tracer particles \citep[see, e.g.,][for a
review]{Monaghan1992}. The particle properties are averaged, `smoothed', over
a kernel function, yielding so-called kernel interpolants for the fluid
properties based on a few sampling points. For example, the kernel-interpolant
of a property $\left<A\right>$ is given by \be \left<A(\vec{r})\right> = \int
{\rm d}\vec{r}' A(\vec{r'}) W(|\vec{r}'-\vec{r}|, h), \ee where $h$ is called
the smoothing length and is defined such that the kernel $W$ drops to zero for
$|\vec{r'}| > h$. In a discretized form this equation becomes \be
\left<A(\vec{r}_i)\right> = \sum_j \frac{m_j}{\rho_j} A(\vec{r}_j)
W(|\vec{r}_j-\vec{r}_i|, h_i), \ee where the sum is over all the particles
that lie inside radius $h$. In the {\small GADGET} code, the kernel has the
following standard spline form: \be W(r) = \frac{8}{\pi h^3}\left \{
\begin{array}{ll} 
1 - 6\left(\frac{r}{h}\right)^2 + 6 \left(\frac{r}{h}\right)^3  & \mbox{for $0 \leq \frac{r}{h} \leq \frac{1}{2}$}  \\
\\
2 \left(1 - \frac{r}{h}\right)^3 & \mbox{for  $\frac{1}{2} < \frac{r}{h} \leq 1$} \\
\\
0 & \mbox{for $1 < \frac{r}{h}$} 
\end{array}
\right.
\ee
An important property of the kernel interpolant is that it can also be used to 
obtain a derivative of the reconstructed function, which can be simply approximated by 
\be
\begin{array}{ll}
\nabla_i A(\vec{r}_i) &= \nabla_i \sum_j  \frac{m_j}{\rho_j} A(\vec{r}_j)
W(|\vec{r}_j-\vec{r}_i|, h_i) \\
\\
 &= \sum_j  \frac{m_j}{\rho_j} A(\vec{r}_j) \nabla_i
W(|\vec{r}_j-\vec{r}_i|, h_i).
\end{array}
\ee Starting from a density estimate in the form \be \rho_i = \sum_j m_j
W(|\vec{r}_j-\vec{r}_i|, h_i), \ee this allows one to calculate pressure
gradients, and from this, equations of motion for the gas elements which
represent the Euler equations. The particular formulation for the equation of
motion we use here is based on the `entropy-formulation' of SPH discussed by
\citet{Springel2002}.

\subsection{Obtaining anisotropic second order derivatives with a kernel
  interpolant}

Discretization of the diffusion term in the RT transfer equation in SPH poses
some difficulties. We are basically confronted with the task to find an
efficient and accurate approximation to terms of the form \be \frac{\partial
  ^2 Q_{\alpha \beta}}{\partial x_s \partial x_k}, \ee where $Q_{\alpha
  \beta}$ is the product of the local Eddington tensor $\vec{h}$ and the
relative photon density $\tilde n_\gamma$ . Simply differentiating a kernel
interpolant twice is not a good solution, as this becomes very noisy because
the kernel-interpolant of SPH is only second-order accurate. On the other
hand, the discretization of the Laplacian discussed by \citet{J2004} does not
work either, as it only works for the isotropic case.

We now describe the solution we have found for this problem, which basically
consists of the task to approximate the second order partial derivatives of an
element $Q_{\alpha \beta}(\vec{x})$ of the matrix $\vec{Q}(\vec{x})$ with a
kernel-interpolant. We consider a Taylor-series for $Q_{\alpha
  \beta}(\vec{x}_j)$ in the proximity of $Q_{\alpha \beta}(\vec{x}_i)$, i.e.
\be
\begin{array}{ll}
 & Q_{\alpha \beta}(\vec{x}_j) - Q_{\alpha
  \beta}(\vec{x}_i) = \nablavec Q_{\alpha \beta}\biggr|_{\vec{x}_i}
\cdot  (\vec{x}_j - \vec{x}_i) +  \\

& \frac{1}{2} \sum_{s,k}\frac{\partial ^2 Q_{\alpha \beta}}{\partial
  x_s \partial x_k}\biggr|_{\vec{x}_i} ( \vec{x}_j - \vec{x}_i )_s( \vec{x}_j
-
\vec{x}_i )_k +   \\

& {\mathcal{O}}(\vec{x}_j-\vec{x}_i)^3 .
\end{array}
\label{eq1}
\ee
Let us use the short-hand notation 
$\vec{x}_{ij} = \vec{x}_j - \vec{x}_i$ and $W_{ij} =
W(|\vec{x}_j - \vec{x}_i|)$, 
where  $W(r)$ is the
SPH smoothing kernel.
Neglecting higher order terms, we multiply the above expansion  with
\be
  \frac{(\vec{x}_{ij})_l \; W_{ij,m}}
  {|\vec{x}_{ij}|^2} ,
  \label{eqn:multiply}
\ee 
and integrate over all $\vec{x}_j$. Here $(\vec{x}_{ij})_l$ is the $l$-th
component of the vector $\vec{x}_{ij}$, and $W_{ij,m}$ is the partial
derivative of $W_{ij}$ with respect to the $m$-component of $\vec{x}_i$.  In
particular, this means we have 
\be
\begin{array}{ll}
 W_{ij,m} &= (\nablavec_i W_{ij})_m =
-(\nablavec_j W_{ij})_m = \frac{\partial W(|\vec{x}_{ij}|)}{\partial
  (\vec{x}_i)_m}\\ 
\\
&= -W'(|\vec{x}_{ij}|) \frac{(\vec{x}_{ij})_m}{|\vec{x}_{ij}|} .
\end{array}
\ee
We now find that 
\be \int \frac{(\vec{x}_{ij})_k \; (\vec{x}_{ij})_l
  \;W_{ij,m}} {|\vec{x}_{ij}|^2} \, {\rm d}{\vec{x}_j} = 0 , 
\ee 
for all
combinations of $k$, $l$ and $m$. This is because there is always at least one
single component of $\vec{x}_{ij}$ left so that the integral vanishes by
symmetry. As a result, the first order term of our integrated Taylor expansion
drops out.

We now consider the second order term, where we encounter the expression \be
T_{sklm} = \int \frac{(\vec{x}_{ij})_s \; (\vec{x}_{ij})_k \; (\vec{x}_{ij})_l
  \; (\vec{x}_{ij})_m \; W'(|\vec{x}_{ij}|)} {|\vec{x}_{ij}|^3} \,
{\rm d}{\vec{x}_j} .\ee There are a number of different cases. If $l$ and $m$
are equal, then $s$ and $k$ must also be equal, otherwise the integral
vanishes. So here we would have three possible contributions to a $s,k$-sum,
corresponding to the three coordinates that $s$ and $k$ can assume. If $l$ and
$m$ are unequal, then we must either have $s=l$ and $k=m$, or have $s=m$ and
$k=s$. So here there are two contributions to a $s,k$-sum in this case.
Evaluating the integral $T_{sklm}$ for these cases gives: \be T_{sklm} =
\left\{
\begin{array}{cl}
-\frac{3}{5} & \mbox{if $l=m$ and $s=k=l=m$},\\
\\
-\frac{1}{5} & \mbox{if $l=m$ and $s=k$, but $s \ne l$},\\
\\
-\frac{1}{5} & \mbox{if $l\ne m$, and $s=l$, $k=m$}\\ 
             & \mbox{                 or $s=m$, $k=l$}, \\
\\
0 & \mbox{in all other cases}.
\end{array}
\right.  \ee Note that we can pick $l$ and $m$ freely when we multiply
the Taylor expansion with the term (\ref{eqn:multiply}) and integrate
over it. In particular, we can also use several different choices one
after the other and then form a linear combination of the
results. This can in fact be used to isolate any of the second
derivatives of the Hessian matrix of $Q_{\alpha \beta}$. Let us assume
for example that we want to calculate the second derivative of
$Q_{\alpha \beta}$ with respect to $x_0$.  Choosing $l=m=0$, then the
three choices $k=s=0$, $k=s=1$ and $k=s=2$ all give terms that
contribute to the integral over the expansion. These are: \be
\begin{array}{l}
2  \int \frac{Q(\vec{x}_j) -
    Q(\vec{x}_i)}{| \vec{x}_{ij}|^2}\, (\vec{x}_{ij})_0 W_{ij,0}
  \,{\rm d}{\vec{x}_j} = \\ 
\\
\;\;\;\;\; \frac{3}{5} \frac{\partial^2 Q}{\partial
    x_0^2}
+ \frac{1}{5} \frac{\partial^2 Q}{\partial x_1^2}
+ \frac{1}{5} \frac{\partial^2 Q}{\partial x_2^2} .
\end{array}
\ee
Here $Q$ is to be understood as $Q=Q_{\alpha \beta}$ for brevity. Based on
this, we can now isolate the desired partial derivative by forming a linear
combination:
\be
\begin{array}{l}
 \frac{\partial^2 Q}{\partial x_0^2}
=
  2 \int \frac{Q(\vec{x}_j) -
    Q(\vec{x}_i)}{| \vec{x}_{ij}|^2}\, \times \\ 
\\
\;\;\;\;\;
\left[ 2
(\vec{x}_{ij})_0 W_{ij,0} - \frac{1}{2} (\vec{x}_{ij})_1 W_{ij,1}
  - \frac{1}{2} (\vec{x}_{ij})_2 W_{ij,2} 
\right]
\,{\rm d}{\vec{x}_j}.
\end{array}
\ee
In a similar fashion, we can obtain a mixed partial derivative in the
following way:
\be
\begin{array}{l}
 \frac{\partial^2 Q}{\partial x_0 \partial x_1}
=
   2 \int \frac{Q(\vec{x}_j) -
    Q(\vec{x}_i)}{| \vec{x}_{ij}|^2}\, \times \\ 
\\
\;\;\;\;\;
\left[
\frac{5}{4} (\vec{x}_{ij})_0 W_{ij,1} + \frac{5}{4} (\vec{x}_{ij})_1 W_{ij,0}
\right]
\,{\rm d}{\vec{x}_j}.
\end{array}
\ee

Formulae for all other second-order partial derivatives can be obtained
from these expressions by cyclic permutation. Also, they are valid for each of the
matrix elements $Q_{\alpha \beta}$. 

Using these results, we can now turn to obtaining an expression for the sum of the second
derivatives, as needed in the anisotropic diffusion equation. Based on the
above, we can write the desired expression in the compact form:
\be
\left. \frac{\partial^2 Q_{\alpha \beta}}{\partial x_{\alpha} \partial
  x_{\beta}}
\right|_{\vec{x}_i}
=
  2 \int \frac{  \vec{x}_{ij}^{\rm T}\, [  \vec{\tilde Q}(\vec{x}_j) -
    \vec{\tilde Q}(\vec{x}_i) ]\, \nablavec_i W_{ij} } {| \vec{x}_{ij}|^2}  
  \,{\rm d}{\vec{x}_j}.
\label{eqnsoq}
\ee
Here we defined a new matrix $\vec{\tilde Q}$ through the matrix
elements of the original matrix $\vec{Q} = (Q_{\alpha \beta}$), in the
following way:\be \tilde \vec{Q} = \frac{5}{2}\vec{Q} -
\frac{1}{2}\rm{Tr}(\vec{Q})\vec{I}\label{anis1} . \ee

Inspection of this result highlights one interesting issue that could
potentially become a numerical stability problem in certain
situations.  The matrix $\tilde \vec{Q}$ is not guaranteed to
correspond to a positive definite quadratic form when it is used in
the SPH discretization form of equation (\ref{eqnsoq}). If the
radiation transfer is very anisotropic, the matrix $\tilde \vec{Q}$
can contain negative diagonal elements and thus gives rise to an
`anti-diffusive' behaviour in the discretized radiation transfer
equation, where radiation is transported from a particle of lower
radiation intensity to one with higher radiation intensity. It is not
clear right away whether this will lead to numerical stability
problems of the radiative diffusion treatment, but it could.

In case this is a problem, one way to avoid it would be to somehow
suppress transport of radiation opposite to the direction of the
gradient of the radiation intensity between a particle pair. Another
way is to add in an isotropic component to $\tilde \vec{Q}$ such that
\be \tilde \vec{Q}^* = \alpha \tilde \vec{Q} + (1-\alpha)
\frac{\vec{I}}{3} \label{anis2} . \ee Here the idea is to make $\tilde
\vec{Q}^*$ slightly more isotropic, such that $\tilde \vec{Q}^*$
becomes positive definite again. In order to guarantee this, we need
to assign $\alpha = \frac{2}{5}$. Form equations (\ref{anis1}) and
(\ref{anis2}) we can see that this `anisotropy-limited' matrix is then
actually $\tilde \vec{Q}^* = \vec{Q}$. One interpretation of this
result is that the unmodified matrix $\vec{Q}$ mediates diffusion
which is a mix of $2/5$ of the `correct' anisotropic diffusion and
$3/5$ of isotropic diffusion.  In some of our tests we will compare
results from both formulations of the matrix. We will refer to $\tilde
\vec{Q}^*$ as the `anisotropy-limited' tensor and to $\tilde \vec{Q}$
as the `fully-anisotropic' tensor.

Note that in the case where $\vec{Q}$ is diagonal and proportional to the
identity matrix, equation (\ref{eqnsoq}) reduces to the isotropic result for a
scalar function derived by \citet{J2004} for the thermal conduction problem.
The important point about equation (\ref{eqnsoq}) is that it involves only a
first order derivative of the kernel function. As a result, it can be
discretized straightforwardly in the usual SPH way, where the integration is
replaced by a sum over all neighboring SPH particles within the kernel volume.

\subsection{Discretization of the anisotropic diffusion term in SPH}

As we have shown in detail in the previous section, kernel-interpolated 
second-order derivatives of some tensor $Q_{\alpha\beta}$ can be obtained as 
in equation (\ref{eqnsoq})
 for
a suitably defined modified tensor $\tilde Q$. 
Furthermore, we note the identity
\be
\frac{\partial}{\partial x} \left(\frac{1}{s}  \frac{Q}{\partial
  y} \right) = \frac{1}{2} \left(\frac{\partial^2}{\partial x \partial
  y} \frac{Q}{s} - Q\frac{\partial^2}{\partial x \partial
  y}\frac{1}{s} + \frac{1}{s}\frac{\partial^2 Q}{\partial x \partial
  y} \right)
\ee
and thus inserting equation (\ref{eqnsoq}) we obtain
\be
\begin{array}{l}
\frac{\partial}{\partial x_{\alpha}} \left(\frac{1}{s}  \frac{Q_{\alpha \beta}} {\partial
  x_{\beta}} \right) =  \\
\\ \;\;\;\;\; 2 \int \frac{  \vec{x}_{ij}^{\rm T}\, \frac{1}{2}\left( \frac{1}{s_i}
 + \frac{1}{s_j} \right)[  \vec{\tilde Q}(\vec{x}_j) -
    \vec{\tilde Q}(\vec{x}_i) ]\, \nablavec_i W_{ij} } {| \vec{x}_{ij}|^2}  
  \,{\rm d}{\vec{x}_j} \, .
\end{array}
\ee
For our application, let us denote the correspondingly modified Eddington
tensor as $\tilde h^{ij}$.  We can then write down an SPH discretization of the
diffusion part of the radiation transfer equation. This can be expressed as:
\be
\frac{\partial \tilde n^i_\gamma}{\partial t}=
2\,\sum_j \frac{c}{\kappa_{ij}}
\frac{\vec{x}_{ij}^{\rm T}
\left[\tilde n^j_\gamma \vec{\tilde h}_j  - \tilde n^i_\gamma
 \vec{\tilde h}_i \right]\, \nablavec_i W_{ij}}
{\vec{x}_{ij}^2}
\,\frac{m_j}{\rho_{ij}} .
\label{eqX11}
\ee Here \be \frac{1}{\kappa_{ij}} = \frac{1}{2}\left[\frac{1}{\kappa_i} +
  \frac{1}{\kappa_j}\right]\ee is a symmetric average of the absorption
coefficients of the two particles $i$ and $j$, and $\rho_{ij}$ is a
symmetrized density. It is however important to be careful about how exactly
the symmetrizations are done in practice, because this can affect the
performance of the scheme if there are particles of varying mass. In
particular, we would like to use a formulation where the conservation of the
number of photons is guaranteed in this case as well. If possible, we would
also like to obtain a formulation where the effective coupling matrix is
symmetric, because this is a prerequisite for using certain, particularly
efficient solution methods from linear algebra, such as the conjugate gradient
(CG) method.

The photon conservation property is best analyzed by switching to variables
that directly encode the photon number of each particle. Let us define for
this purpose the quantity \be N_i = m_i \tilde n_\gamma^i
\label{eqX12}
\ee 
for each particle. The real photon number of a particle is actually 
$N_\gamma= n_\gamma m / \rho = m\, \tilde n_\gamma X_{\rm H}/m_p$, but
we here drop the constant $X_{\rm H}/m_p$ in the definition
(\ref{eqX12}), for notational simplicity.
Multiplying Eqn.~(\ref{eqX11}) through with $m_i$ gives now
\be
\frac{\partial N_i}{\partial t}= 
2\,\sum_j \frac{c}{\kappa_{ij}\rho_{ij}}
\frac{\vec{x}_{ij}^{\rm T}
\left[m_i N_j \vec{\tilde h}_j  - m_j N_i
 \vec{\tilde h}_i \right]\, \nablavec_i W_{ij}}
{\vec{x}_{ij}^2}.
\ee
Note that we can also write this as
\be
\frac{\partial N_i}{\partial t}= 
\sum_j (w_{ij} N_j  - w_{ji} N_i),
\label{eqX13}
\ee
where
\be
w_{ij}\equiv
\frac{c}{\kappa_{ij}\rho_{ij}}
\frac{\vec{x}_{ij}^{\rm T}
m_i\vec{\tilde h}_j \nablavec_i W_{ij}}
{\vec{x}_{ij}^2} .
\ee
From the
formulation in equation (\ref{eqX13}) we easily see that the total photon
number, $\sum_i N_i$, is conserved,
but in general the matrix $w_{ij}$ is not symmetric. 
Even though other
linear solvers may work,
this would prevent
us from safely applying the CG scheme
to calculate a solution for an backwards-Euler timestep of equation
(\ref{eqX13}), as the required implicit solution involves the inversion of a
matrix that linearly depends on $w_{ij}$ (see below).

To fix this problem, we also symmetrize the mass-weighted Eddington tensor in
(\ref{eqX13}), which results in the following 
final form of the anisotropic diffusion
equation that we use for our numerical implementation: 
\be \frac{\partial
  N_i}{\partial t}= \sum_j w_{ij} (N_j - N_i), \ee where $w_{ij}$ is now redefined
in a symmetric form: \be w_{ij}\equiv \frac{2\,c\,
  m_{ij}}{\kappa_{ij}\rho_{ij}} \frac{\vec{x}_{ij}^{\rm T} \vec{\tilde h}_{ij}
  \nablavec_i W_{ij}} {\vec{x}_{ij}^2}. \ee

We may also include the sink term, which yields \be \frac{\partial
  N_i}{\partial t}= \sum_j w_{ij} (N_j - N_i) -
c\hat{\kappa}_iN_i\label{eqnRTfinal}. \ee This equation is still symmetric and
can thus also be treated with the CG method, as we explain in more detail in
the next subsection.

\subsection{Time integration of the radiative transfer equation}

As discussed earlier, we use an operator-split approach for the time
integration of the radiation transfer equation, in which we effectively treat
the time integration of the source term and that of the transport through
anisotropic diffusion and the absorption through the sink terms as separate
problems. In fact, we extend the operator-split idea also to the
hydrodynamical evolution of the system, i.e.~we alternate the timestepping of
the diffusion equation with that of the Euler equations that describe the
dynamical evolution of the gas. In the following, we first discuss the time
integration of the diffusion and sink part, which is the most complicated part
in our scheme.

It is well known that explicit time integration schemes of the diffusion
equation becomes easily numerically unstable, unless a very small timestep is
used. To ensure numerical stability, we therefore adopt an implicit method,
namely the simple `backwards Euler' scheme, which provides sufficient accuracy
for the diffusion problem. To advance equation (\ref{eqnRTfinal}) for one
timestep, we therefore want to solve the equation \be
\begin{array}{ll}
 N_i^{n+1} =& N_i^n + \Delta t \tilde s_i m_i + \sum_j \Delta t\,
w_{ij}(N_j^{n+1} - N_i^{n+1})\\
\\ &- \Delta t c \sigma_0 n_{\rm HI}N_i^{n+1} \, ,
\end{array}
\label{eqY1}
\ee where $N_i^{n+1}$ are the new photon numbers at the end of the timestep,
and $N_i^{n}$ are the ones at the beginning of step $n$. The last term in this
equation encodes the photon loss term, which we also integrate implicitly. We
note that the source term, on the other hand, can be simply advanced with an
explicit Euler step.  

The equations (\ref{eqY1}) are in fact a large, sparsely populated linear
system of equations that can be written in the generic form \be \vec{A}\vec{x}
= \vec{b},
\label{Ax}
\ee where $\vec{A}$ is a coefficient matrix, $\vec{b}$ is a vector of known
values and $\vec{x}$ is a vector of unknown values.  For our application, the
components of vector $\vec{b}$ are $b_i = N_i^n + \Delta t \tilde s_i m_i$,
and the matrix elements are given by \be A_{ij} = \delta_{ij} \left( 1 +\sum_k
  \Delta t\, w_{ik} + \Delta t c \sigma_0 n_{\rm HI} \right) - \Delta t\,
w_{ij} \, ,
\label{eqY2}
\ee where the indexes $i$ and $j$ run over all the SPH particles. The solution
vector of the linear problem defines the new photon numbers at the end of the
step, $x_i = N_i^{n+1}$.

There are many different approaches for solving linear systems of equations,
but the huge size of the matrix $\vec{A}$ in our problem (which is equal to
the particle number squared) makes many standard approaches that rely on
storing the whole matrix $\vec{A}$ impractical. Fortunately, the matrix
$\vec{A}$ is only sparsely populated because in each row only approximately
$\sim N_{\rm sph}$ elements are non-zero, those that describe the coupling of
a particle to its neighbors. Sparse systems of this type can often be solved
well with iterative schemes. We use such an iterative scheme in our work, the
conjugate gradient (CG) method.

The conjugate gradient approach applies successive corrections to a trial
solution that is used as a starting point. With every iteration, the solution
becomes better. Since the corrections added in each of the steps are all
orthogonal to each other, the rate of convergence of this method is often
quite high, this is why we think it is a promising iteration scheme for the
problem at hand.  For reference, a derivation of the well-known CG method is given in
Appendix \ref{AppendixB} and Appendix \ref{AppendixC}. However, a prerequisite
for the applicability of the conjugate gradient method is that the matrix
$\vec{A}$ is positive definite and symmetric. The symmetry is evident from our
formulation and the matrix is positive definite since $w_{ij}\ge 0$, and thus
$\sum_{ij}x_iA_{ij}x_j \ge 0 \, \forall \, x_i$.

To find a solution for the new photon number field, we iterate with the CG
scheme until the difference between two successive approximations to $\vec{x}$
has dropped to a small percentage of $|\vec{x}|$.  Note that the expensive
parts in the calculation of one iteration are the matrix-vector
multiplications. For each particle, they reduce to sums over all of its SPH
neighbors, which is equivalent to an ordinary SPH loop, similar in
computational cost to, e.g., the SPH density estimation. Since in our parallel
code some of the SPH neighbours of a particle can be stored on other
processors, this step also involves communication. 

In practice, we start the iteration with the current photon
  distribution for $\vec{x}$, which is usually a fairly good starting point,
  since the expected solution for the photon distribution does not differ
  significantly from the previous state, except in the vicinity of the
  sources. In the absence of a radiation field we set the vector to zero. The
  closer the guessed values for the vector $\vec{x}$ are to the solution, the
  faster the algorithm converges. One could also start from a random photon
  distribution, which will not affect the solution, but will slow down the
  algorithm and is therefore not a desirable choice. The number of iterations
required to reach convergence depends on the condition number of the matrix
$\vec{A}$, where the condition number is defined as $\lambda_{\rm max} /
\lambda_{\rm min}$, the ratio of the largest to the smallest eigenvalue of the
matrix. A large condition number slows down the convergence rate of iterative
solvers of linear systems of equations.  However, the number of required
iterations can be reduced by {\em preconditioning} the matrix $\vec{A}$. For
this purpose we employ the simple Jacobi preconditioner. More specifically, we
modify our matrix equation as follows, \be \vec{C}^{-1}\vec{A}\vec{x} =
\vec{C}^{-1}\vec{b} , \ee where the matrix $\vec{C}$ is the Jacobi
preconditioning matrix. It is defined as \be C_{ij} = A_{ii}\delta_{ij} , \ee
and has the rather simple inverse form \be C_{ij}^{-1} =
\frac{\delta_{ij}}{A_{ii}} .\ee Applying the Jacobi preconditioner to the
matrix $\vec{A}$ basically means to divide the matrix $\vec{A}$ by its
diagonal, which is simple to implement and to parallelize. While the
associated reduction of the condition number improves the convergence speed,
it would be desirable to find still better preconditioners that are more
effective in this respect.

\subsection{Implicit time integration of the chemical network}

The chemical network we need to solve is described by equation
(\ref{eqB}). Even though it contains only a simple time derivative, the
numerical integration can be rather tricky because very large changes in the
number of ionising photons can {\em suddenly} occur from timestep to
timestep. A stable integration with a simple explicit scheme therefore
requires a very small timestep, of the other of $10^{-5}$ the typical
dynamical timestep of the simulation. This would make the solution of the
chemical network very expensive, and it is highly desirable to find another
method that speeds up the computation.

To this end we use an implicit scheme for the evolution of the chemical
network.  Our variant of this approach
discretizes  equation (\ref{eqB}) in time  as follows (here the upper
index $n$ denotes the integration time step): 
\be
\frac{\tilde n_{\rm HII}^{n} - \tilde n_{\rm HII}^{n-1}}{\Delta t} = c \,
\sigma_0 n_{\rm H}\,\tilde n_{\rm  HI}^{n} \tilde n_\gamma^{n}
- \alpha\, n_{\rm H} \, (\tilde n_{\rm HII}^{n})^2, 
\label{chem2}
\ee where we have substituted $\tilde n_{e} = \tilde n_{\rm HII}$. This
quadratic equation is easy to solve. We have found this scheme to be quite
accurate and robust even for moderately large timesteps.

\subsection{Photoheating and cooling with a multi-frequency scheme}\label{heating}

\begin{figure*}
\includegraphics[height=0.4\textwidth]{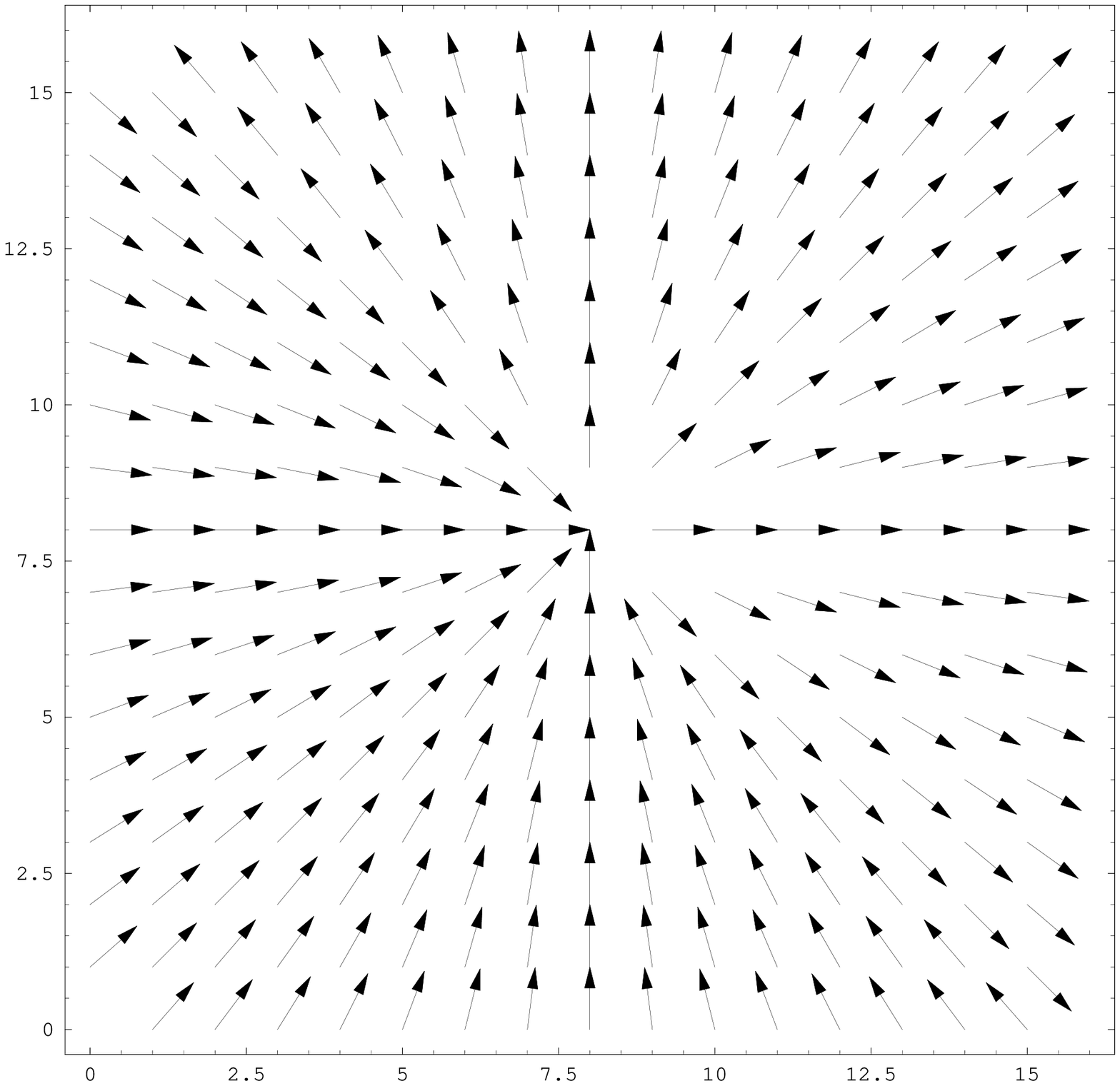}
\includegraphics[height=0.4\textwidth]{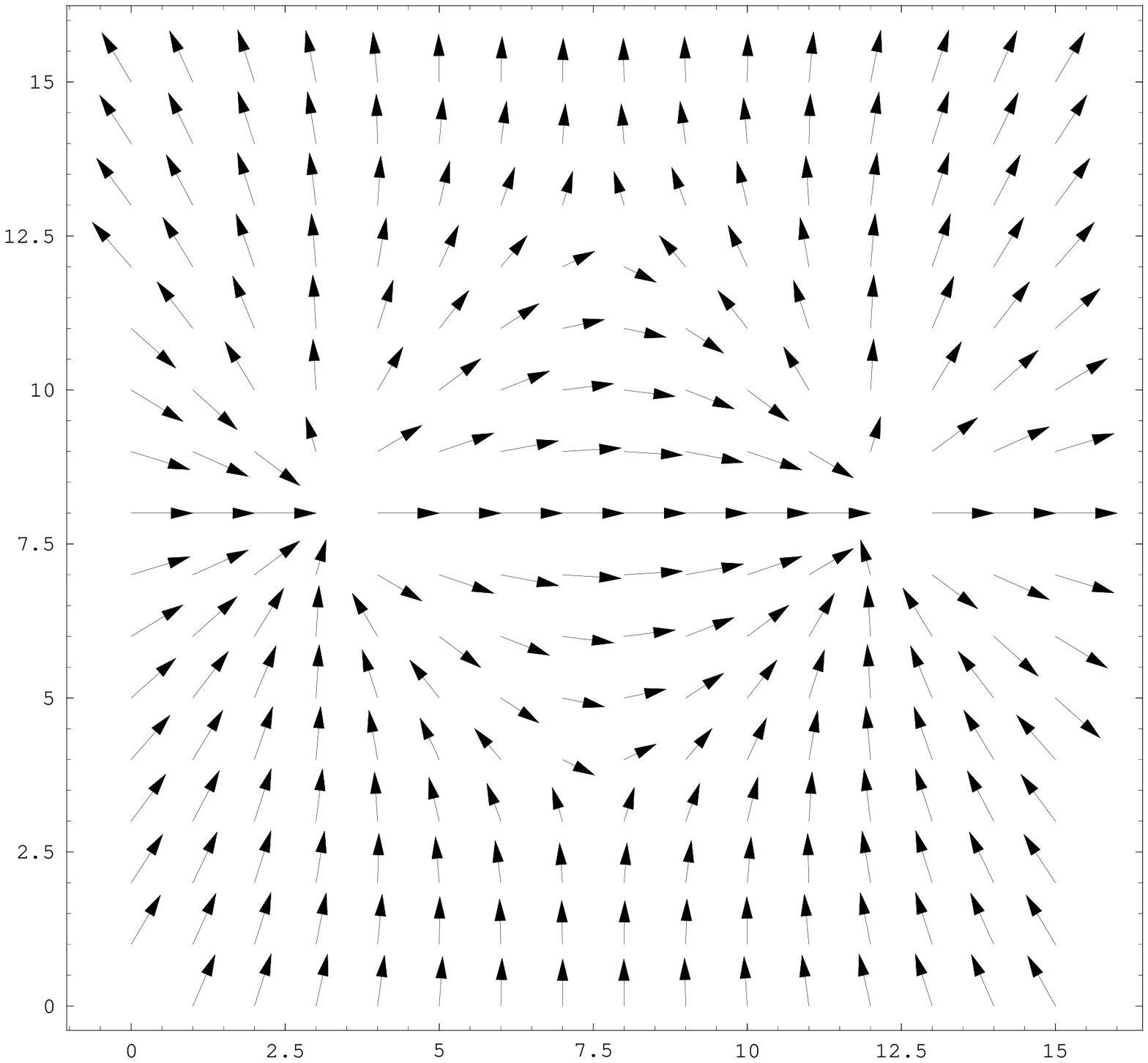}
\caption{Eigenvectors of the Eddington tensor for a single source
  (left panel), and for two sources (right panel), calculated with our
  tree-code extensions of the {\small GADGET-3} code. Both vector
  fields match the expectations. Note that the directions of the
  vectors can be turned $180^{\rm o}$ without affecting the direction
  of the transport of radiation, an effect due to the symmetric nature
  of the Eddington tensor.
\label{fig:edd}}
\end{figure*}

Since our code tracks monochromatic radiation, we need to provide a
multi-frequency description that allows photoheating of the irradiated
gas. Such approaches have already been introduced by \citet{MM} and are often
used in the literature \citep[e.g.][]{AT2007}.  The frequency spectrum of the
source is approximated by a black body spectrum, which allows the use of a
frequency independent photoionisation cross-section $\tilde \sigma$ defined as
\be \tilde \sigma = \int_0^\infty d\nu \frac{4\pi \sigma_\nu J_\nu}{h\nu}
\times \left( \int_0^\infty d\nu \frac{4\pi J_\nu}{h\nu} \right)^{-1}.
\label{eqn:sigma}
\ee
Similarly, a frequency averaged photon energy $\tilde \epsilon$ (above the
hydrogen ionisation energy) can be defined. Thus the hydrogen photoheating rate
from equation (\ref{eqn:realgamma}) can be approximated by
\be
\Gamma = n_{\rm HI}\, c\, \tilde \epsilon \, \tilde \sigma \,
n_\gamma,
\label{eqn:gamma}
\ee
where $c$ is the speed of light and $n_\gamma$ is the number density of
ionising photons.

In order to evolve the temperature of the gas correctly we consider
cooling processes such as recombination cooling,
collisional ionisation cooling, collisional excitation cooling and
Bremsstrahlung cooling. All rates have been adopted from
\citet{Cen1992} and are summarized in Appendix \ref{AppendixA}. As a
combination of heating and cooling terms the
temperature of each particle is then evolved as follows
\be
T^{n+1} = T^n + \Delta\, t \, \frac{2}{3k_{\rm B}n_{\rm H}}\left[
\Gamma - \Lambda \right],
\label{temperature}
\ee
where the upper index $n$ denotes the timestep of the simulation, $k_{\rm B}$ is the
Boltzmann constant, $n_{\rm H}$ is the hydrogen number density,
$\Gamma$ is the heating rate and $\Lambda$ is total cooling
rate, evaluated for the new temperature at the end of the step.

\subsection{The source terms}

As given in equation (\ref{eqA}), the source term in the RT equation is
$\tilde s_\gamma$ in our formulation. This represents the number of photons
emitted per unit time and per hydrogen atom. In our cosmological applications
we usually want to represent this term by the stars that have formed in the
simulation. To more accurately account for the short-lived massive stars, we
can also use star-forming SPH particles as sources, converting their
instantaneous star formation rates into an ionising luminosity.
Alternatively, we can also consider harder sources like Active Galactic Nuclei
(AGN), if they are followed in the simulation. The time integration of the
source function is unproblematic, and can be done with a simple explicit
scheme on the dynamical timestep of the simulation. We only need to assign the
source luminosities in a conservative fashion to the nearest SPH particles,
provided they are not already gas particles anyway.

In some of our test problems considered in Section~\ref{SecResults}, the
source term is prescribed as a certain number $\dot N_\gamma$ of ionising
photons per second, independent of the mass of the source. In this case we
have \be \tilde s_\gamma = \frac{\dot N_\gamma}{n_{\rm H}\, V} = \dot N_\gamma
\frac{m_p}{m} ,\ee where $V = m/\rho$ is a measure of the volume of the SPH
particle with mass $m$ that hosts the source. If we choose to distribute the
source function to more than one particle, we use the `scatter' approach. In
this case the source function has the form \be \tilde s_\gamma = \dot N_\gamma
\sum_j\frac{m_j}{\rho_j}W(r)\frac{m_p}{m_j} ,\ee where the sum is over the SPH
neighbors of the emitting particle.

\subsection{Eddington tensor calculation}\label{ETcalc}

An important quantity in our formulation of the RT problem are the local
Eddington tensors, which we estimate based on an optically thin approximation,
as defined by equations (\ref{edd1}) and (\ref{edd2}).  The $1/r^2$ dependence
of the contribution of each source suggests a calculation method similar to
that of gravity -- via a hierarchical tree algorithm. To this end we extend
the gravitational tree code in {\small GADGET-3} with additional data
structures. For each node of the tree, we also calculate the total luminosity
and the luminosity-weighted centre-of-mass. Depending on whether we consider
the star particles as sources, the star-formation rate of gas particles, or
also black hole particles, this can involve different types of particles.

The individual elements of the Eddington tensor are then computed by walking
the tree in a way exactly analogous to the procedure for calculating
gravitational forces. If a tree node appears under a small enough angle as
seen from the point of interest, all its sources can be represented by the
total luminosity of the node. Otherwise, the tree node is opened, and the
daughter nodes are considered in turn. As a result, we obtain a multipole
approximation to the local radiation pressure tensor, with a typical accuracy
of $\sim 1\%$, provided a similar node-opening threshold value is used as for
collisionless gravitational dynamics. However, since for the Eddington tensor
only the direction of the radiation pressure tensor ultimately matters, the
final accuracy of the Eddington tensor is even better, and more than
sufficient for our purposes. A very important property of this
  calculational method is that its overall speed is essentially independent of
  the number of sources that are present since the calculation is done
  simultaneously with the tree-walk that computes gravity, and involves only
  few additional floating point operations. This is quite different from the
widely employed ray-tracing or Monte-Carlo schemes for radiation transfer,
where the calculation cost may scale linearly with the number of sources.

As an example, Figure \ref{fig:edd} shows the eigenvectors of the Eddington
tensor for two different source configurations, calculated with our modified
version of the {\small GADGET-3} code. The left panel is for a single
source. The vectors point radially outward or inward from the source, as
expected. Note that the directions of the vectors can be turned by $180^{\rm
  o}$ without changing the radiation transport, because the tensor is
symmetric. The panel on the right hand side shows the vector field around two
sources, in the plane of the stars. The field is a dipole in this case and
matches the expectations well.

\begin{figure*}
\includegraphics[width=0.4\textwidth]{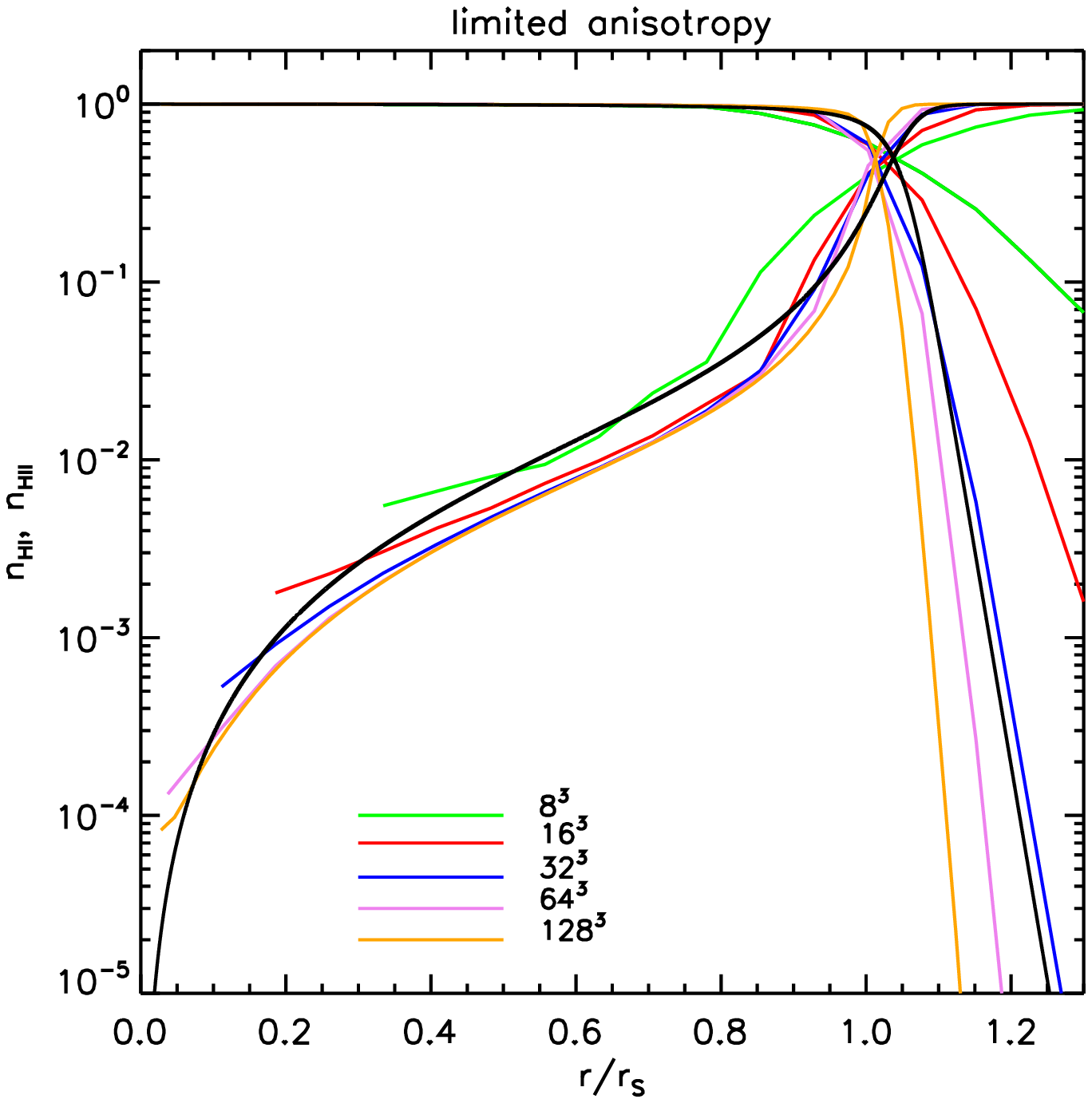}
\includegraphics[width=0.4\textwidth]{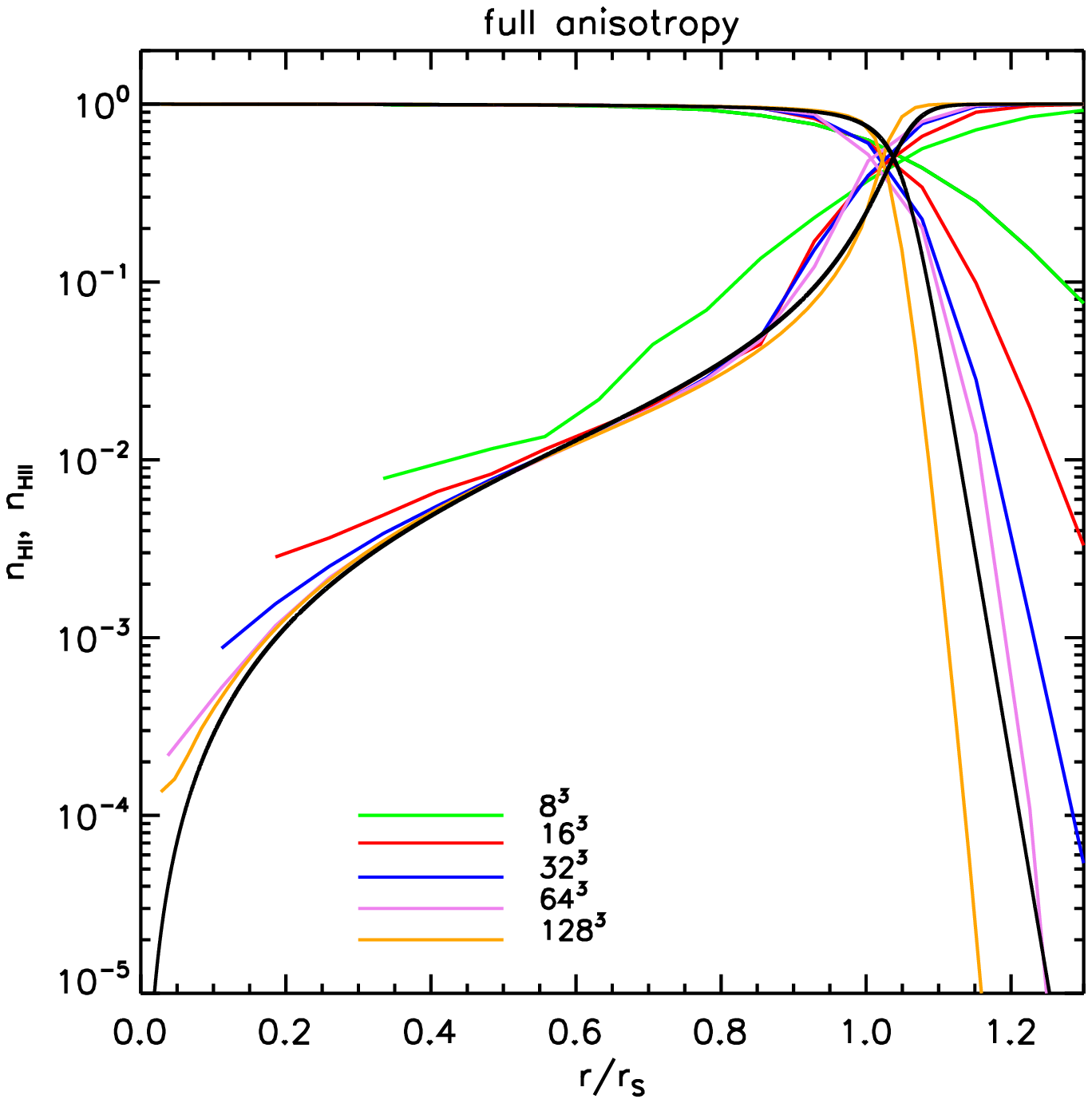}
\caption{Resolution comparison of spherically averaged ionised and
  neutral fractions as a function of radial distance from the source,
  normalized by the Str\"omgren radius $r_{\rm S}$. Results are shown
  for an anisotropy-limited (left panel) and fully anisotropic (right
  panel) Eddington tensor formulation at time $t=500\, {\rm Myr}\,
  \sim \,4 t_{\rm rec}$. The compared resolutions are $8^3$, $16^3$,
  $32^3$, $64^3$ and $128^3$ particles.  The black lines show the
  analytical solution, integrated radially outward from the source as
  in equation (\ref{exact}).  In both formulations the accuracy
  clearly increases with resolution and saturates when the highest
  number of particles is reached. 
  In all cases, the final Str\"omgren
  radius agrees with the analytical predictions. The
  anisotropy-limited Eddington tensor formalism gives very accurate
  predictions for the ionised fraction in the regions outside the
  Str\"omgren radius ($r>r_{\rm S}$), but fails to give a correct
  value for the inner parts of the ionised regions. The fully
  anisotropic Eddington tensor formalism, however, predicts accurate
  values in both regions.
\label{fig:ETall}}
\end{figure*}

\begin{figure*}
\includegraphics[width=0.4\textwidth]{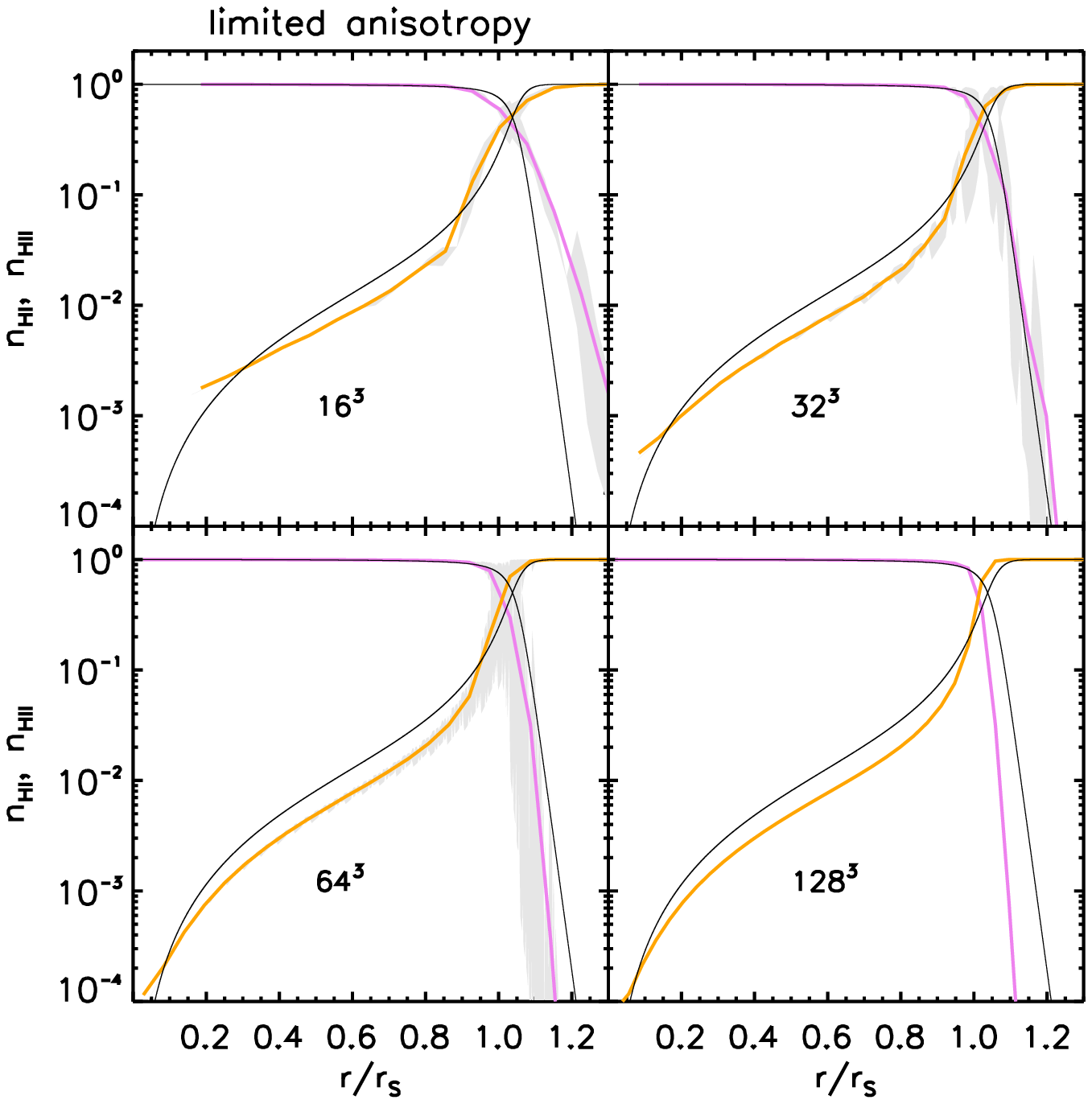}
\includegraphics[width=0.4\textwidth]{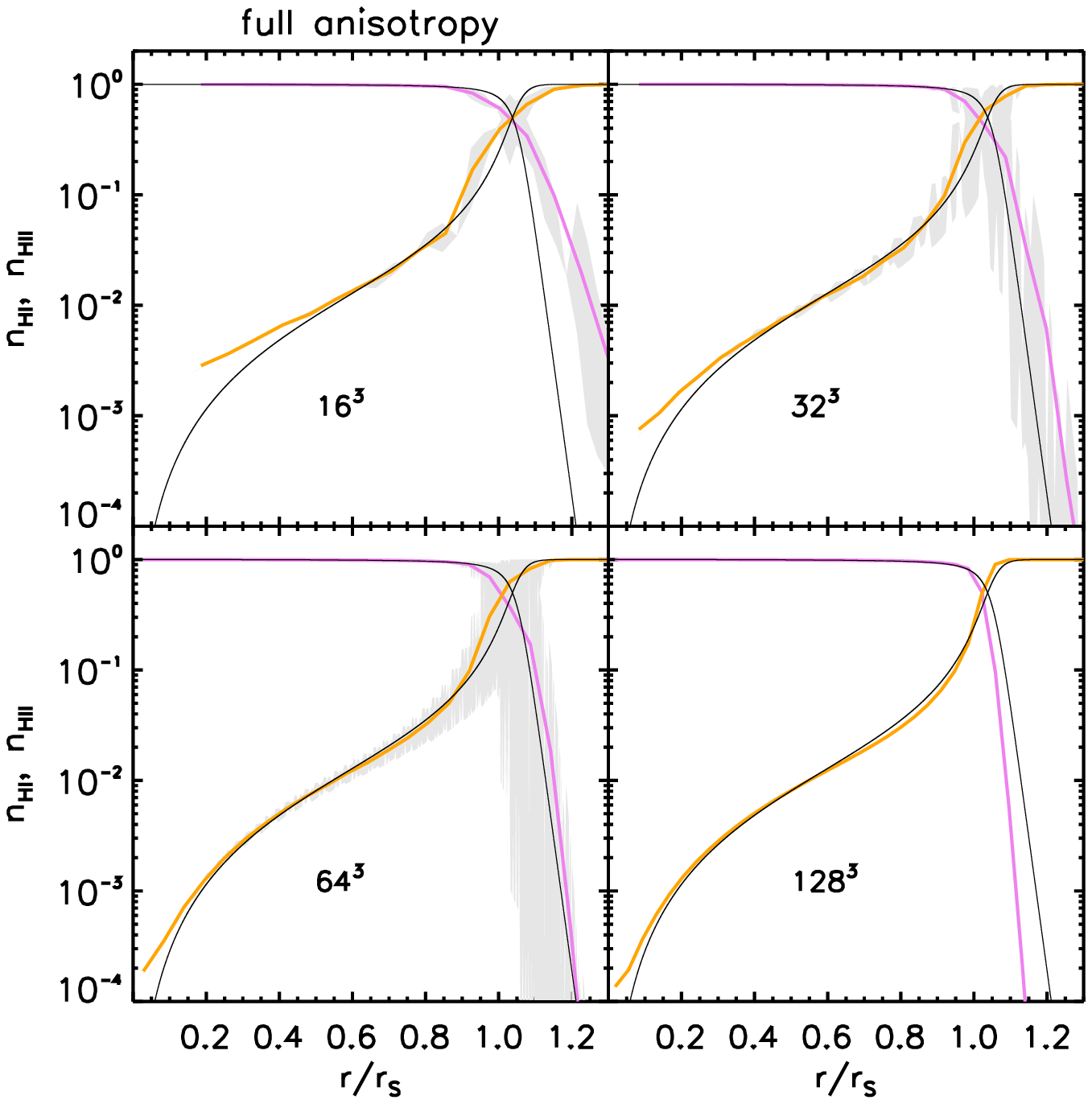}
\caption{Resolution comparison of the scatter (grey areas) of the
  spherically averaged ionised and neutral fraction as function of
  radial distance from the source, normalized by the Str\"omgren
  radius $r_{\rm S}$. Results are shown for an anisotropy-limited
  (left panel) and fully anisotropic (right panel) Eddington tensor
  formulation at time $t=500\, {\rm Myr}\, \sim \,4 t_{\rm rec}$. The
  compared resolutions are $16^3$, $32^3$, $64^3$ and $128^3$
  particles.  The black lines show the analytical solution, integrated
  radially outward from the source as in equation (\ref{exact}). The
  orange lines show the spherically averaged neutral fraction and the
  violet lines the spherically averaged ionised fraction. The range of
  the scatter does not change significantly with resolution, since it
  is due to the intrinsic diffusive nature of SPH and the inaccuracies
  of the SPH density estimate. The highest resolution simulation has
  no scatter due to the high accuracy of the density estimates. The
  scatter in the fully anisotropic Eddington tensor formalism is
  larger due to the larger anisotropy in the diffusion terms of the RT
  equation.
\label{fig:ETmulti}}
\end{figure*}

\begin{figure*}
\includegraphics[width=0.4\textwidth]{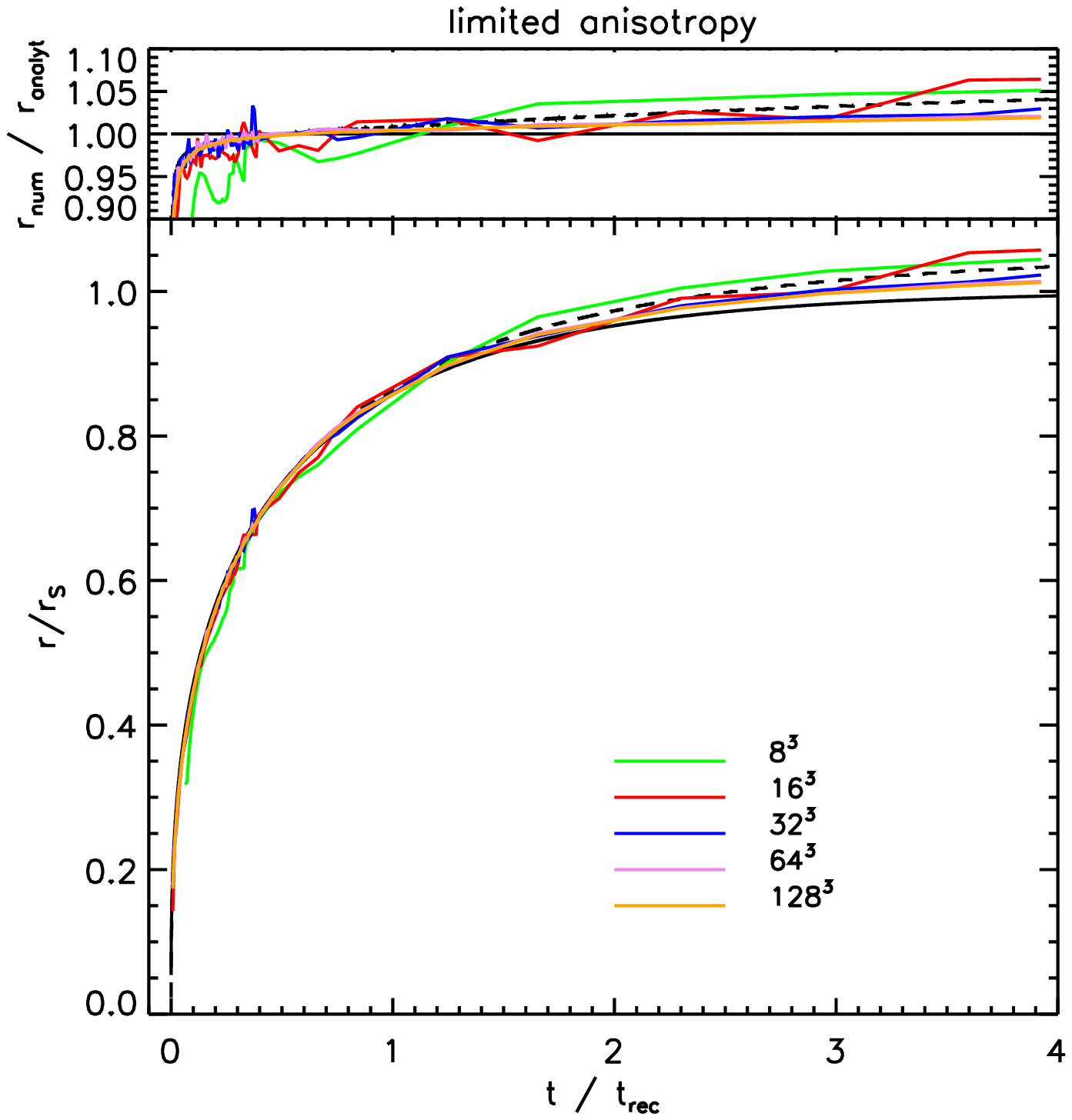}
\includegraphics[width=0.4\textwidth]{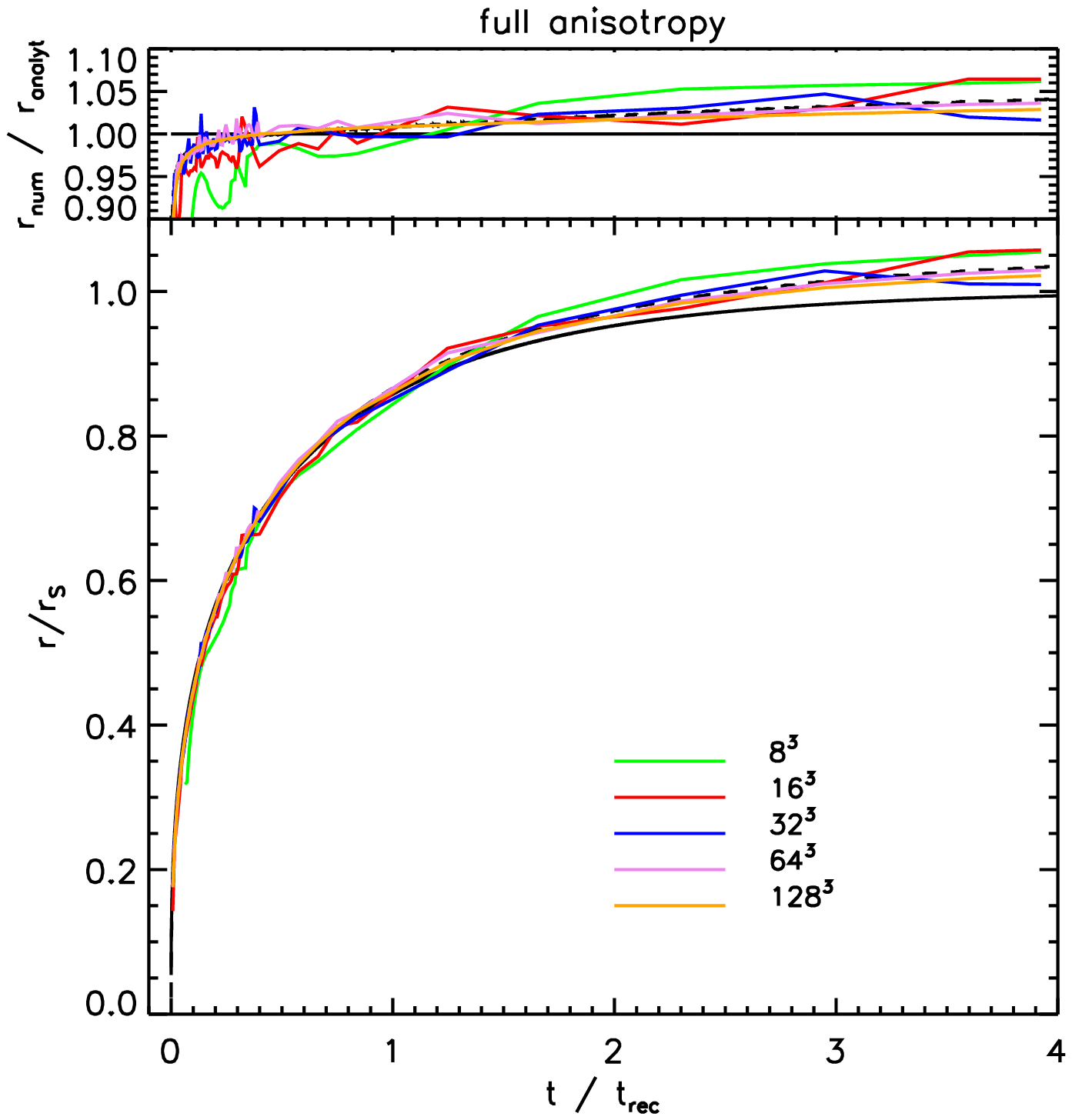}
\caption{I-front expansion for an anisotropy-limited (left panel) and
  a fully anisotropic (right panel) Eddington tensor formulation with
  $16^3$, $32^3$, $64^3$ and $128^3$ particles. The dashed line is the
  exact solution obtained from equation (\ref{exact}) and the solid
  one as obtained from equation (\ref{rI}). Both results agree very
  well with the analytical predictions, but the fully anisotropic
  Eddington tensor formulation shows better results and smaller
  relative error for the I-front position.
\label{fig:Ifrontall}}
\end{figure*}

\subsection{Flux limited diffusion}

If $n_\gamma$ is the density of photons, then the maximum photon flux $f$ that
can occur is limited by the speed of light to $f=c\,n_\gamma$. This physical
limit for the possible photon flux can in principle be violated under certain
conditions when the diffusive approximation to the photon flux, \be f^j = -
\frac{1}{\hat{\kappa}_\nu} \frac{1}{a}\frac{\partial n_\gamma h^{ij}}{\partial
  x^i} , \ee is used. In treatments of radiative transfer in the isotropic
diffusion approximation one therefore often invokes so-called flux limiters that are
designed to enforce the condition \be f \le c n_\gamma\ee by damping the
estimated flux when needed.

In our anisotropic diffusion treatment, we have also implemented an optional
limiter that serves the same purpose.  We observe the maximum flux constraint
with the help of a parameter $R$, which is a function of the gradient of the
photon density \be R \equiv \frac{|\nabla n_\gamma|}{\kappa n_\gamma} .\ee We
then define a flux limiter of the form \be\lambda(R) =
\frac{1+0.1R}{1+0.1R+0.01R^2}\, ,\ee where $\lambda(R)\rightarrow 0 \mbox{ as
} R \rightarrow \infty$. The detailed form of the analytic
  expression used for the flux limiter is arbitrary as long as it ensures a
  smooth transition between the two limiting states. We have chosen this form
  since it is widely used in other numerical RT codes, for example, a similar
  version is used by \citet{Whitehouse2004}. The flux limiter $\lambda$ is
then introduced into the diffusion part of equation (\ref{eqA}) as follows:
\be \frac{\partial \tilde n_\gamma}{\partial t}= c \frac{\partial}{\partial
  x_j}\left( \frac{\lambda}{\kappa}\frac{\partial \tilde n_\gamma
    h^{ij}}{\partial x_i} \right) - c\,\kappa\, \tilde n_\gamma + \tilde
s_\gamma .  \ee

However, we note that superluminal propagation of photons is usually not a
problem in the ionisation problems we are interested in. Here the speed of the
ionisation fronts is not limited by the speed of light, but rather by the
luminosity of the sources and the density of the absorbing
medium. Nevertheless, the flux limiter can also be usefully employed as a
means to control the behavior of the ionisation front (I-front) propagation in
dense media. Due to the specific dependence of the $R$ parameter, the photon
propagation can effectively be limited in high-density regions, where the
intensity gradient becomes very large.

\subsection{Notes on the performance of the code}

An important consideration in the development of RT codes is the
  calculational cost of the implemented scheme, as this determines whether the
  method is sufficiently fast to allow a coupling with hydrodynamic simulation
  codes.  It is difficult to make general statements about the computational
  cost of our new RT scheme as this depends strongly on the particular
  physical problem that it is applied to.  Arguably of most interest is a
  comparison of the speed of the method to other parts of the simulation code
  when applied to the problem of cosmological structure formation with a
  self-consistent treatment of cosmic reionisation.  We here report
  approximate numbers for the speed of our new method in this situation, based
  on work in preparation that studies this problem.

First we note that in our implementation, when gravity is also integrated,
  the computation of the Eddington tensor incurs negligible costs, since it
  is done together with the gravity. Therefore, we find that in our
  implementation the increase of the computational cost with respect to the
  other relevant code modules (gravity, SPH density and SPH hydrodynamical
  forces) is primarily a function of the number of iterations required at each
  timestep to construct the implicit solution of the anisotropic diffusion
  equation. One iteration is approximately as costly as one SPH hydro
  computation, and the average number of iterations required ranges from
  typically 10 up to 200 in the most extreme cases. However, the RT equation
  is integrated only on a relatively coarse timestep, whereas the
  hydrodynamics is done also for many more smaller sub-steps. This reduces the
  effective cost of the whole RT calculation to several times the total cost
  of the hydrodynamics calculations. In practice, we measure a slow-down of the
  simulation code by a factor of order of 2 to 5 when the radiative transfer
  is included. While this is non-negligible, it does not seriously impact the
  ability to carry out large cosmological simulations. We also note that
  further optimizations in the radiative transfer algorithms, perhaps through
  an improved pre-conditioner, may reduce the cost of the RT calculation
  in the future.

\section{Testing the code}  \label{SecResults}

In the following we present several basic tests for our new radiative
transfer code. Where possible, we compare our results with analytical
solutions or with results from other simulations \citep[from the RT
  code comparison study by][]{Iliev2006b}. In section \ref{isothermal}
we discuss our results for the classic test of the isothermal
expansion of an ionised sphere in a homogeneous and static density
field. In section \ref{time} we study the effects of different
timesteps on the accuracy of our simulations, while in section
\ref{two} we evolve two nearby sources with interacting ionised
spheres. Then in section \ref{nonisothermal} we repeat the single
ionised sphere expansion test, but this time allowing the temperature
to evolve. In section \ref{shadow} we present a shadowing test, where
a dense clump is placed in the way of a plane-parallel I-front.
Finally, in section \ref{cosmo} we evolve the radiation transport in a
static cosmological density field.

\subsection{Isothermal ionised sphere expansion}\label{isothermal}

\begin{figure*}
\includegraphics[width=0.9\textwidth]{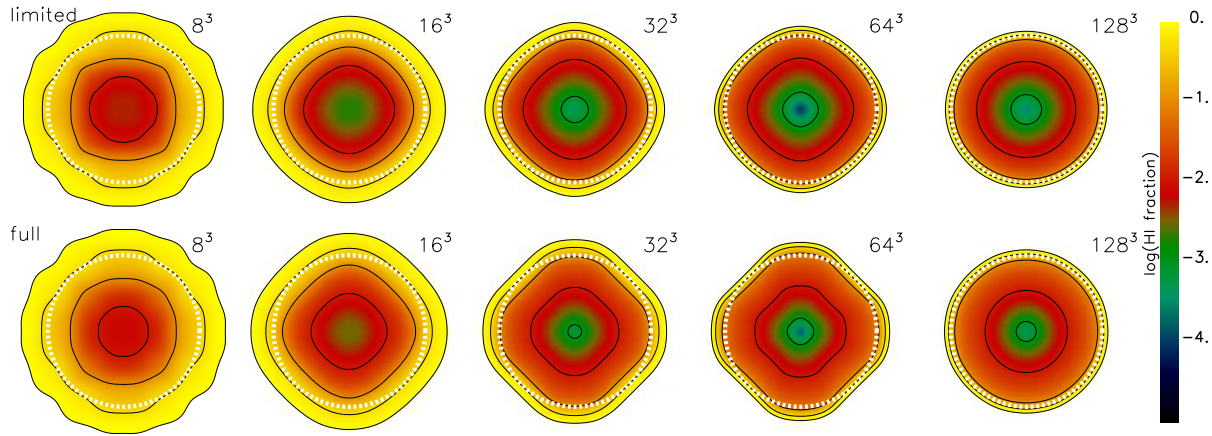}
\caption{Slices of the neutral fraction through the ionised sphere at
  the position of the source. The upper row shows the results from the
  anisotropy-limited Eddington tensor formulation and the lower row
  from the fully anisotropic Eddington tensor formalism. Five
  different resolutions are compared: $8^3,\, 16^3,\, 32^3,\, 64^2 \,
  {\rm and} \, 128^3$ particles. The contours mark neutral fractions
  of $\tilde n_{\rm HI}=0.9,\, 0.5,\, 0.1,\, 0.01\,{\rm and}\,
  0.001$. The white circles give the radius of the Str\"omgren sphere.
  The geometrical distribution of the SPH particles affects the shape
  of the ionised sphere. The spheres are elongated in the $x$- and
  $y$-direction of the Cartesian grid, where the particle spacing is
  smaller and the SPH-kernel interpolant leads to slightly different
  couplings as in diagonal directions. This effect is stronger for the
  anisotropic Eddington tensor formulation.\label{fig:surface}}
\end{figure*}

The expansion of an I-front in a static, homogeneous and isothermal
gas is the only problem in radiation hydrodynamics that has a known
analytical solution and is therefore the most widely used test for RT
codes. A monochromatic source emits steadily $\dot{N}_\gamma$ photons
with energy $h\nu=13.6\, \rm eV$ per second into an initially neutral
medium with constant gas density $n_{\rm H}$. Then the Str\"omgren
radius, at which the ionised sphere around the source has reached its
maximum radius, is defined as \be r_{\rm S} =
\left(\frac{3\dot{N}_\gamma}{4\pi\alpha_{\rm B} n_{\rm
    H}^2}\right)^{1/3} ,
\label{rs1}
\ee where $\alpha_{\rm B}$ is the recombination coefficient. This
radius is obtained by balancing the number of emitted photons by the
number of photons lost due to recombinations along a given line of
sight. If we assume that the I-front is infinitely thin, i.e.~there is
a discontinuity in the ionisation fraction, then the expansion of the
Str\"omgren radius can be solved analytically and the I-front radius
$r_{\rm I}$ is given by \be r_{\rm I} = r_{\rm S}[1-\exp(-t/t_{\rm
    rec})]^{1/3},
\label{rI}
\ee
where
\be
t_{\rm rec} = \frac{1}{n_{\rm H}\alpha_{\rm B}}
\ee
is the recombination time and $\alpha_{\rm B}$ is the recombination coefficient.

The neutral and ionised fraction as a function of radius of the stable
Str\"omgren sphere can be calculated analytically
\citep[e.g.][]{OF2006} from the equation
\be
\frac{\tilde n_{\rm HI}(r)}{4\pi r^2}\int {\rm d}\nu\,\dot{N_\gamma}(\nu)\,
e^{-\tau_\nu(r)} \, \sigma_\nu\,=\,\tilde n_{\rm HII}^2(r) \, n_{\rm H}
\, \alpha_{\rm B},
\label{eqn:fraction}
\ee
where $\tilde n_{\rm HI}$ is the neutral fraction, $\tilde n_{\rm
HII}$ is the ionised fraction and
\be
\tau_\nu(r)\,=\, n_{\rm H} \, \sigma_\nu \, \int_0^r  {\rm d}r' \, \tilde
n_{\rm HI}(r').
\ee
Moreover, considering spherical symmetry and a point source we can
solve analytically for the photon density radial profile $n_\gamma(r)$,
yielding
\be
n_\gamma(r) = \frac{1}{c}\frac{\dot N_\gamma}{4\pi r^2}\,{\rm exp} \left\{
-\int_0^r\kappa(r')\, {\rm d}r'  \right\}.
\label{exact}
\ee
From this we obtain ionised fraction profiles
$\tilde n_{\rm HII}(r)$ for the whole evolution time.

The Str\"omgren radius obtained by direct integration of
equation (\ref{eqn:fraction}) differs from the one obtained from
equation (\ref{rs1}) because it does not approximate the ionised
region as a sphere with constant, but with varying ionised
fraction. We compare our results with both analytical solutions.

For definiteness, we follow the expansion of the ionised sphere around a
source that emits $\dot N_\gamma = 5 \times 10^{48} \, \rm photons \,
s^{-1}$. The surrounding hydrogen number density is $n_{\rm H} = 10^{-3} \,
\rm cm^{-3}$ at a temperature of $T = 10^4 \, \rm K$. At this temperature, the
case B recombination coefficient is $\alpha_{\rm B} = 2.59 \times 10^{-13} \,
\rm cm^3 \, s^{-1}$. Given these parameters, the recombination time is $t_{\rm
  rec} = 125.127 \, \rm Myr$ and the expected Str\"omgren radius is $r_{\rm S}
= 5.38 \, \rm kpc$. We impose periodic boundary conditions in order
  to make sure that the density field is effectively infinite and uniform. We
  note that this does not affect our RT calculation since the Eddington tensor
  is computed non-periodically.

We present results from the fully-anisotropic and the
anisotropy-limited Eddington tensor formalism simulations by comparing
first the final state of the ionised sphere and then the evolution of
the I-front. Figure \ref{fig:ETall} shows the spherically averaged
ionised and neutral fraction as a function of radial distance from the
source, normalized by the Str\"omgren radius $r_{\rm S}$, at time
$t=500\, {\rm Myr}\, \sim 4\, t_{\rm rec}$. In this case we compare
the resolution effects on the accuracy of our numerical predictions by
using simulations with $8^3$, $16^3$, $32^3$, $64^3$ and $128^3$
particles, corresponding to mean spatial resolutions of $2.0$ kpc,
$1.0$ kpc, $0.5$ kpc and $0.25$ kpc. In both formalisms the accuracy
increases with resolution and the profiles converge for the higher
resolutions. We also note that the anisotropy-limited Eddington tensor
formalism gives very accurate predictions for the ionised fraction in
the regions outside the Str\"omgren radius ($r>r_{\rm S}$), but fails
to give correct values in the inner parts of the ionised sphere. The
fully anisotropic Eddington tensor formalism, on the other hand,
predicts the ionisation state in both regions quite accurately.

We compare also the scatter of the ionised and neutral fraction
profiles. Figure \ref{fig:ETmulti} shows the scatter (gray areas) of
the spherically averaged ionised and neutral fraction profiles for
four different resolutions ($16^3$, $32^3$, $64^3$ and $128^3$
particles) at the end of the I-front expansion. All results agree well
with the analytical radius of the Str\"omgren sphere. The range of the
scatter does not change significantly with resolution, since it is due
to the intrinsic diffusive nature of SPH and the inaccuracies of the
SPH density estimate. This means that we obtain a density scatter of
about 0.01\% and thus introduce fluctuations in the gas density, which
result in fluctuations in the hydrogen densities of the SPH particles
and thus of the ionised and neutral fractions.  The SPH density
scatter in the $128^3$ particle simulation is zero (thanks to the use
of a Cartesian grid -- but note that in real-world dynamical
applications some density scatter is unavoidable) and thus there is no
scatter in the ionised and neutral fractions.  Moreover, the scatter
in the fully anisotropic Eddington tensor formalism simulations is
larger than in the other formalism simulations due to the larger
retained anisotropy in the diffusion term of the RT equation.

The evolution of the I-front expansion is shown in Figure
\ref{fig:Ifrontall}, comparing the two formalisms at different
resolutions.  The results from both formalisms agree very well with
the analytical predictions and the accuracy increases with
resolution. The fully anisotropic Eddington tensor formalism
simulations show better results and smaller relative error for the
I-front position. In both formalisms the error stays within 5\% of the
analytical solution and traces the analytical result obtained by
direct integration of equation (\ref{exact}).

However, we note that the geometrical distribution of the SPH
particles we used in our simulations introduces slight deviations from
perfect sphericity into the shape of the ionised region. This reflects
the Cartesian grid of particles used for these tests, an effect that
can be clearly seen in the shapes of the ionised regions displayed in
Figure \ref{fig:surface}. The spheres are elongated in the $x$- and
$y$-directions of the Cartesian grid, where the particle spacing is
smaller and the SPH-kernel interpolant weights the nearest neighbours
slightly differently than in off-axis directions. This discreteness
effect is stronger for the anisotropic Eddington tensor formulation.

Considering all our results that compare the limited and fully
anisotropic Eddington tensor formalisms, we use in all our further
tests and simulations only the fully anisotropic formulation because
it shows more accurate results. It turns out that this formulation is
also robust, i.e.~it does not show stability problems due to its
`anti-diffusive' terms when used in conjunction with an implicit
solver, at least we have not experienced such problems in our test
calculations.

\subsubsection{Timestep comparison}\label{time}

\bfig
\includegraphics[width=0.9\columnwidth]{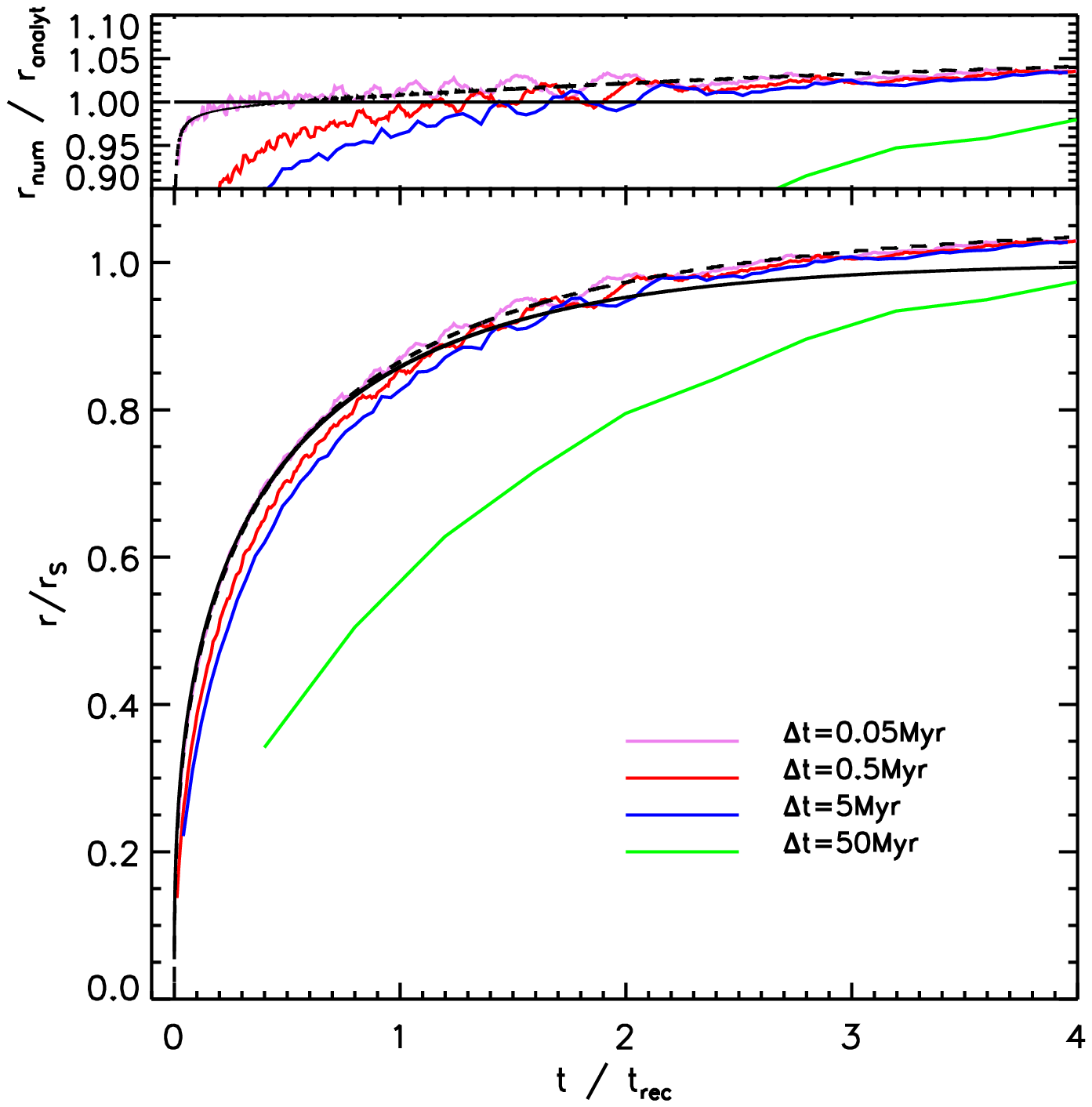}
\caption{I-front expansion as a function of the recombination
  time. Results are shown for four different simulation timesteps:
  $\Delta t=0.05,\, 0.5,\, 5\, {\rm and}\, 50\, \rm Myr$. The dashed
  line is the exact solution obtained from equation (\ref{exact}) and
  the solid one is obtained from equation (\ref{rI}). The smallest
  timestep simulation agrees very well with the analytical
  solution. As the timestep increases, the results in the early phases
  of the expansion become more inaccurate. However, after about two
  recombination times, the I-front radius catches up with the
  analytical solution. The simulation with timestep $\Delta t=50\, \rm
  Myr$ is very inaccurate, but in the end of the expansion the I-front
  radius is still within 5\% of the analytical solution.
\label{fig:Ifronttime}}
\efig

In order to test the accuracy of our RT scheme we perform simulations
of $64^3$ particle resolution with different fixed timesteps:
$\Delta\, t\, =\, 0.05,\, 0.5,\, 5\, {\rm and}\, 50\, {\rm
  Myr}$. Applying the von Neumann stability criterion for an explicit
integration of the diffusion part of our RT equation (\ref{eqnRTMom}),
we find a bound of the timestep equal to \be \Delta t \le
\frac{1}{2}\frac{\kappa (\Delta x)^2}{c},\ee where $\Delta x$ is the
mean spatial resolution, $c$ is the speed of light and $\kappa =
n_{\rm HI}\, \sigma$ is the absorption coefficient.  At the I-front
the assumed neutral fraction is $\tilde n_{\rm HI} = 0.5$ and thus the
absorption coefficient is $\kappa = 3.15 \times 10^{-21} \, \rm
cm^{-1}$, resulting in an upper limit for the timestep $\Delta t \le
10^{-3}\, \rm Myr$. However, this limit on the timestep is only a
reference point for our results. Because we use an implicit scheme
that is stable for all timestep sizes, we are fortunately not bound by
this timestep limit and can in principle use much larger timesteps,
subject only to the condition that the final accuracy reached is still
acceptable.

We compare results for our simulations with the four different
timestep sizes in Figure \ref{fig:Ifronttime}. The smallest timestep
that we use is ten times as big as the analytical upper limit for an
explicit scheme, yet its results agree perfectly with the analytical
solution. For the other simulations the timestep sizes increase
progressively by factors of $10$, and the numerical results start to
deviate from the analytical calculation. The changes are largest in
the early phase of the I-front expansion. As the source ``suddenly''
switches on, very small time steps are needed in order to achieve good
accuracy in the beginning, where the gradients in the photon density
are very large. But later, after a couple of recombination times, the
numerical results approach the analytical solution even for coarse
timesteps and follow it until the expansion of the I-front ends. The
$\Delta t = 50 \, \rm Myr$ simulation is initially very inaccurate,
but note that its Str\"omgren radius is still within 5\% of the
analytical result. Therefore, our method manages to essentially
produce correct Str\"omgren radii of the ionised spheres for all
considered timesteps.

\subsubsection{Two nearby sources}\label{two}

\bfig
\includegraphics[width=0.8\columnwidth]{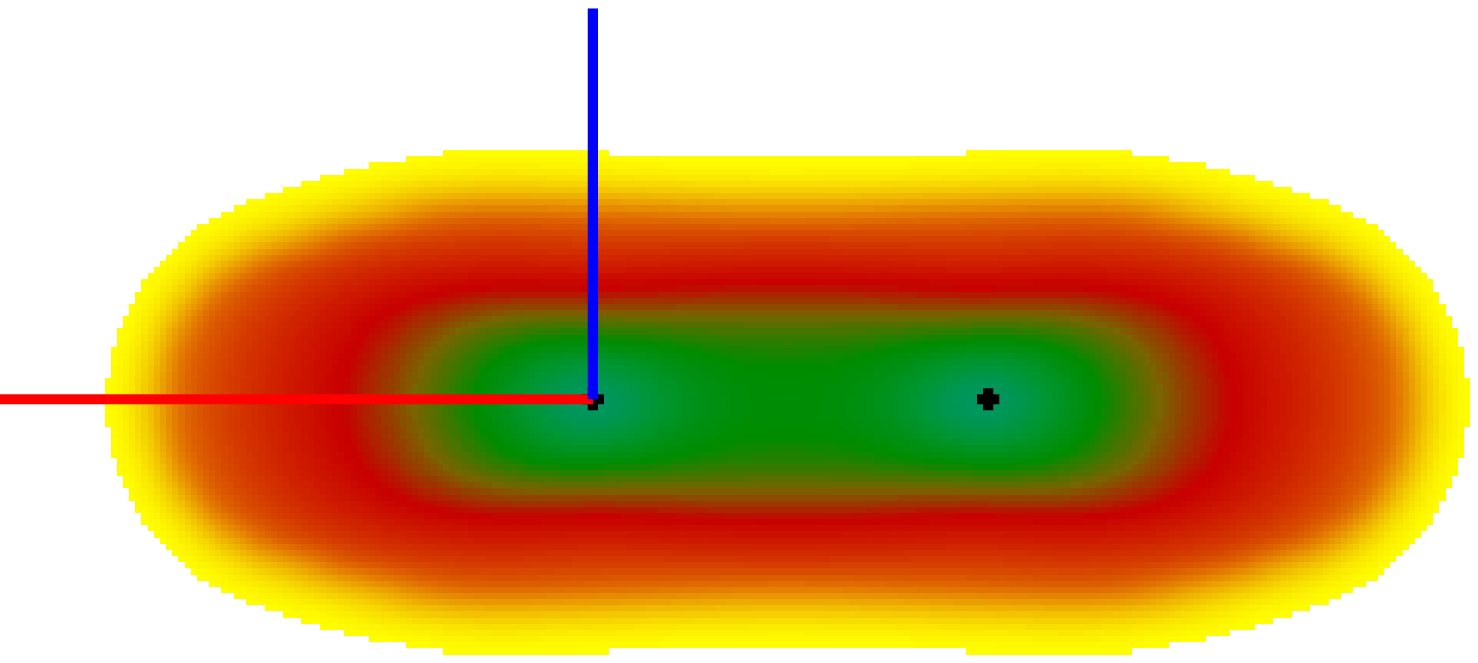}
\includegraphics[width=0.1\columnwidth]{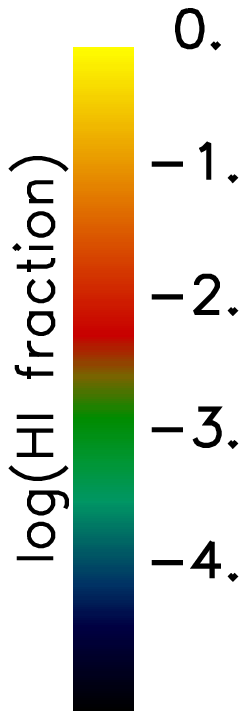}
\caption{Ionised fraction in a plane of two equally luminous
  sources. The positions of the sources ($8\,\rm kpc$ apart) are
  marked with black crosses. The snapshot is taken at time $t=500\,
  \rm Myr$, when the expansion of the ionised regions has stopped, and
  the region where the Str\"omgren spheres overlap is approximately
  $3\,\rm kpc$ wide. There is a clear elongation along the axis
  connecting the two sources, as also described by
  \citet{GA2001}. Figure \ref{fig:ratio} shows the evolution of the
  I-front along the aligned (red) and perpendicular (blue) directions
  with respect to the axis through the sources.
\label{fig:two}}
\efig

\bfig
\includegraphics[width=0.9\columnwidth]{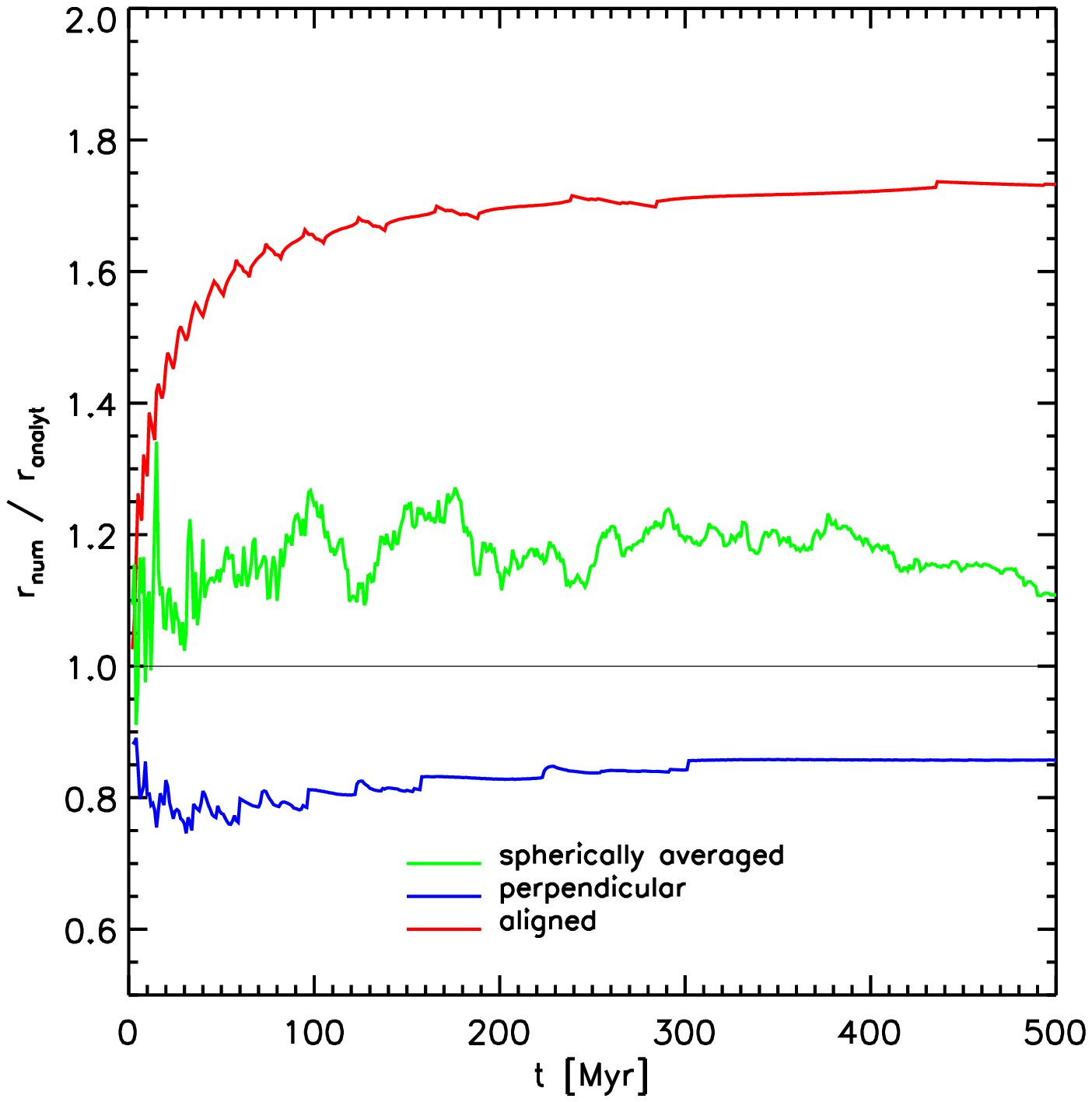}
\caption{I-front expansion around a source in a double system,
  normalized by the analytical solution from equation (\ref{rI}), as a
  function of time. The red line is the I-front in the direction
  aligned with the two sources and the blue line is for the orthogonal
  direction. The green line shows the spherically averaged I-front
  position. As observed by \citet{GA2001}, the ionised spheres are
  elongated along the axis of the two sources and compressed in the
  perpendicular direction. The spherically averaged I-front position
  is within 20\% of the analytical expectation.
\label{fig:ratio}}
\efig

In our next test we follow the expansion of ionised regions around two
nearby sources, where we expect to see inaccuracies due to the
optically thin assumption used for estimating the Eddington tensors.
Both sources emit $\do N_\gamma=5 \times 10^{48} \, \rm photons \,
s^{-1}$ and are $8\rm \, kpc$ away from each other. The number density
of the surrounding static and uniform hydrogen gas is $n_{\rm H} =
10^{-3} \, \rm cm^{-3}$, at a temperature of $T = 10^4 \, \rm K$. From
tests conducted by \citet{GA2001} we expect that the ionised regions
are not spherical, but rather elongated along the axis through the
sources. This effect results from the calculation of the Eddington
tensor, whose values along the symmetry axis are estimated high and
boost the diffusion in this direction, while reducing it in the
perpendicular direction.

In Figure~\ref{fig:two}, we show the neutral fraction for this test in
a slice in the plane of the sources, taken at time $t=500\, \rm Myr$
when the expansion of the regions has stopped. The expected elongated
shape of the ionised regions is clearly visible.  In Figure
\ref{fig:ratio} we show the time evolution of three characteristic
radii of the expanding ionised regions: one radius is measured in a
direction aligned with the axis through the sources, one is measured
perpendicular to it, and the third is a spherically averaged
radius. We note that we do not expect the
radii to match the analytical prediction from equation (\ref{rI})
exactly since the approximations there are valid only for a single
ionised region expansion, but we here use the obtained value to
compare the expansion of the ionised regions around two nearby
sources. As expected, the aligned radius is always larger than the
analytical result, while the perpendicular radius is smaller. However,
the spherically averaged radius of the expanding region stays within
20\% of the analytical value.  We conclude that the optically thin approximation to estimate
Eddington tensors can in certain situations introduce errors in the
shapes of ionised bubbles, but these errors should be quite moderate
or negligible in situations where the Eddington tensor is dominated by
a bright nearby source, which is probably generic in many scenarios
for cosmological reionisation.

\subsection{Ionised sphere expansion with varying temperature
\label{nonisothermal}}

\begin{figure*}
\includegraphics[width=0.4\textwidth]{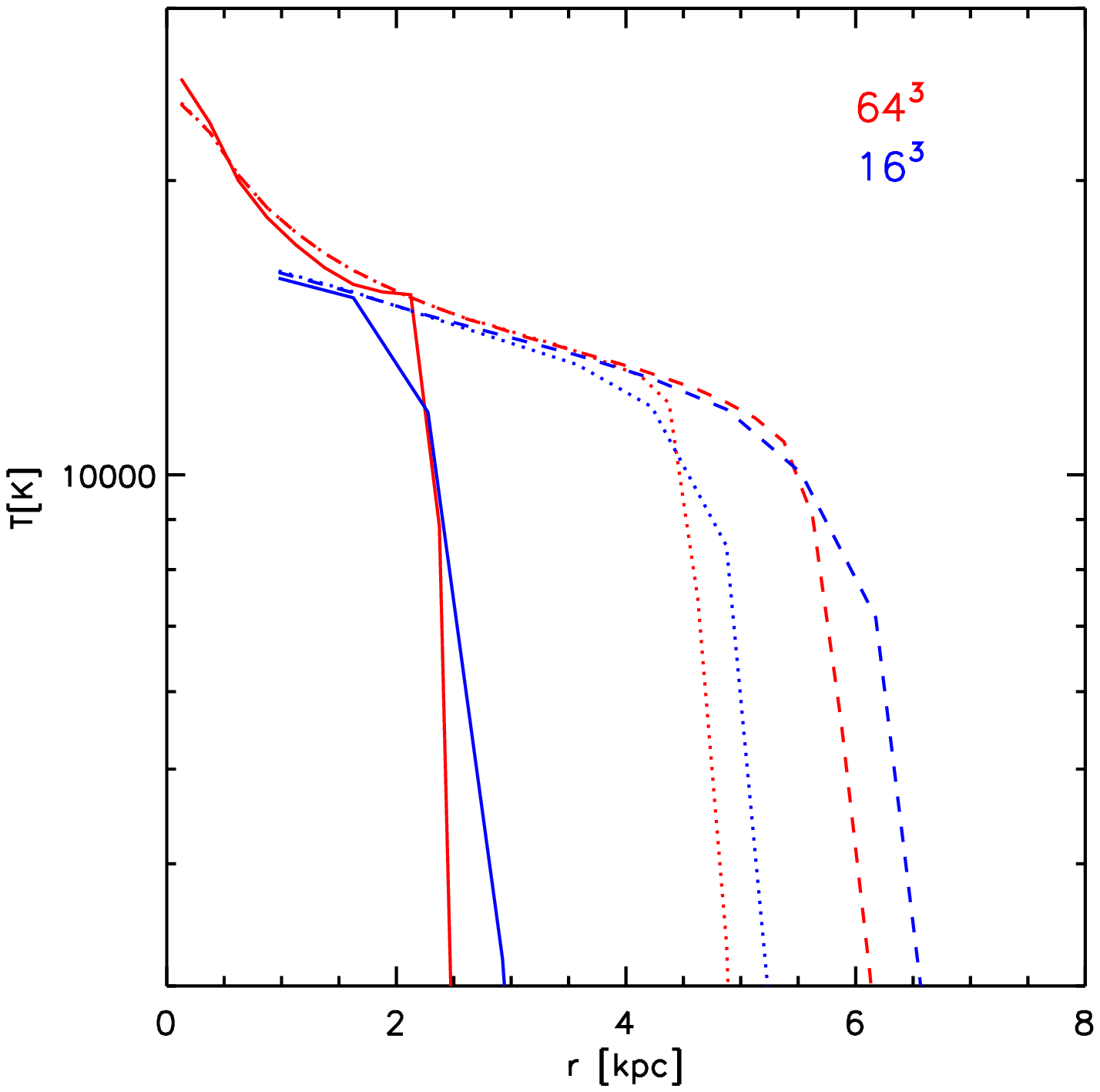}
\includegraphics[width=0.4\textwidth]{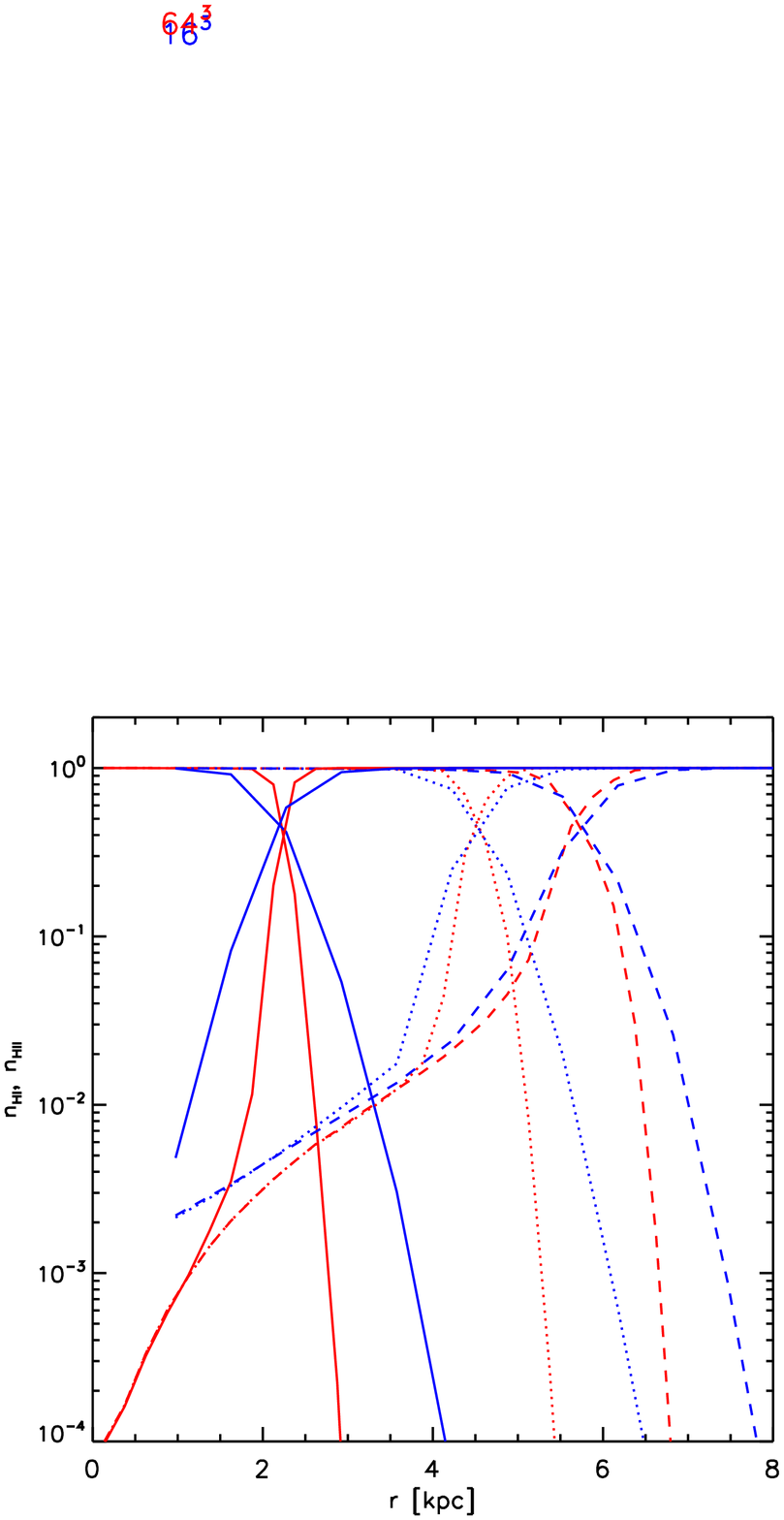}
\caption{Time evolution of temperature (left panel) and ionised and
  neutral fraction (right panel) profiles of an expanding ionised
  sphere for two different resolutions: $16^3$ (blue) and $64^3$ (red)
  particles. Results at times $t=10,\, 100\, {\rm and}\, 500 \, \rm
  Myr$ are shown in solid, dotted and dashed lines, respectively. The
  temperature profiles inside the ionised sphere converge for both
  resolutions at all times. The position of the I-front agrees well
  for both resolutions at all times.
\label{fig:nf}}
\end{figure*}

\begin{figure*}
\includegraphics[width=0.3\textwidth]{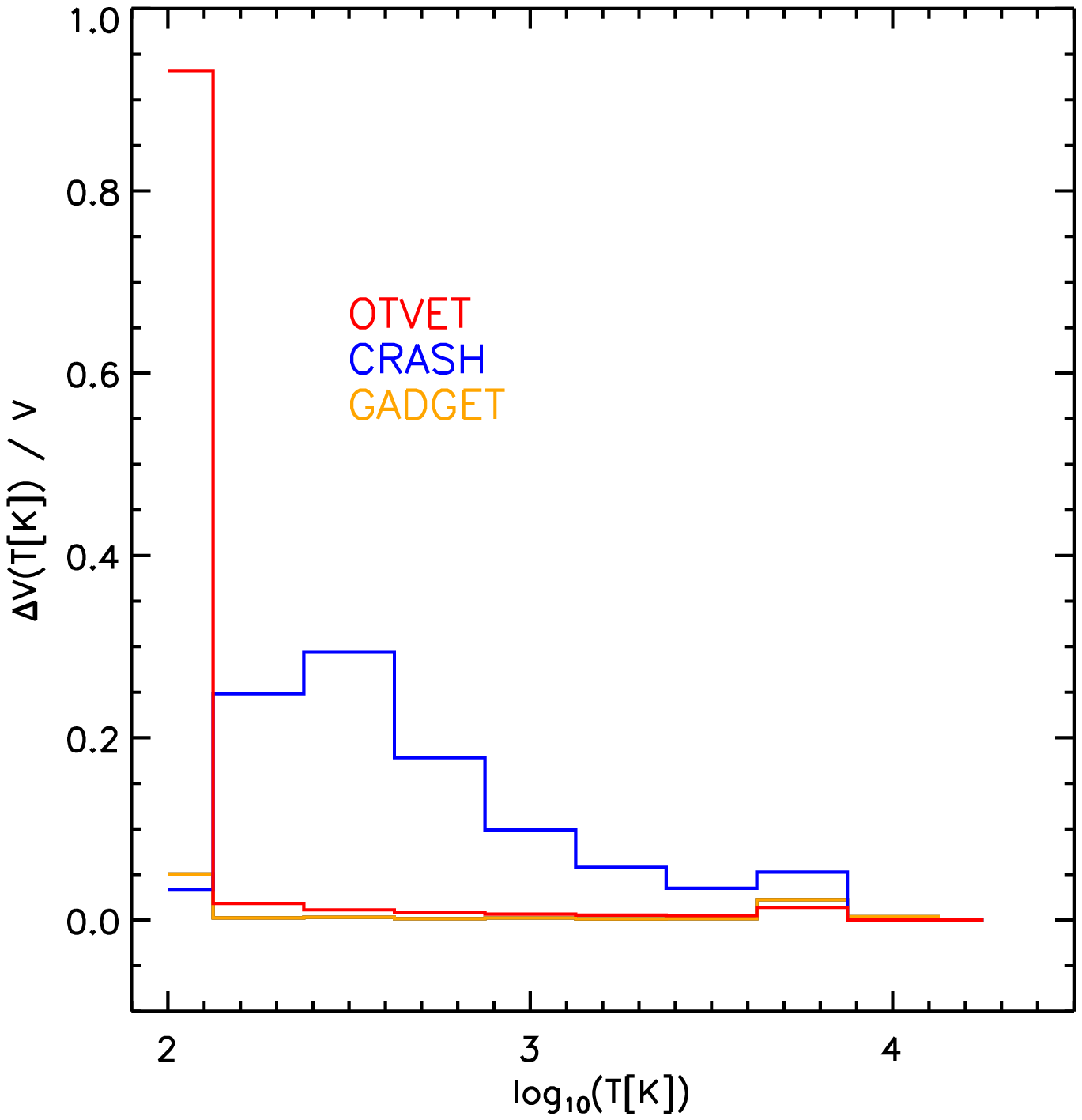}
\includegraphics[width=0.3\textwidth]{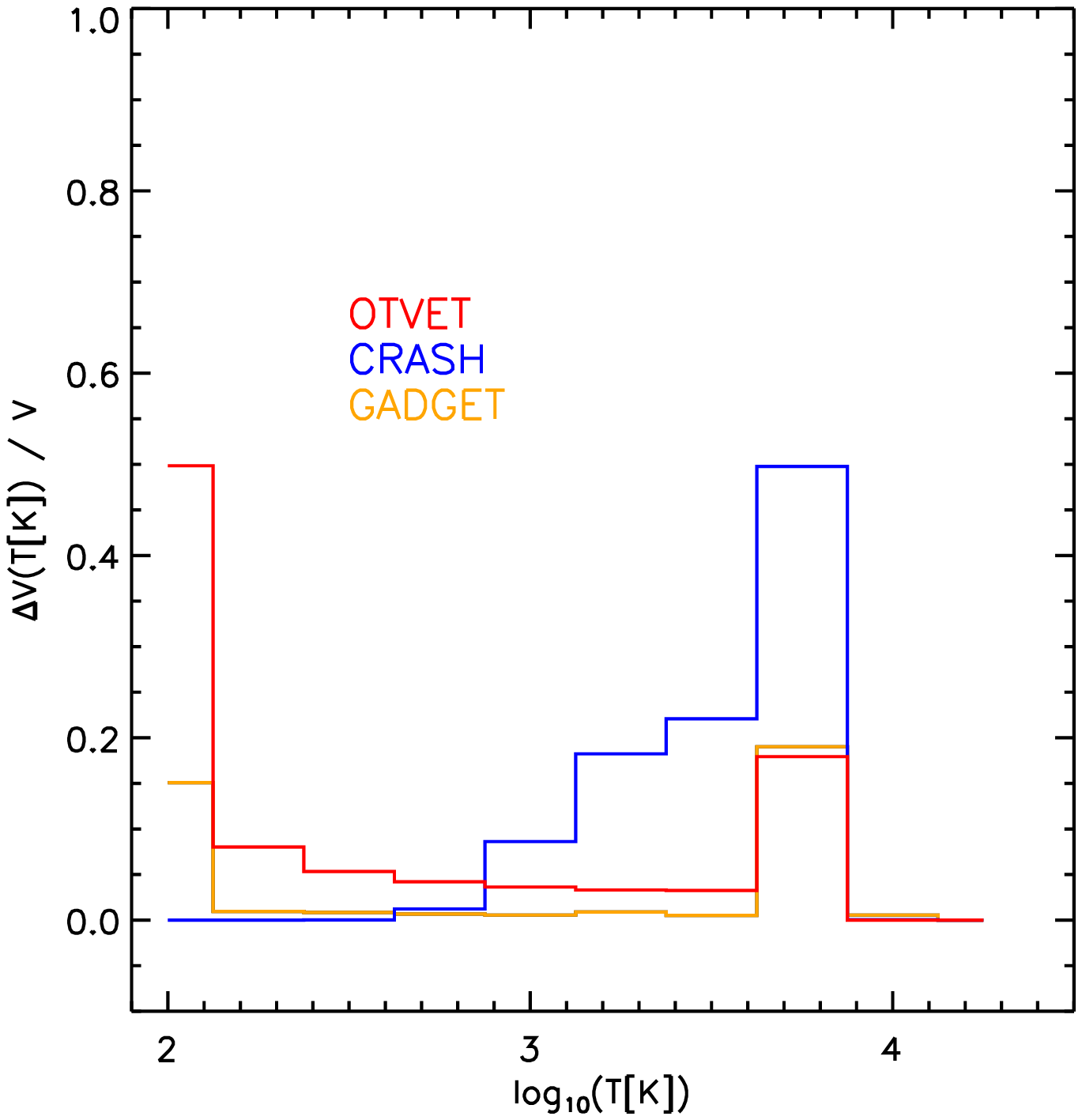}
\includegraphics[width=0.3\textwidth]{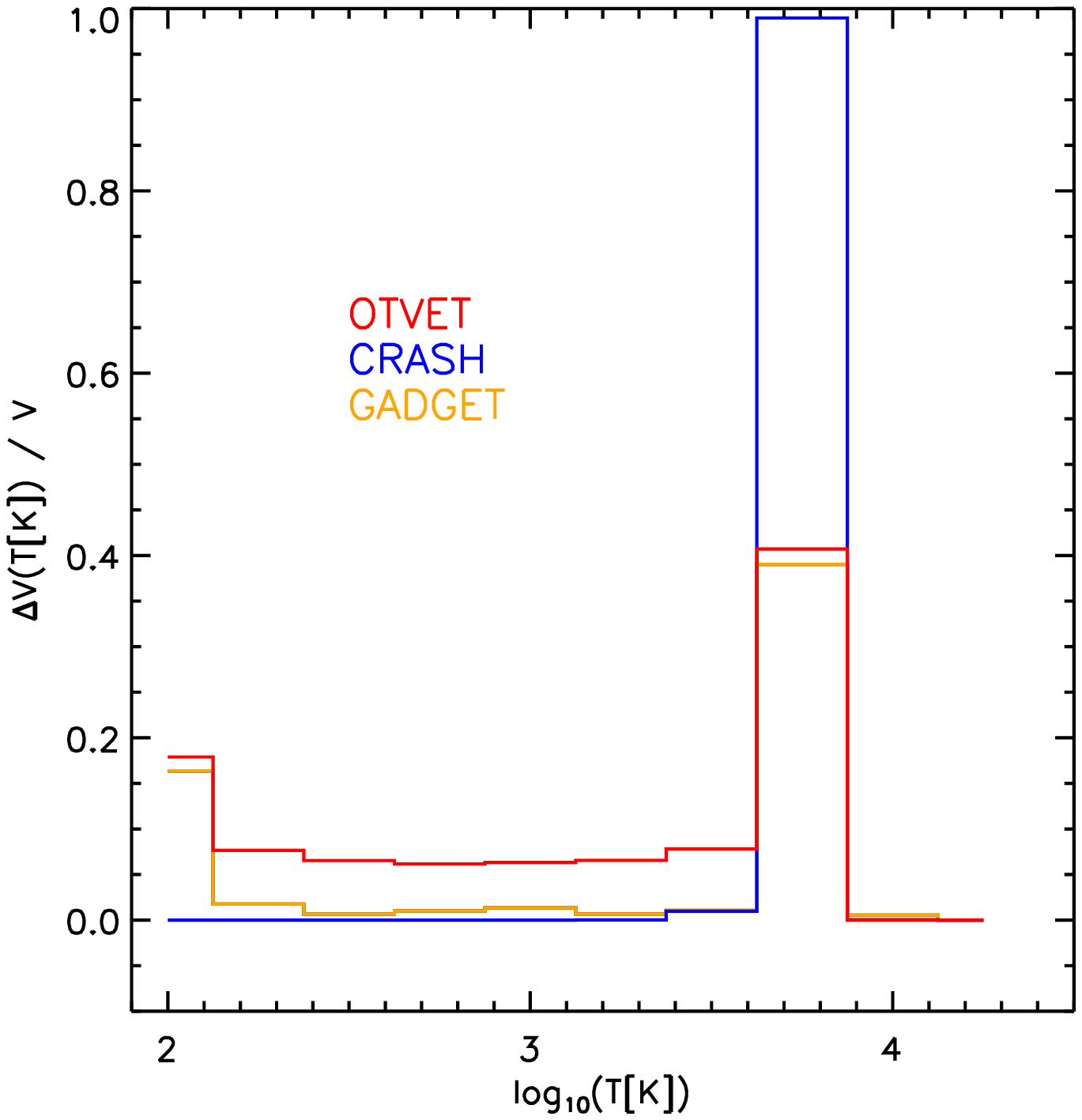}
\caption{Volume fraction of the temperature for a comparison between
  {\small CRASH}, {\small OTVET} and {\small GADGET} at three
  different times $t=10,\, 100\, {\rm and}\, 500 \, \rm Myr$ (left to
  right). The results of {\small GADGET} and {\small OTVET} are
  comparable, but {\small CRASH} has a harder spectral distribution
  and employs multiple frequency bins
  and thus gives slightly different results. Differences between
  {\small GADGET} and {\small OTVET} may also be due to different
  resolutions ({\small OTVET} has $0.05\, \rm kpc$, while for {\small
    GADGET} $\sim 0.25\, \rm kpc$).
\label{fig:histtemp}}
\end{figure*}

\begin{figure*}
\includegraphics[width=0.3\textwidth]{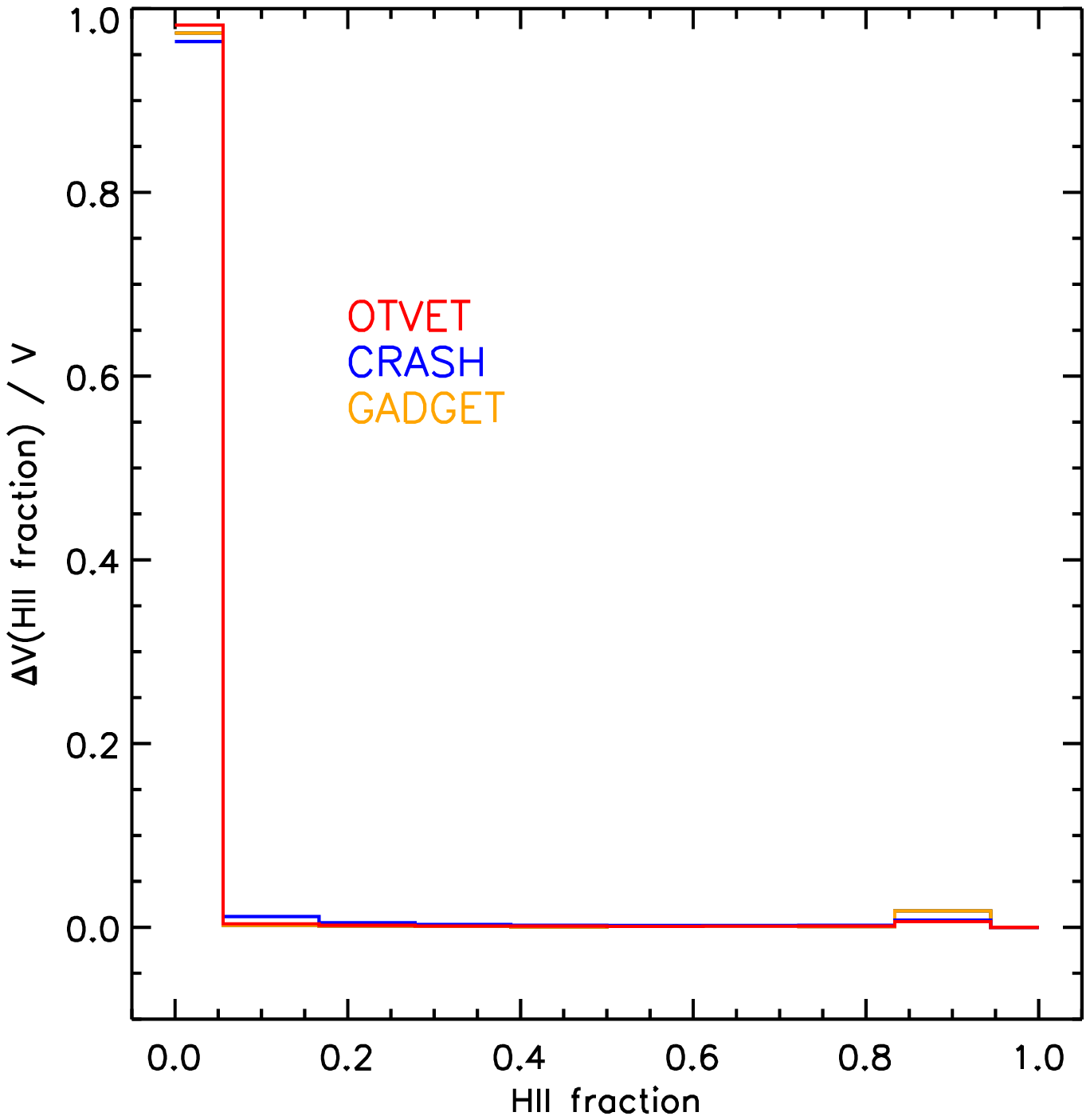}
\includegraphics[width=0.3\textwidth]{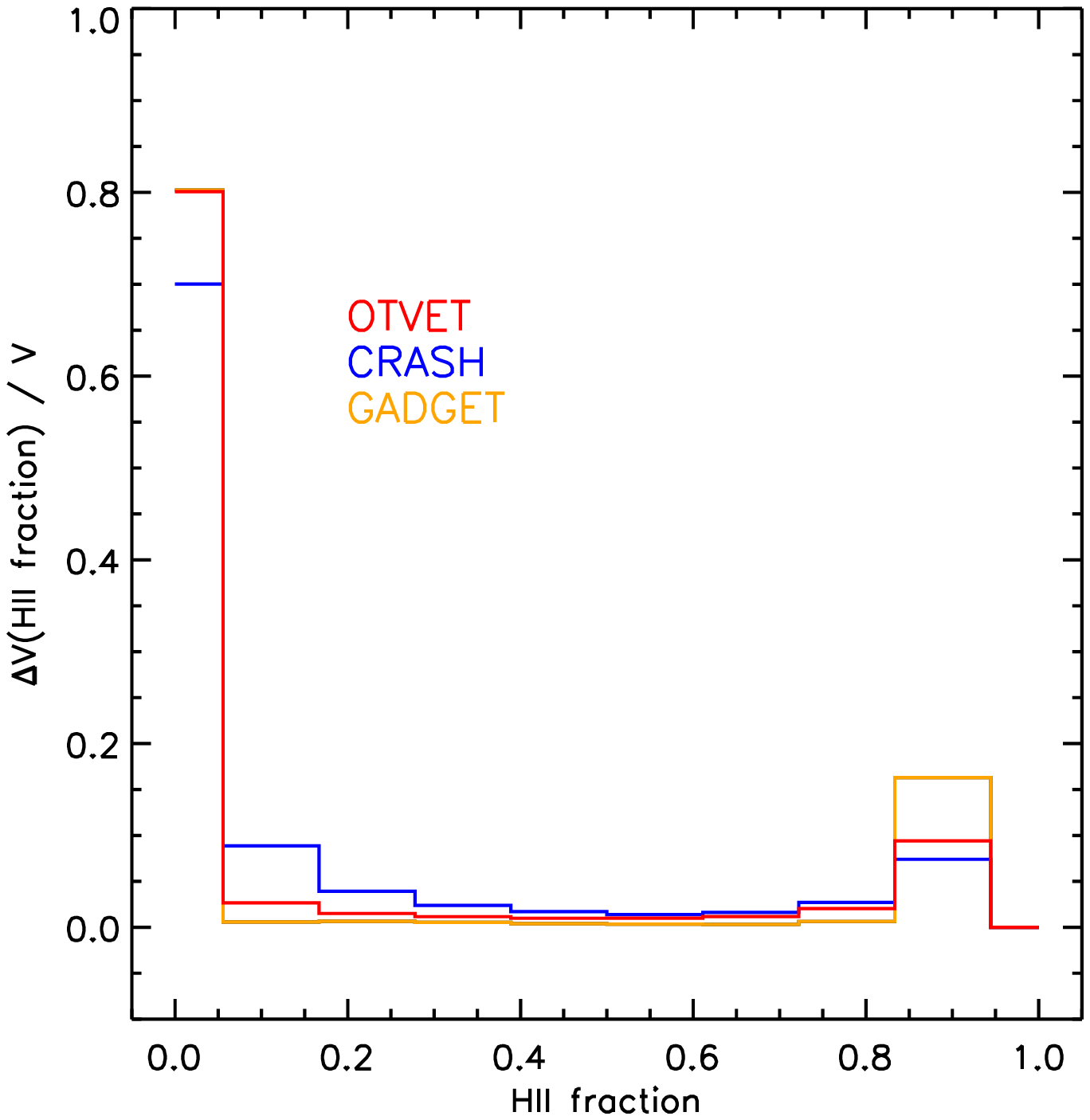}
\includegraphics[width=0.3\textwidth]{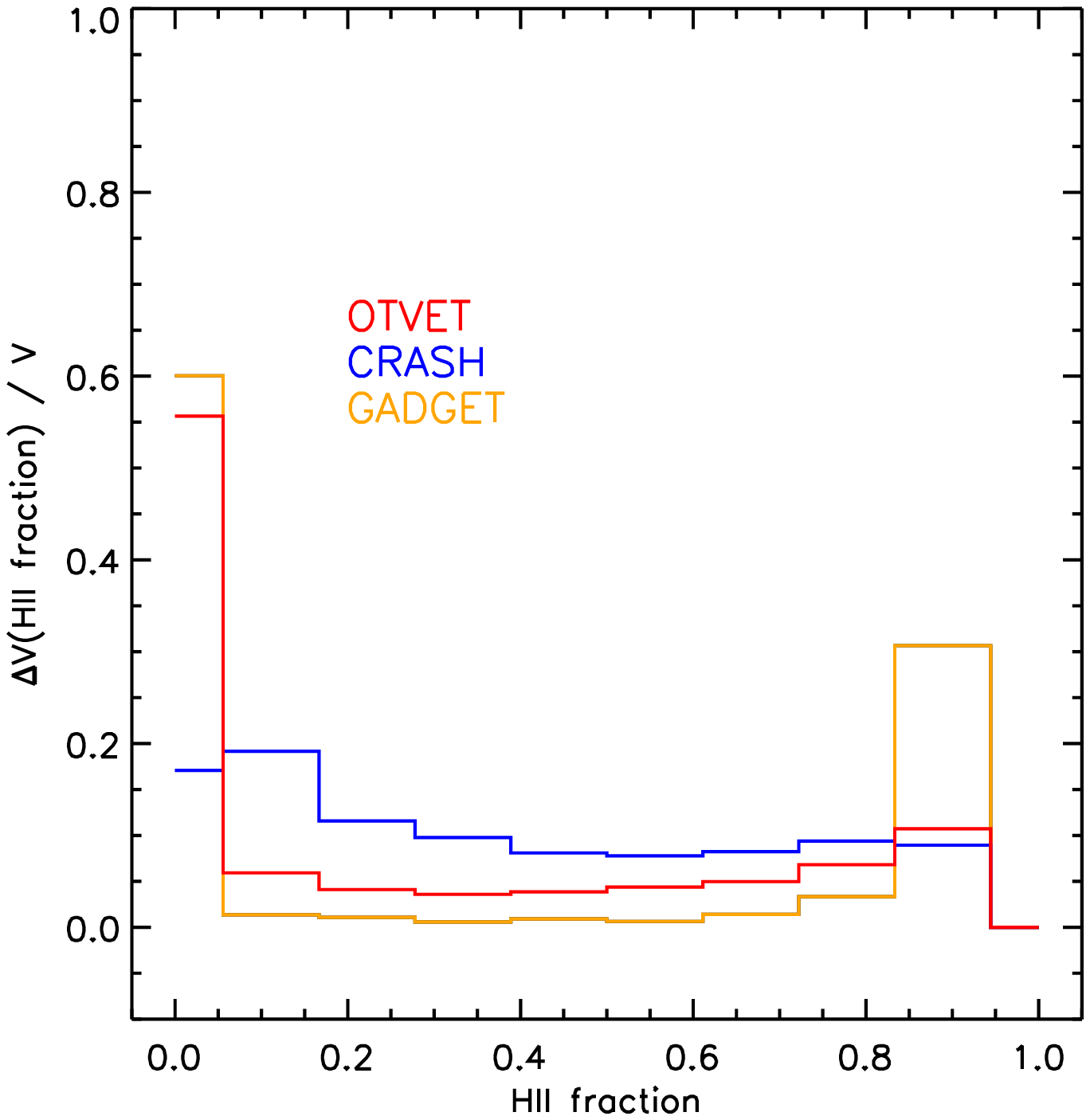}
\caption{Volume fraction of the ionised fraction in a comparison
  between {\small CRASH}, {\small OTVET} and {\small GADGET} for three
  different times $t=10,\, 100\, {\rm and}\, 500 \, \rm Myr$ (left to
  right). With increasing time, {\small GADGET} produces larger
  strongly ionised volume and smaller intermediately ionised volume
  than the other codes, mirrored in the larger gradient of the ionised
  fraction that it produces. These deviations are likely due to  different
  source spectra treatments, temperature structures of the ionised
  spheres, and  different numerical resolutions.
\label{fig:histion}}
\end{figure*}

In this test we use the same setup as in section \ref{isothermal}, but
we initialize the gas temperature with $T_0 = 10^2 \, \rm K$ and let
it evolve due to the coupling to the radiation field. We furthermore
approximate the recombination coefficient $\alpha_{\rm B}$ with \be
\alpha_{\rm B}(T) = 2.59\times 10^{-13}\left( \frac{T}{10^4\, {\rm
    K}}\right)^{-0.7} \, {\rm cm^3 \, s^{-1}} ,\ee and assume a black
body spectrum of temperature $10^5\, \rm K$ for the source, setting
the parameters from equation (\ref{eqn:gamma}) to $\tilde \sigma =
1.63 \times 10^{-18} \, \rm cm^2$ and $\tilde \epsilon = 29.65 \, \rm
eV$. Then we evolve equation (\ref{temperature}) for every particle,
at every time step, considering photoheating, recombination cooling,
collisional ionisation cooling, collisional excitation cooling and
Bremsstrahlung cooling. In this way we test a realistic expansion of
an ionised sphere around a single source.

We test our scheme with two different resolutions: $16^3\, {\rm and}\,
64^4$ particles. The time evolution of the temperature profile in both
tests is shown in the left panel of Figure \ref{fig:nf}.  The
temperature close to the source rises to approximately $2 \times 10^4
\, \rm K$ and then drops down to $10^2\, \rm K$ outside the ionised
region.  From left to right, the profiles are shown at three different
times: $t=10, \, 100, \, {\rm and}\, 500\, \rm Myr$. The results from
the simulations converge inside the ionised sphere. Outside the
ionised region the low resolution simulation produces, as expected, a
smaller slope of the temperature drop further from the ionised sphere
radius. In the right panel of Figure \ref{fig:nf}, we compare the
neutral and ionised fractions at the same times.  Both resolutions
converge at the radius of the ionised sphere.

In order to verify our results we compare them with results obtained
with the codes {\small CRASH} \citep{MFC2003} and {\small OTVET}
\citep{GA2001}, as summarized by \citet{Iliev2006b} in the RT Code
Comparison Project. In Figure \ref{fig:histtemp}, we show a comparison
of the volume fraction of the temperature and in Figure
\ref{fig:histion} we present a comparison of the ionised volume
fractions, at three different times: $t=10, \, 100,\, {\rm and}\,
500\, \rm Myr$. The temperature volume fractions are all somewhat
different due to the very different heating schemes employed. The
results of {\small GADGET} and {\small OTVET} are comparable, but
{\small CRASH} uses a harder spectral distribution and
multiple frequency bins and thus
gives different results. With increasing time {\small GADGET}
produces a larger strongly ionised volume, and smaller intermediately
ionised volume fraction than the other codes, which is also mirrored
in the larger gradient of the ionised fraction that it produces. These
differences are due to the different treatments of the source spectra
that the codes employ, which are in general difficult to
compare. Deviations might also be due to the different temperature
structures of the ionised spheres and the different resolutions of the
codes ({\rm OTVET} and {\small CRASH} used $0.05\, \rm kpc$, {\small
  GADGET} only $\sim 0.25\, \rm kpc$).

\subsection{Shadowing by a dense clump}\label{shadow}

\begin{figure*}
\includegraphics[width=0.3\textwidth]{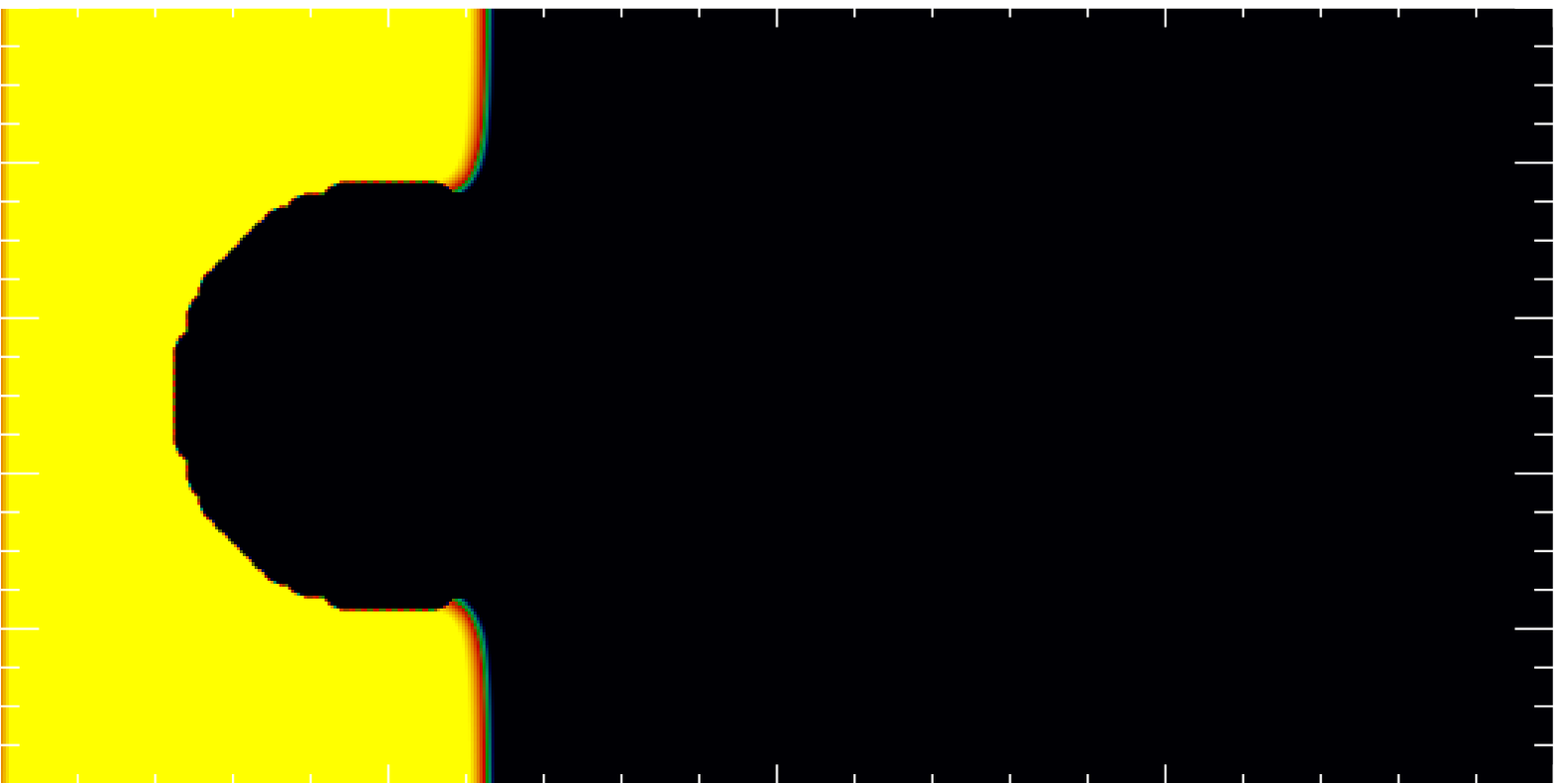}
\includegraphics[width=0.3\textwidth]{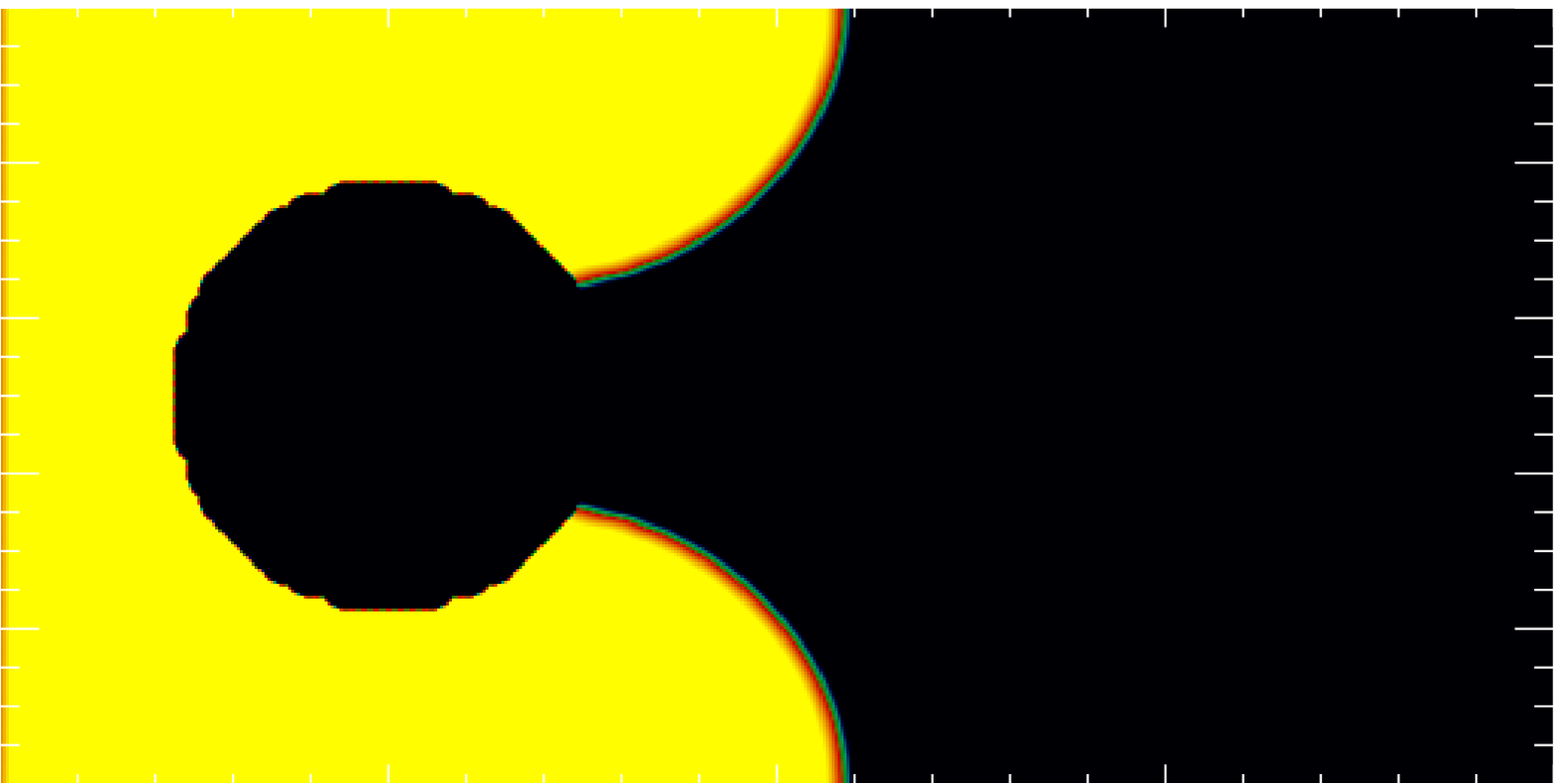}
\includegraphics[width=0.3\textwidth]{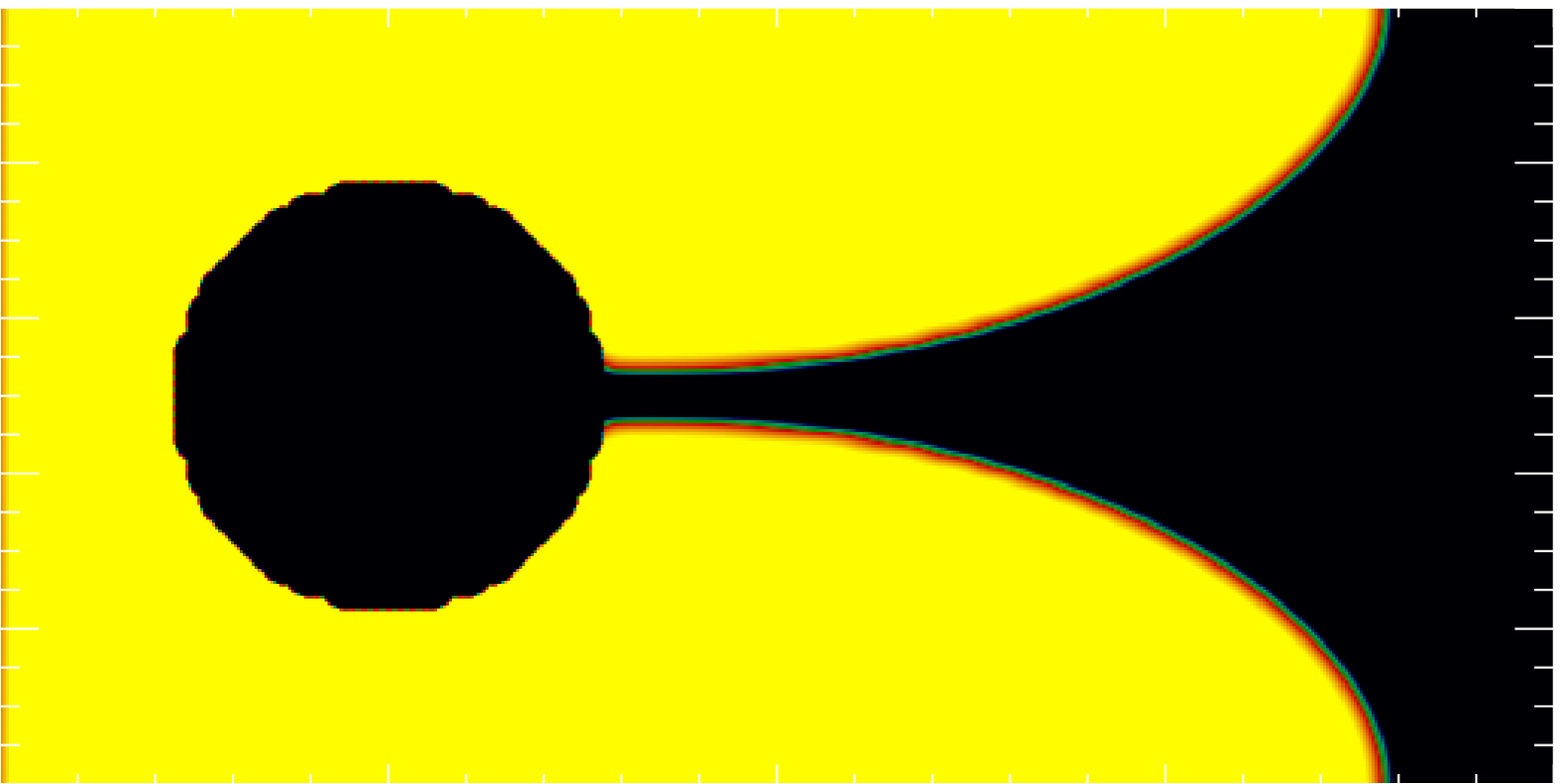}\\
\includegraphics[width=0.3\textwidth]{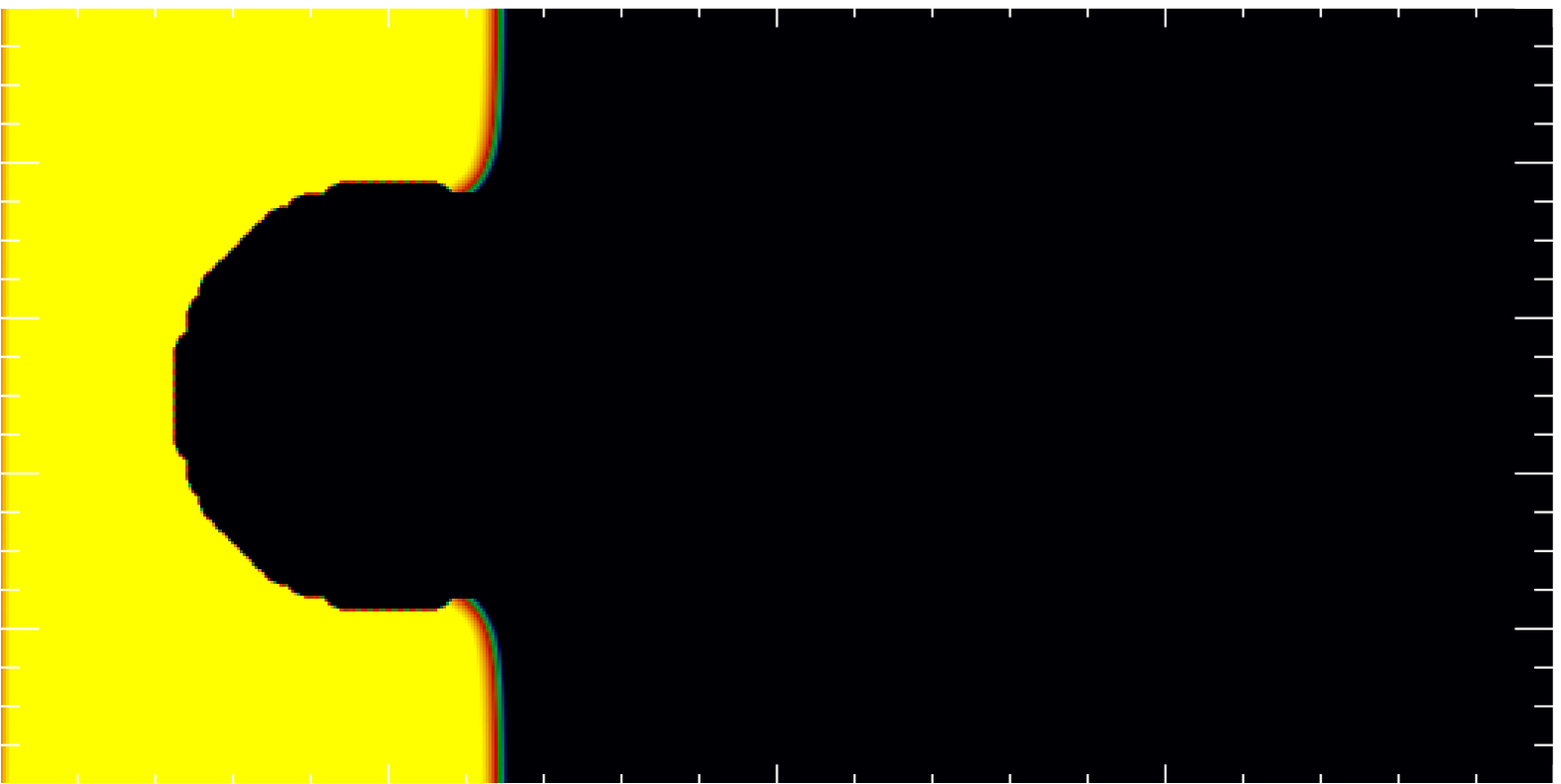}
\includegraphics[width=0.3\textwidth]{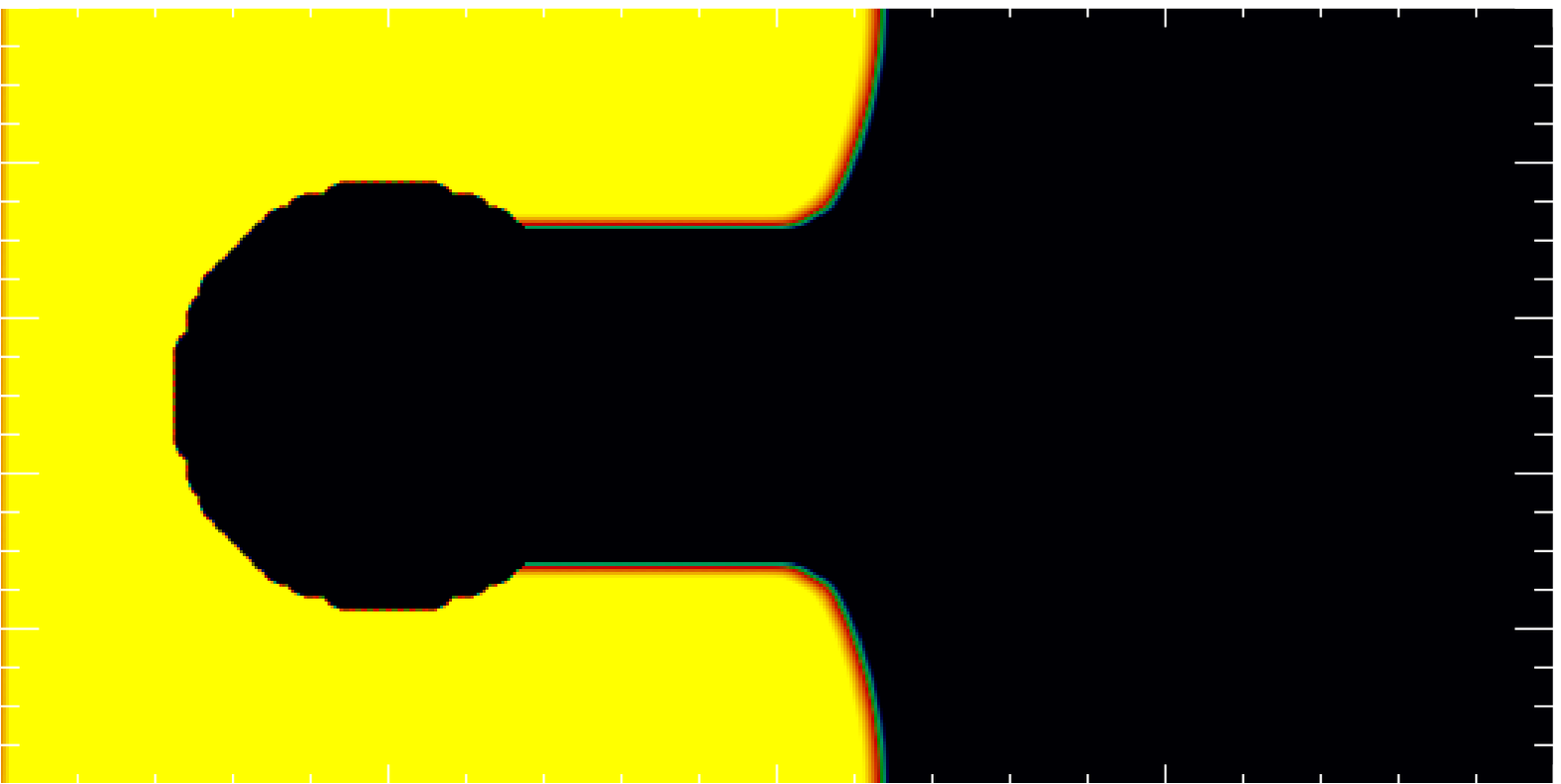}
\includegraphics[width=0.3\textwidth]{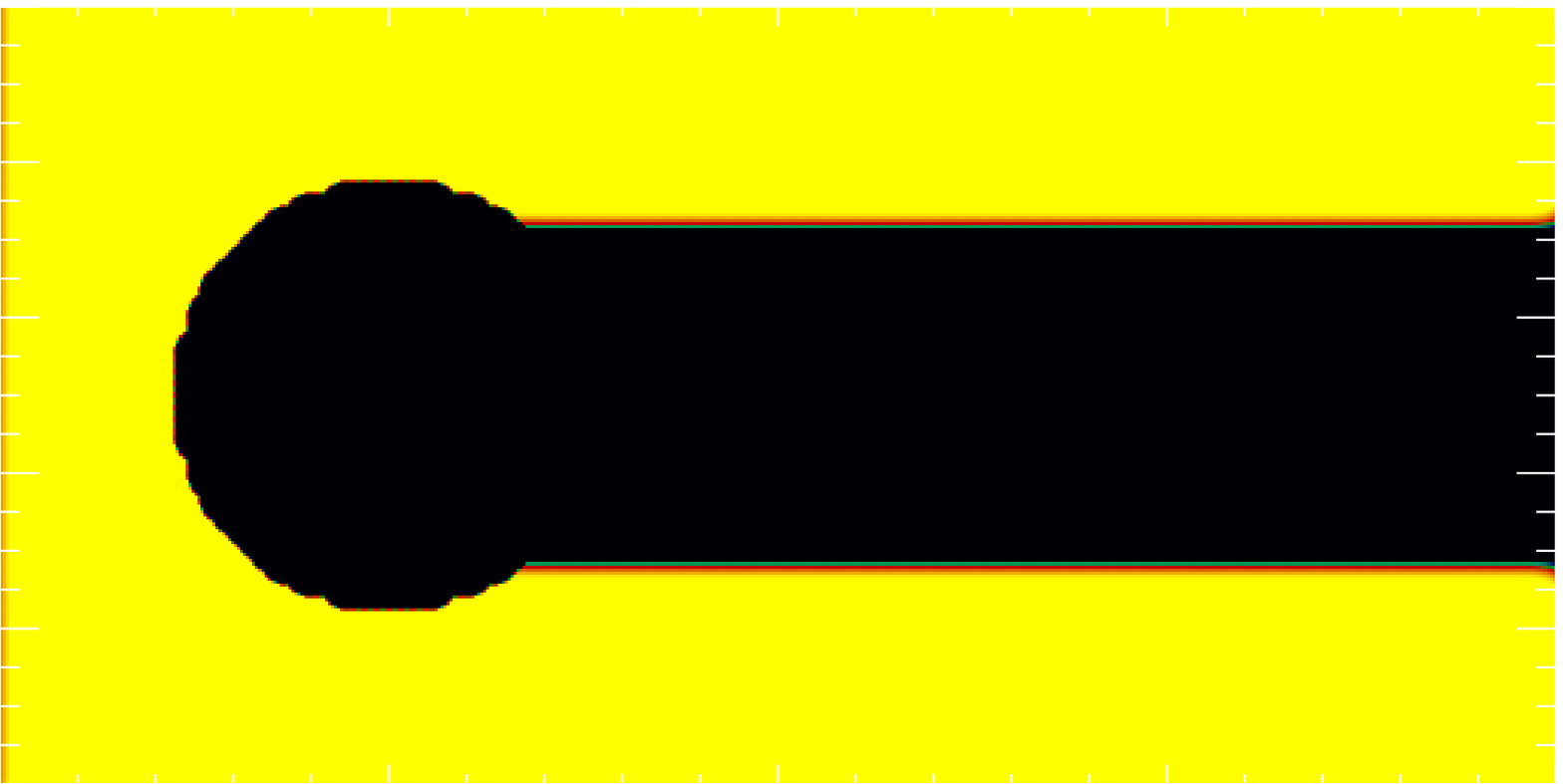}
\caption{Ionised fraction (yellow) in a slice through the middle of
  the simulation box. Time increases from left to right and the three
  different columns of snapshots are taken at $t=0.5$, $1$, and
  $2\,\rm Myr$. In the upper row the Eddington tensor has been
  approximated with the OTVET scheme. In the second row the Eddington
  tensor has been set to zero in the shadowed region, given that the
  dense clump is optically thick, which should yield the correct
  result.  The optically thin Eddington tensor approximation fails to
  produce a sharp shadow since radiation diffuses away from the
  plane-parallel front direction. In the optically thick
  approximation, however, a sharp shadow is obtained, implying that
  the failure of our method to produce a sharp shadow can also be
  blamed on using a non-vanishing Eddington tensor in the shadowed
  region.
\label{fig:shadow}}
\end{figure*}

As a further test problem, we consider the interaction of a
plane-parallel front of ionising photons with a uniform dense cylinder
of neutral gas. The setup of our problem consists of a box with
dimensions $(x, y, z) = (40\,\rm kpc, 10\,\rm kpc, 20\,\rm kpc)$. One
side of which (the $xy$-plane) is aligned with a plane of stars that
produce the ionising photons. A dense cylinder of gas is located $5$
kpc from this sheet-like source and has a radius $r_{\rm C} = 2.5$
kpc. The axis of the cylinder is oriented parallel to the $z$-axis of
the box. The particle resolution is $(N_x, N_y, N_z) =
(256,64,8)$. The hydrogen number density in the cylinder is $10^5$
times the surrounding density of $n_{\rm H}=10^{-3} \rm cm^{-3}$.
There are $512$ stars and each of them emits $\dot N_\gamma =
1.2\times 10^{48}\, \rm photons\, s^{-1}$.

We first present results obtained with an Eddington tensor that mimics
a plane-parallel I-front, of the form \be \vec{\tilde h} = \left(
\begin{array}{ccc}
1 &
0 &
0  \\
0 &
0 &
0  \\
0 &
0 &
0 \\
\end{array}
\right), \ee which represents photon transport only in the $x$-direction. We
also consider two other cases where the Eddington tensor is calculated
differently. In the first case we use the OTVET approximation, where the
tensor is computed assuming all gas is optically thin. In the second case we
account for the fact that the dense clump is optically thick by enforcing that
no radiation is transported into the shadowed region. We achieve this by
setting the Eddington tensor in the shadowed area to zero, such that the
product $h^{ij}J_{\nu}$ vanishes. Note that for a vanishing radiation pressure
tensor, the trace condition for the Eddington tensor needs not to be
fulfilled. This second case is only meant to produce the expected solution
with a sharp shadow.

Figure \ref{fig:shadow} shows the ionised fraction in a cut through
the simulation volume, in the plane $z = z_{\rm box}/2$, at three
different times: $t = 0.5$, $1$, and $2 \, \rm Myr$. In the case of
the OTVET approximation (first row), the dense cylinder fails to
produce a sharp shadow. Instead, the radiation also diffuses around
the obstacle, and is propagated eventually also in directions different from the
$x$-direction, albeit more slowly. However, if we account for the fact
that the dense clump is optically thick (second row), a clear and
sharp shadow is produced, as expected. In Figure \ref{fig:shadow2},
we show the position of the I-front with respect to the centre of the
dense clump as a function of time. It is clear that the I-front moves
faster in the case where the Eddington tensor is approximated closer
to the analytical case, i.e. it is zero in the shadow area.

The inability to create a sharp shadow is a limitation of the OTVET
approximation implemented in SPH, as we have shown above. This
  limitation of codes using a moment method together with an OTVET
  approximation has already been noted by several groups, e.g. \citet{GA2001,
    AT2007}. It appears that despite our attempts to fully account for the
anisotropic diffusion, the diffusion operator we have derived remains quite
diffusive in SPH. This is simply a consequence of the non-vanishing coupling
between particles with separation vectors not perfectly aligned with the
direction of radiation propagation (here the $x$-axis).  Unfortunately, this
deficiency may also have other detrimental effects besides just slowing down
the expansion of the I-front itself. We acknowledge that this can be an
important limitation of our scheme for certain applications, especially when
shielding is common and shadowing is important. Nevertheless, this
  limitation can influence the morphology of the reionisation, but properties
  such as redshift and duration of the process, as well as temperature
  evolution of the gas and the radiation field, should still be accurate.

\bfig
\includegraphics[width=0.9\columnwidth]{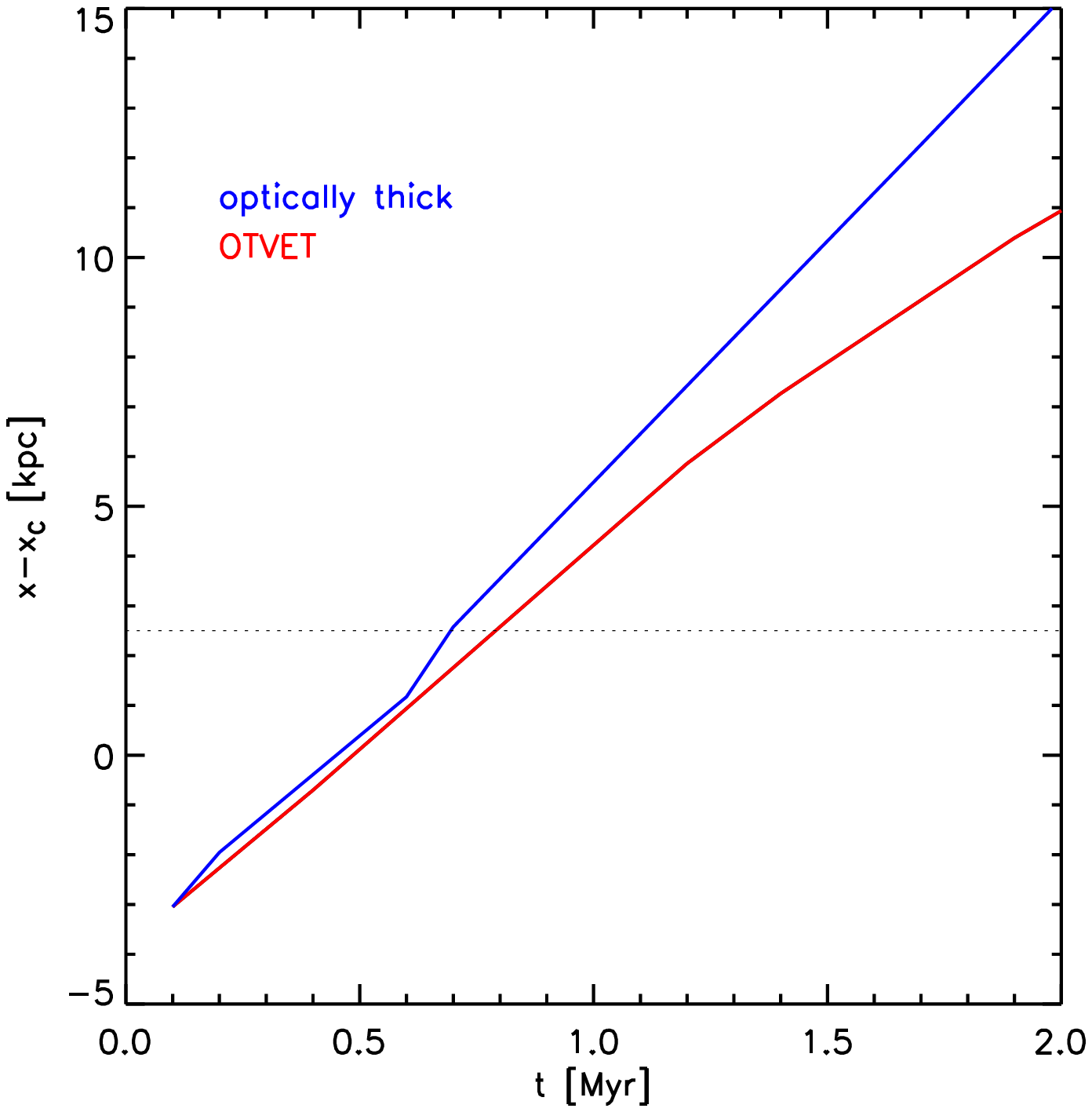}
\caption{Position of the I-front, relative to the centre of the dense
  clump, as a function of time. The I-front moves faster in the case
  where the Eddington tensor is computed accounting for the optically
  thick clump (blue line). In the case where the OTVET scheme is used
  (red line), the I-front expands slower. The dashed line shows the
  position at the end of the clump, where the difference between the
  expansion rates begins to grow.
\label{fig:shadow2}}
\efig

\subsection{Static cosmological density field}\label{cosmo}

\bfig
\includegraphics[width=0.9\columnwidth]{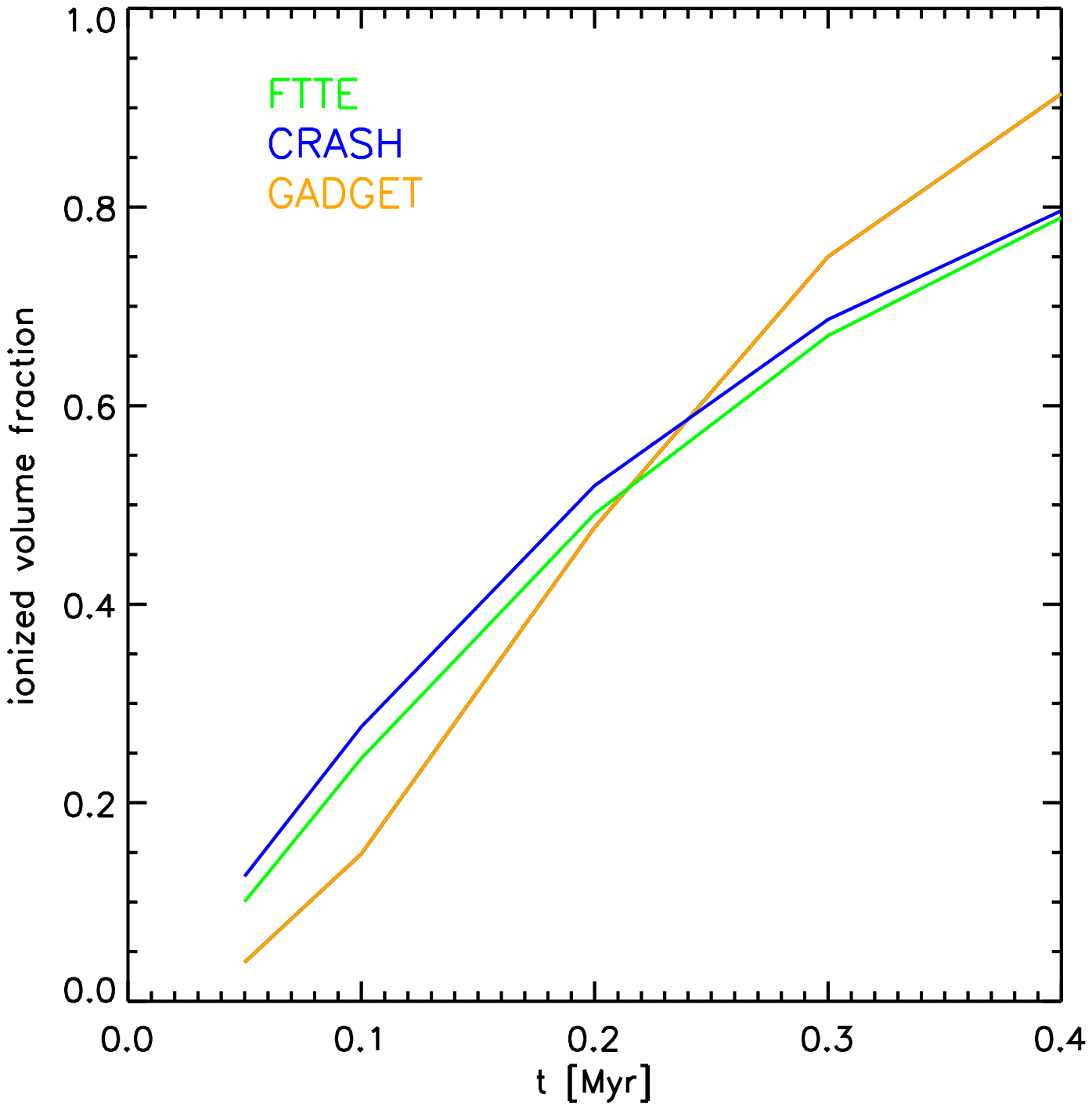}
\caption{Time evolution of the volume-averaged ionised fraction in the
  whole simulation box. in comparison to {\small CRASH} (blue) and
  {\small FTTE} (green), {\small GADGET} (orange) produces lower
  ionised fractions at earlier times and higher ionised fractions at
  later times. The overall trend in the increase of the ionised
  fraction is the same, and the speed of ionisation first accelerates
  and then decelerates. Differences between our results and the ones
  from the Comparison Project are in part also due to differences in
  the initial conditions, as we had to transform the grid-based
  density field to an SPH realization.
 \label{fig:statictimeMANY}}
\efig

In our final and most demanding test calculation we follow hydrogen
ionisation in a realistic cosmological density field, which is taken
to be static for simplicity.  In order to compare our results with
those of the cosmological radiative transfer comparison project
\citep{Iliev2006b} we use the same cosmological box parameters and assign sources
in the same way.  The box with size $0.5\,h^{-1} {\rm comoving \,
  Mpc}$ is evolved with a standard $\Lambda$CDM model with the
following cosmological parameters: $\Omega_{\rm 0}=0.27$, $\Omega_{\rm
  b} = 0.043$, $h=0.7$, until redshift $z=9$. The density field at
this point is considered for our further analysis.

The source distribution is determined by finding halos within the
simulation box with a FOF algorithm and then assigning sources to the
16 most massive ones. The photon luminosity of the sources is \be \dot
N_\gamma = f_\gamma \frac{M\Omega_{\rm b}}{\Omega_{\rm 0}m_{\rm
    p}t_{\rm s}}, \ee  where $M$ is the total halo mass, $t_{\rm s} = 3
\, \rm Myr$ is the lifetime of the source, $m_{\rm p}$ is the proton
mass and $f_\gamma = 250$ is the number of emitted photons per atom
during the lifetime of the source. We find that the total source
luminosity in our simulated box agrees well with the one from
\citet{Iliev2006b}.  For simplicity we also set the initial
temperature of the gas to $100$ K throughout the whole box.

We found that simply mapping the grid cells onto a Cartesian mesh of
  SPH particles with different masses introduces large noise into our RT
  calculation, due to the large variations in the mass of neighbouring
  particles. It is therefore not straightforward to translate the mesh-based
  data of the code comparison project into an equivalent SPH realization, and
  we therefore needed to created our own initial conditions. We note that
  simple methods to create an equal particle mass SPH realization from the
  given grid cells, e.g. through random sampling, tend to introduce large
  amounts of Poisson noise and wash out extrema in the density field.
 
The evolution of the total volume-averaged ionised fraction in the box is very
similar for {\small GADGET} and the Comparison Project codes, as shown in
Figure \ref{fig:statictimeMANY}.  In the beginning of the simulation the total
ionised fraction rises rapidly and then the increase decelerates. {\small
  GADGET} produces an overall lower ionised fraction until approximately
$t=0.2 \, {\rm Myr}$ and higher one at later times. This mismatch is in part
certainly caused by the morphological differences in the initial conditions
that introduce different clumping properties of matter and therefore different
I-front expansion histories. We recall that the true solution of the
  problem is unknown. Given the non-linearity of the system, the differences
  in detail of the initial conditions and cooling rates, and the fact that we
  compare fundamentally different RT schemes, the agreement that we
  obtain is actually very good.

\begin{figure*}
\includegraphics[width=0.25\textwidth, clip]{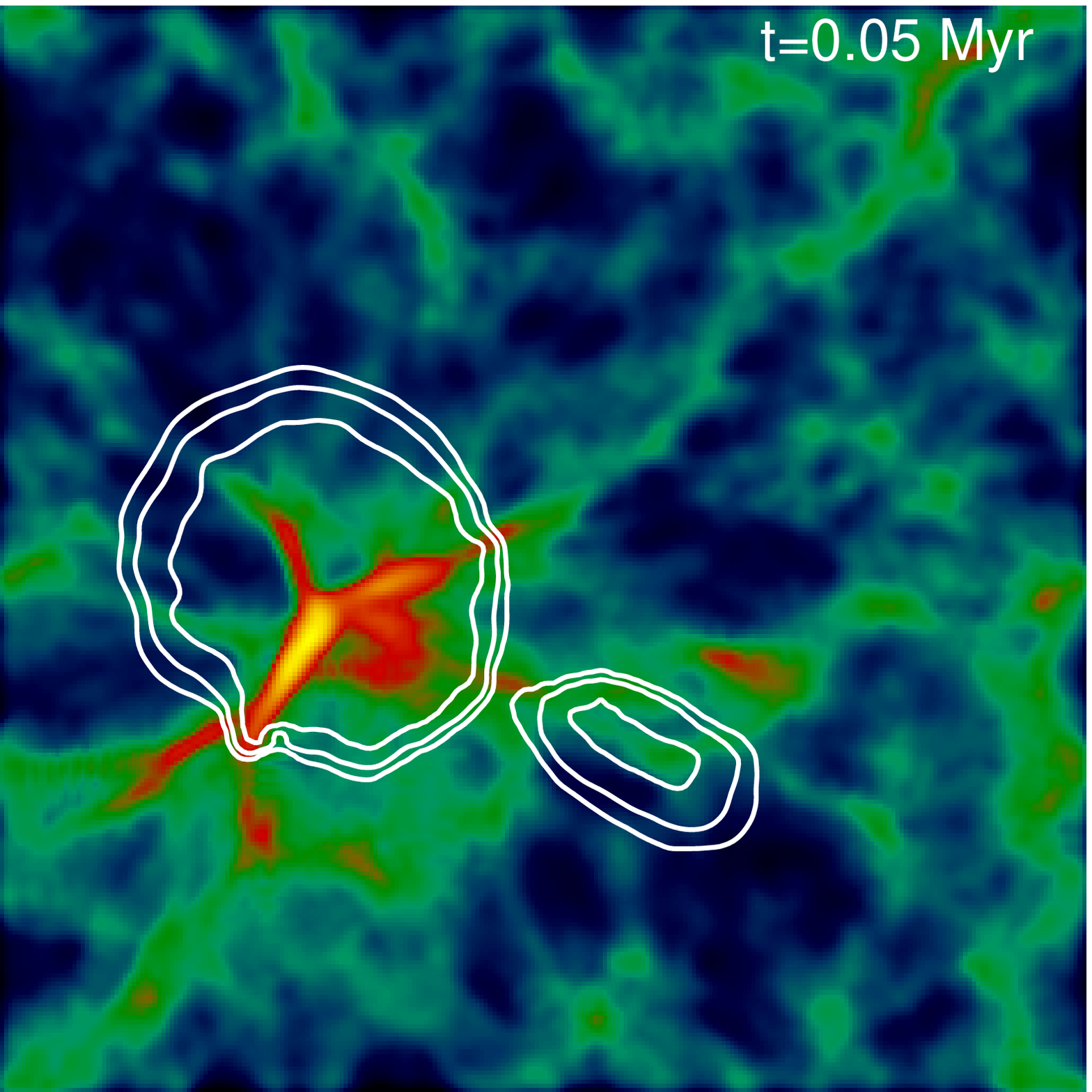}
\hspace{1cm}
\includegraphics[width=0.25\textwidth, clip]{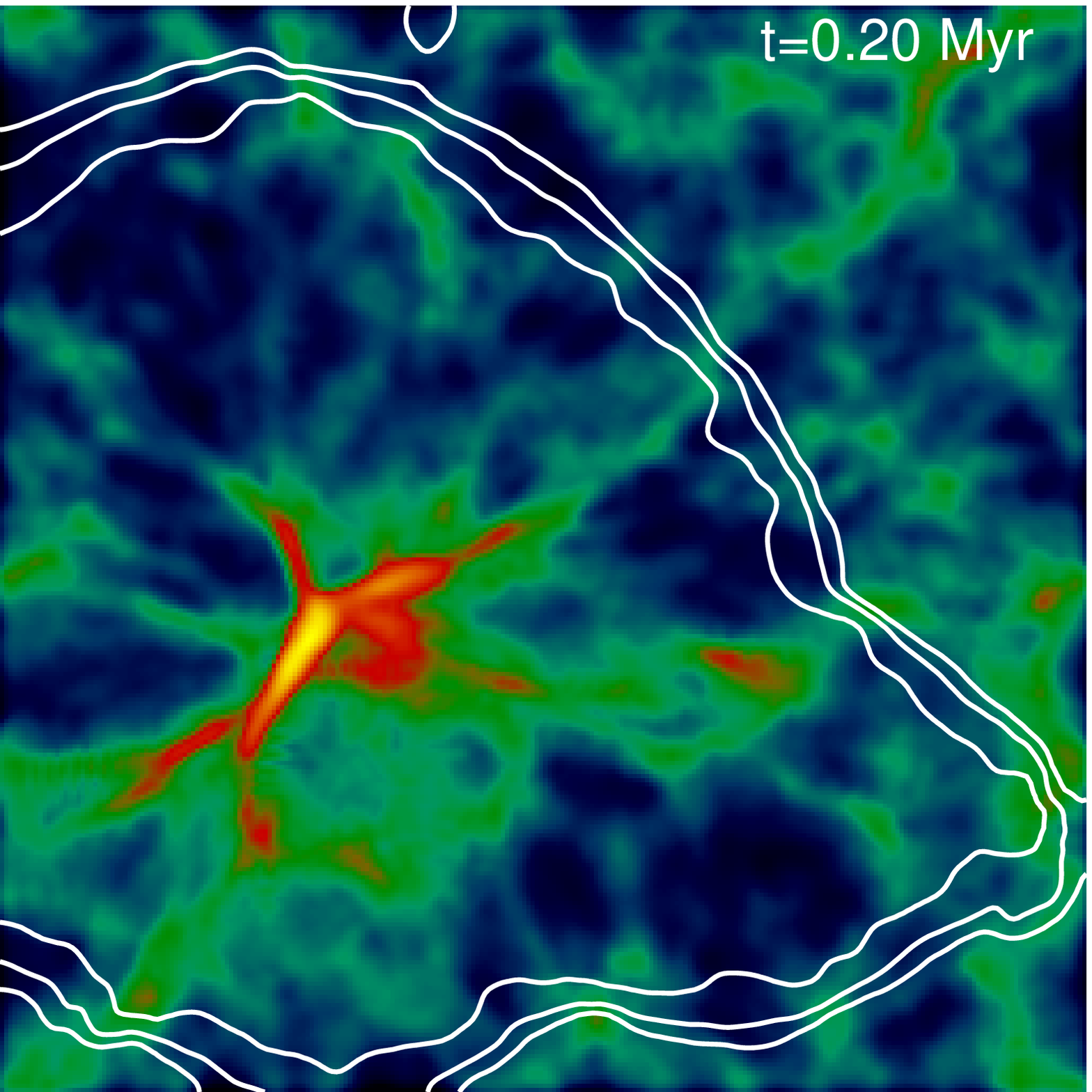}
\hspace{1cm}
\includegraphics[width=0.25\textwidth, clip]{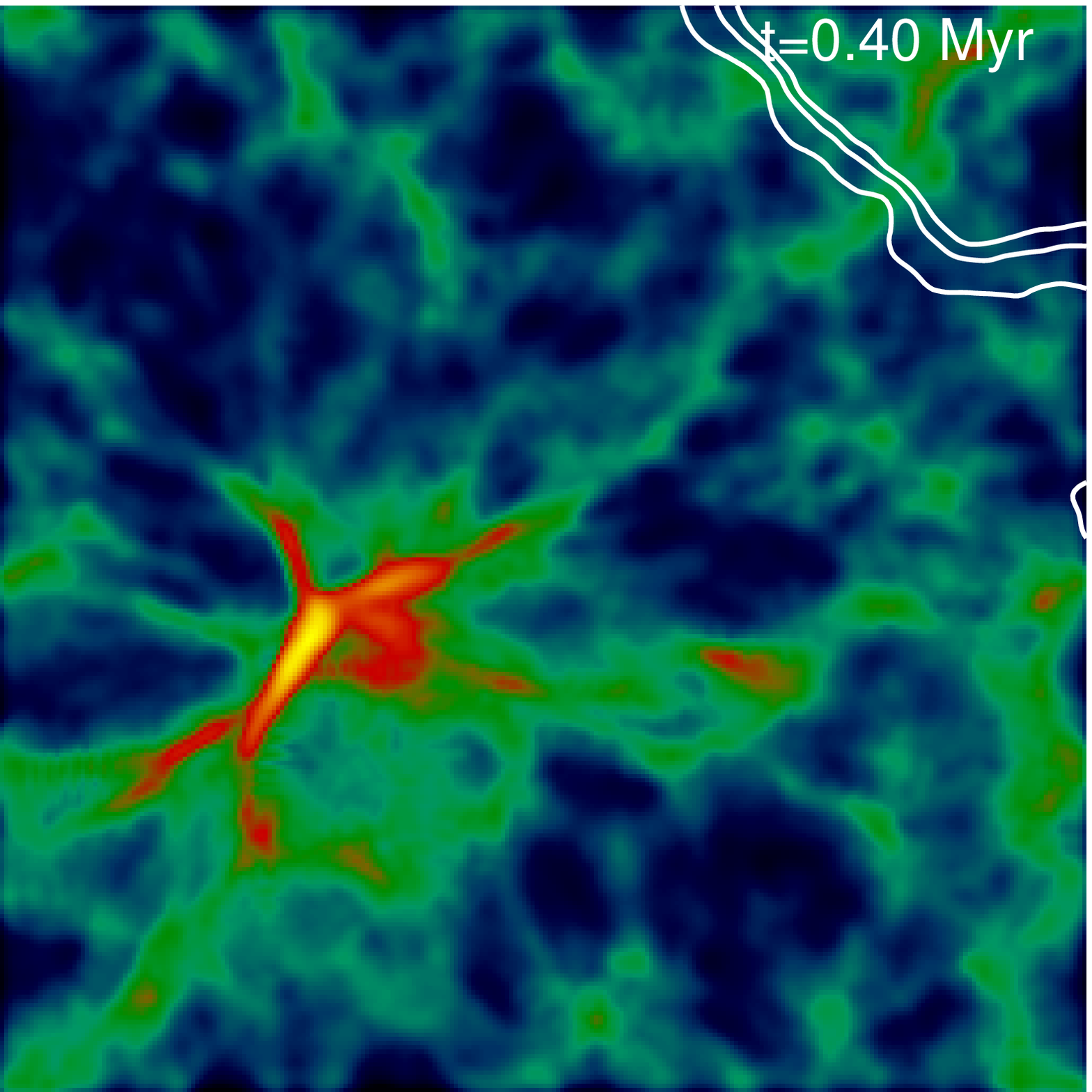}\\
\includegraphics[width=0.25\textwidth, clip]{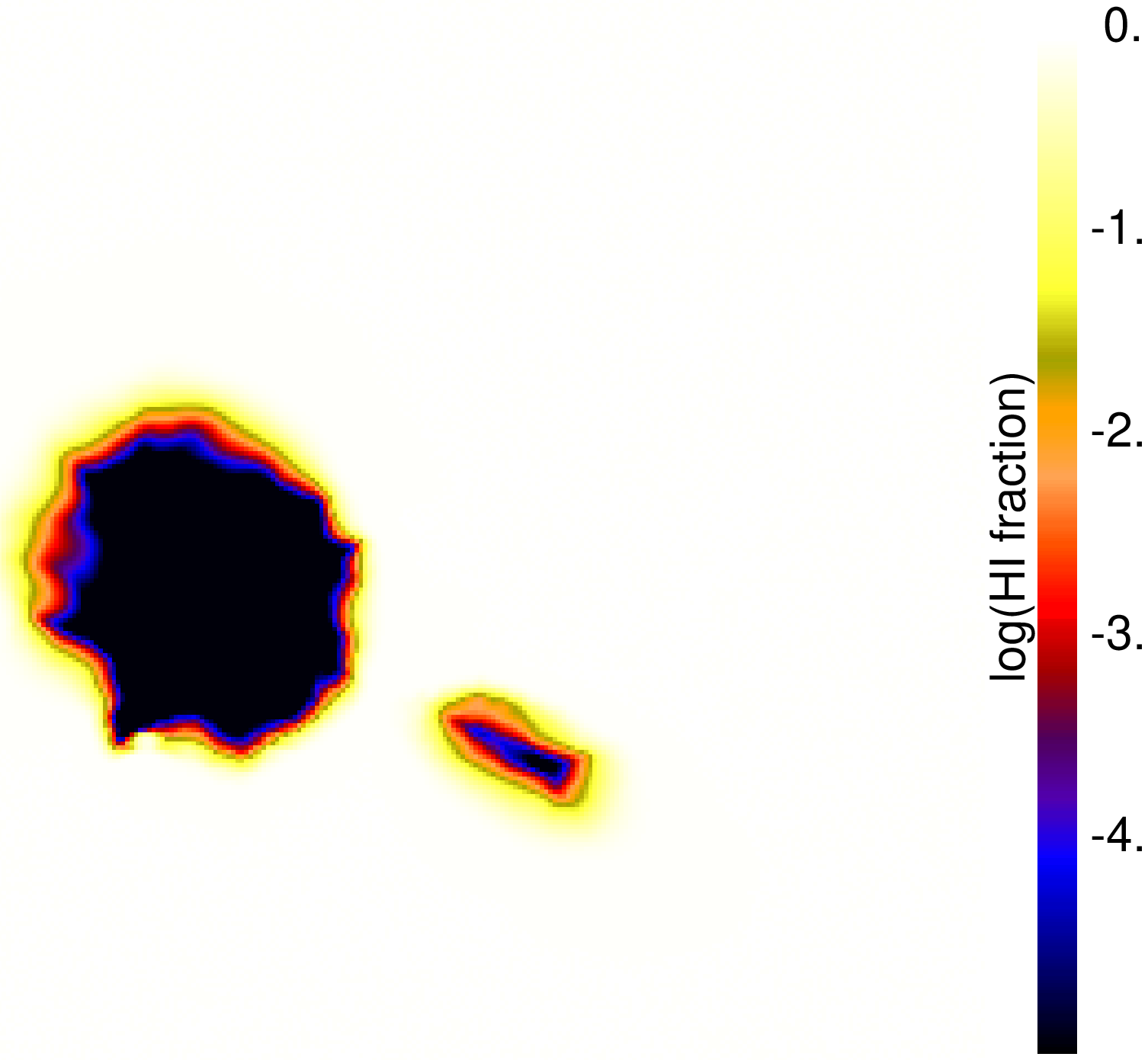}
\hspace{1cm}
\includegraphics[width=0.25\textwidth, clip]{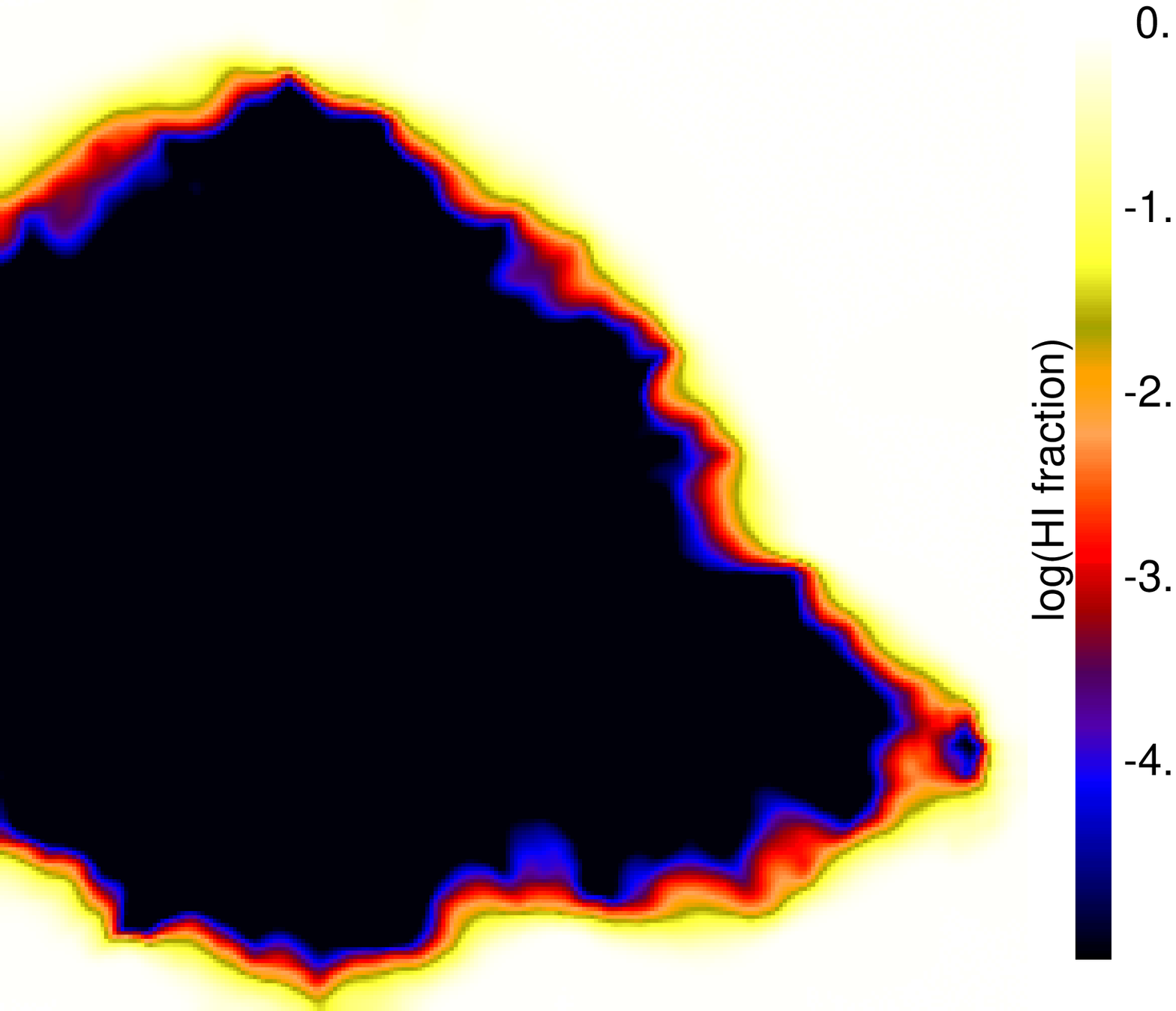}
\hspace{1cm}
\includegraphics[width=0.25\textwidth, clip]{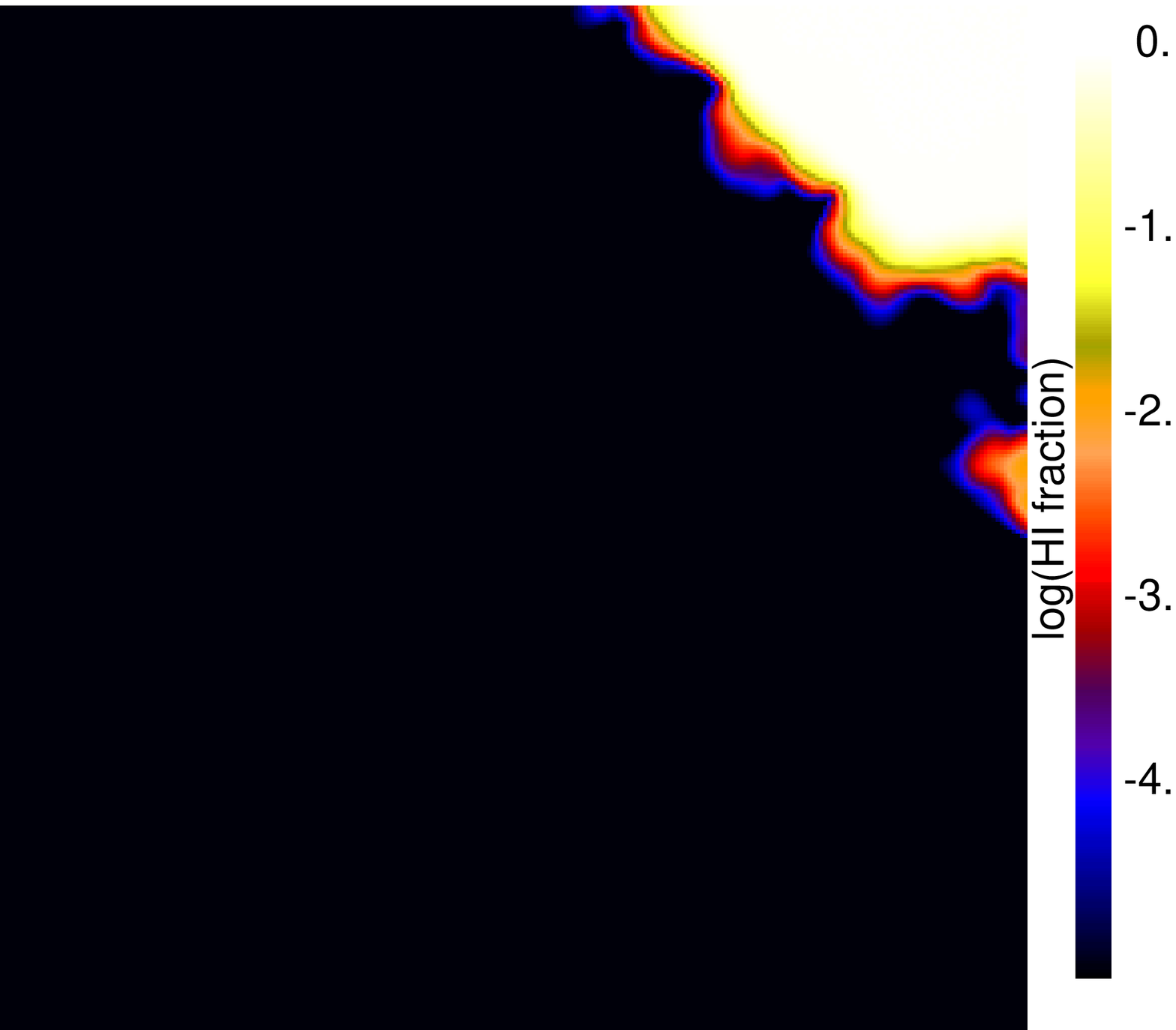}\\
\includegraphics[width=0.25\textwidth, clip]{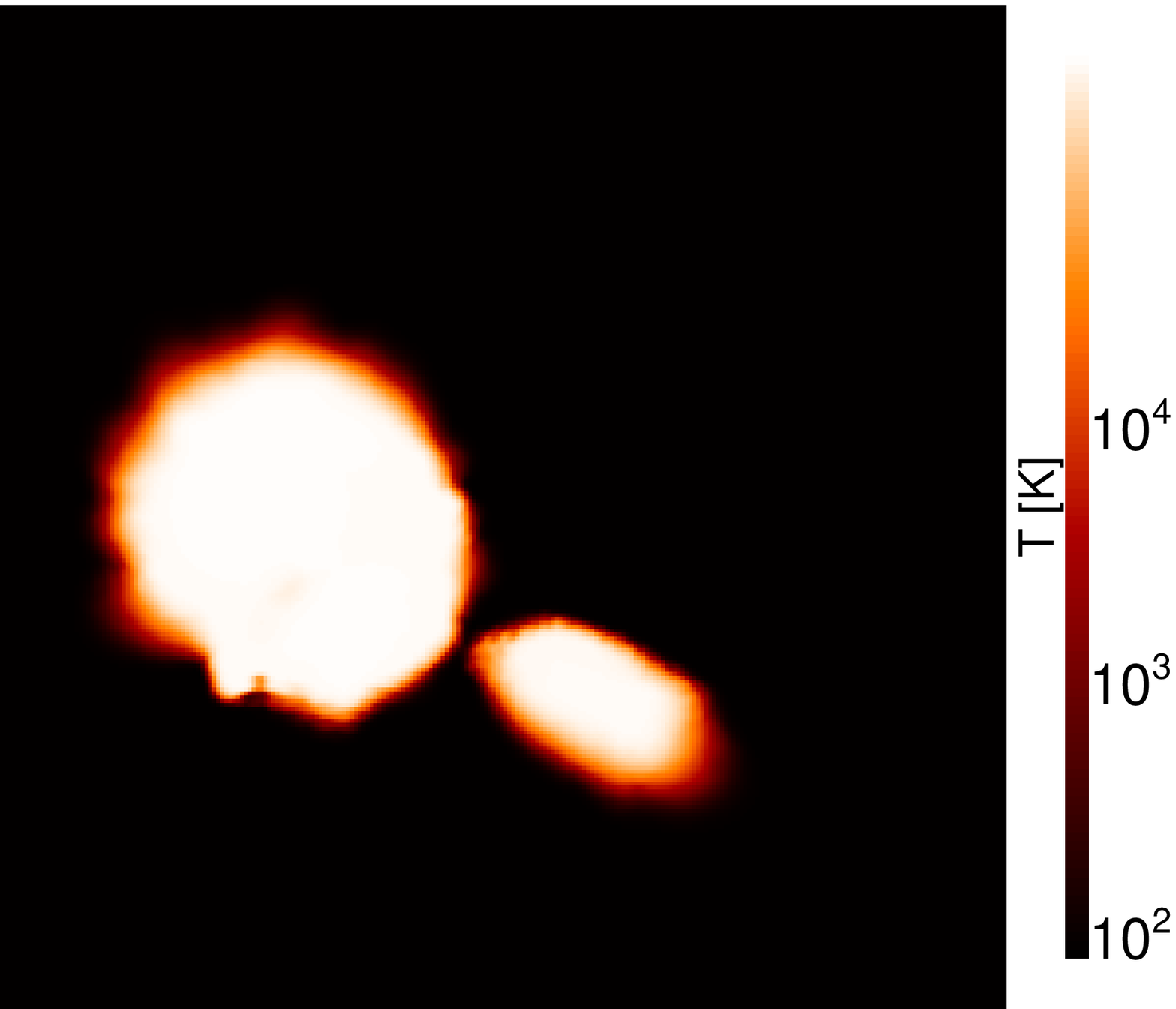}
\hspace{1cm}
\includegraphics[width=0.25\textwidth, clip]{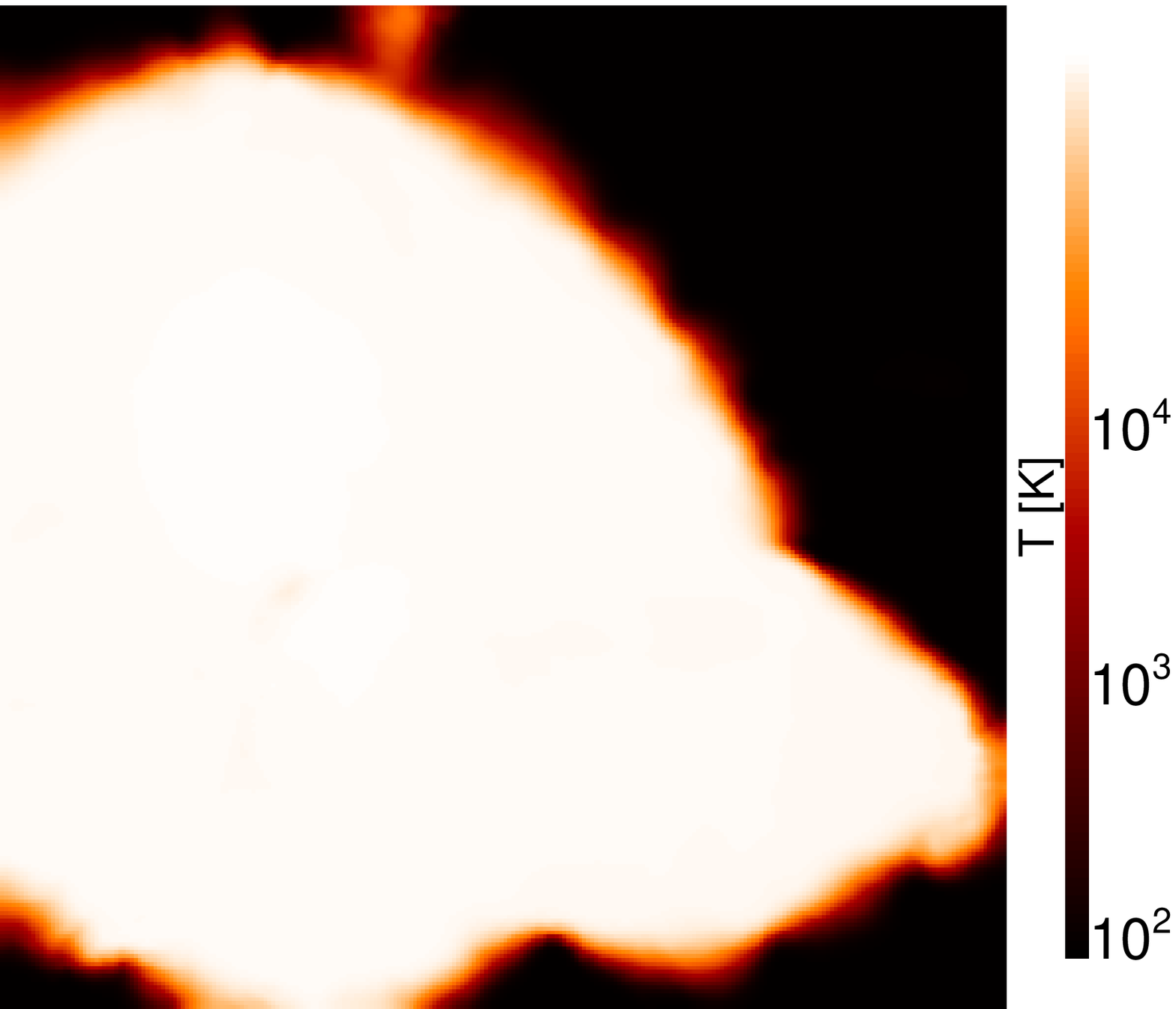}
\hspace{1cm}
\includegraphics[width=0.25\textwidth, clip]{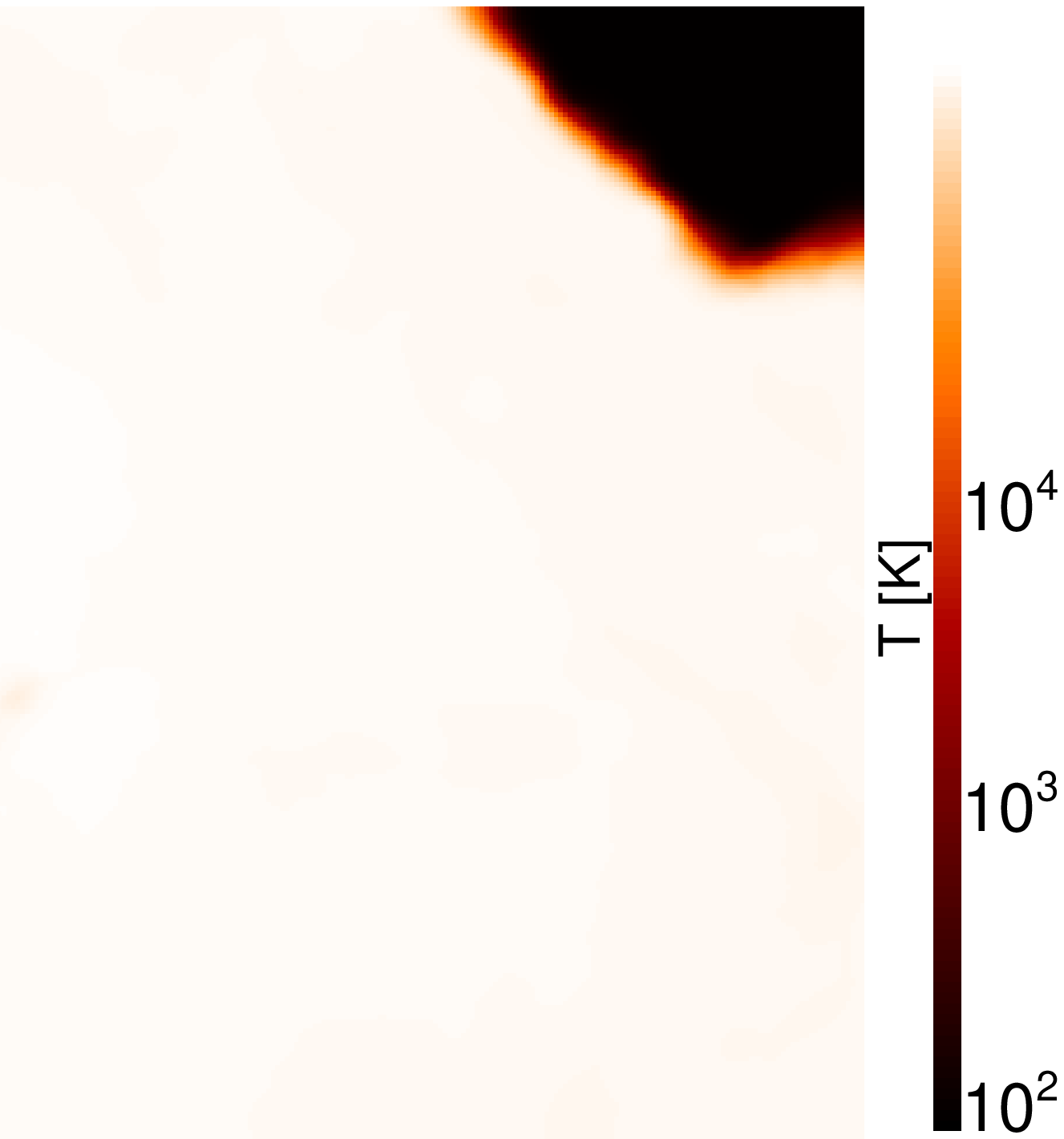}
\caption{The upper row shows contours of neutral fractions equal to
  $0.01, 0.5$ and $0.9$ in a slice through the simulation volume at $z
  = 0.7z_{\rm box}$, through the largest group. The snapshots are
  taken at times $t=0.05$, $0.2$ and $0.4 \, \rm Myr$ (left to right).
  The background shows a slice of the density distribution. The
  ionised regions expand with time as the I-front is trapped at high
  density regions and extends into low density regions. The second row
  shows the neutral fraction in the same slice. The ionised regions
  are uniform with some substructure visible near the front-trapping
  regions, where the I-front is not as diffuse as in low density
  regions. The third row shows the temperature distribution in the
  slice. The temperature in the ionised regions reaches several
  $10^4{\rm K}$ and remains uniform outside these
  regions.  \label{fig:densiontemp}}
\end{figure*}

To illustrate the spatial distribution of the ionised fraction and the
temperature of the gas we show in Figure \ref{fig:densiontemp} slices
through the simulation volume at $z = 0.7z_{\rm box}$ (through the
largest group) at three different times $t=0.05,\, 0.2\, {\rm and}\,
0.4 \, \rm Myr$. The upper row shows contours of the neutral fraction
plotted over a density field, the second row shows a map of the
neutral fraction and the third row shows a map of the temperature of
the gas.  The dense regions trap the I-front and thus produce sharp
gradients of the radiation density. In the under-dense regions
ionisation is more effective and the I-front is extended. Even though
the ionised regions are mostly uniform, traces of the dense structures
that are less ionised can be seen near the front-trapping points. As
shown in the contour maps, the I-fronts are broader in the low density
regions and thiner in the high density regions. The temperature in the
ionised regions reaches several $10^4\,{\rm K}$ and is uniform. It
remains unchanged outside, where no photons are present.

\begin{figure*}
\includegraphics[width=0.25\textwidth, clip]{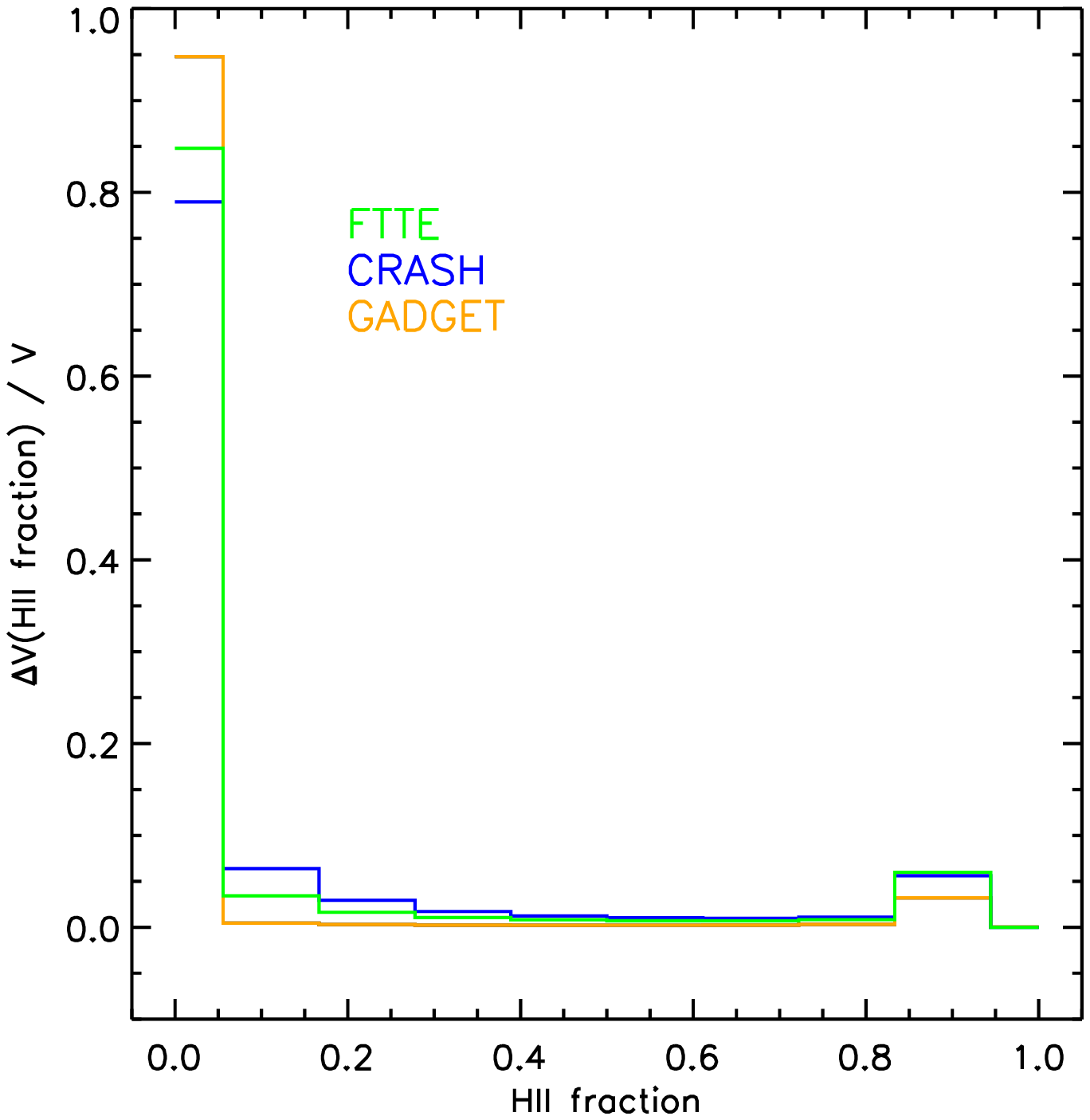}
\hspace{1cm}
\includegraphics[width=0.25\textwidth, clip]{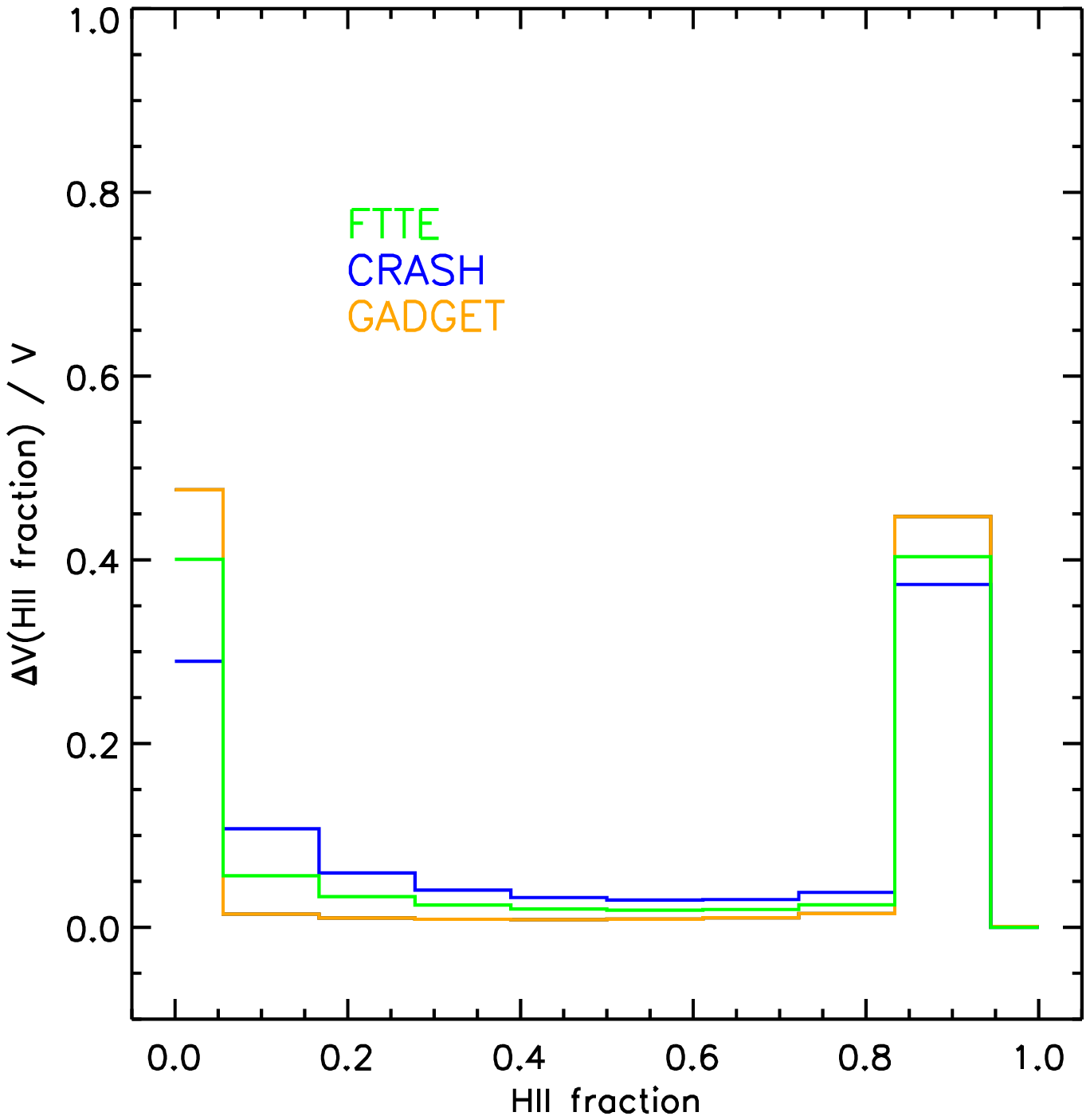}
\hspace{1cm}
\includegraphics[width=0.25\textwidth, clip]{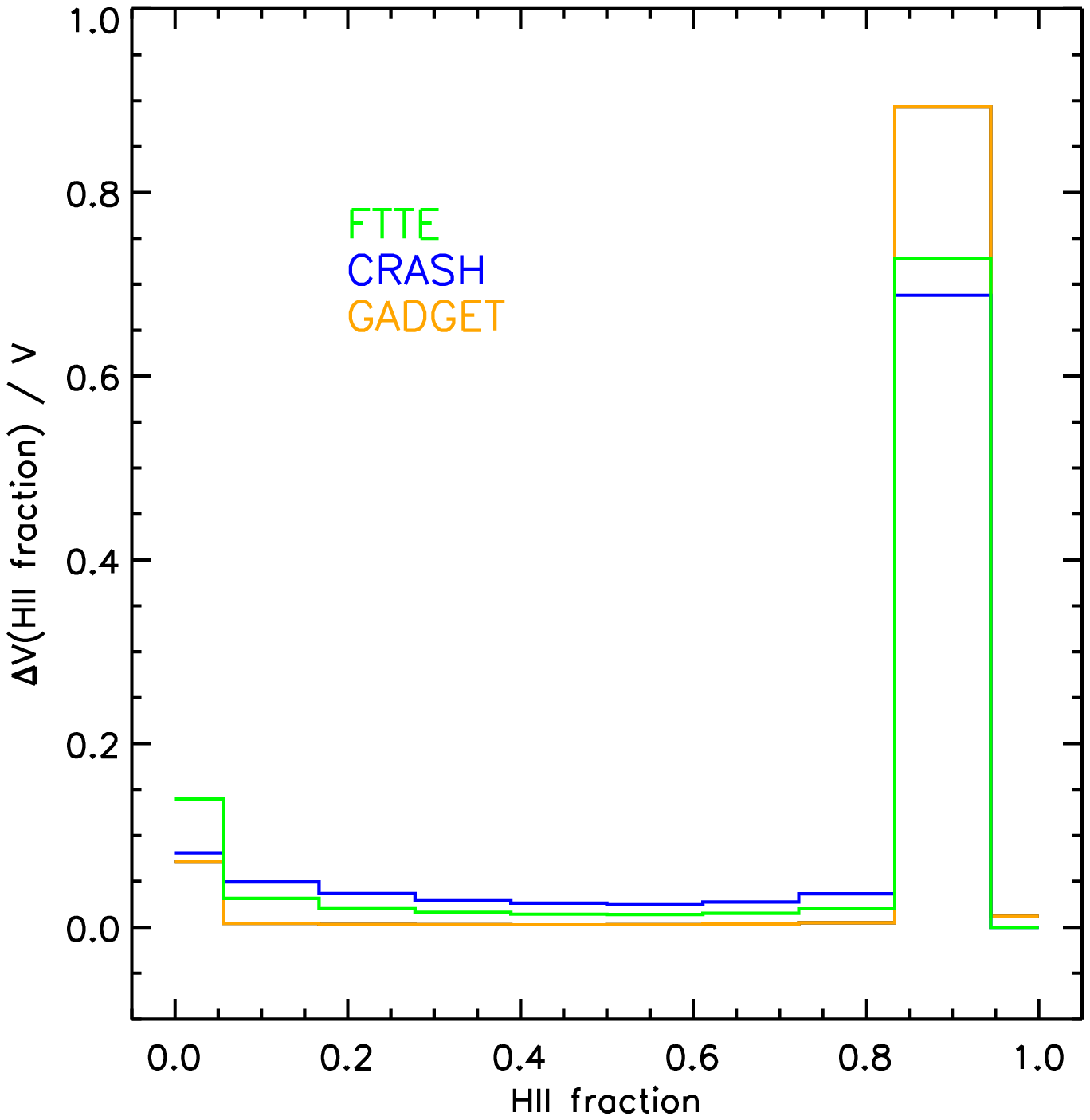}
\caption{Volume fraction of the ionised fraction at three different
  times $t=0.05$, $0.2$, and $0.4 \, \rm Myr$ (left to right). Results
  from {\small GADGET} are compared with results from {\small CRASH}
  and {\small FTTE} from \citet{Iliev2006b}. All codes match in the
  shape of the histograms, but {\small GADGET} gives a lower
  intermediately ionised fraction.
 \label{fig:histioncosmo}}
\end{figure*}

\begin{figure*}
\includegraphics[width=0.25\textwidth, clip]{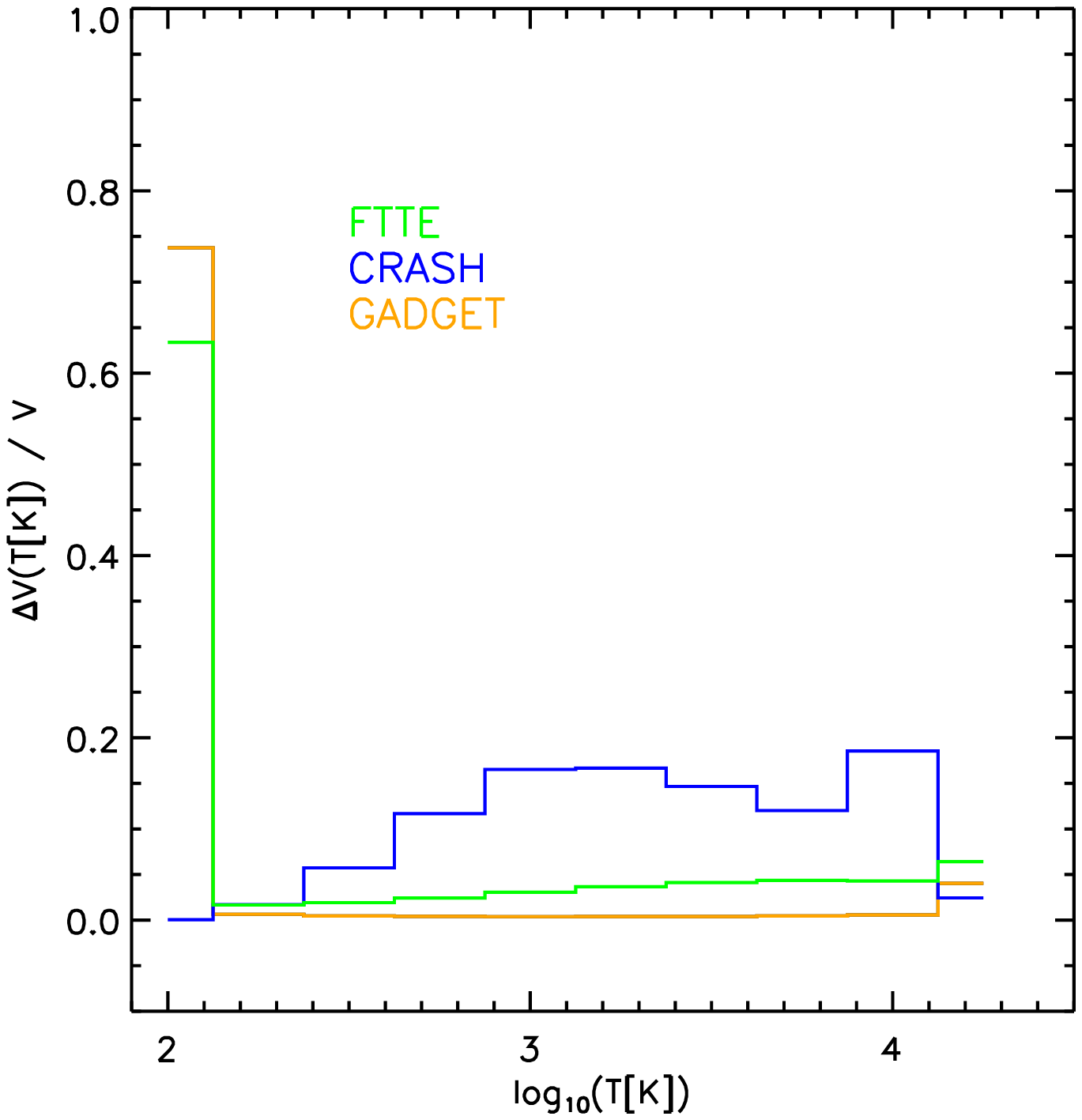}
\hspace{1cm}
\includegraphics[width=0.25\textwidth, clip]{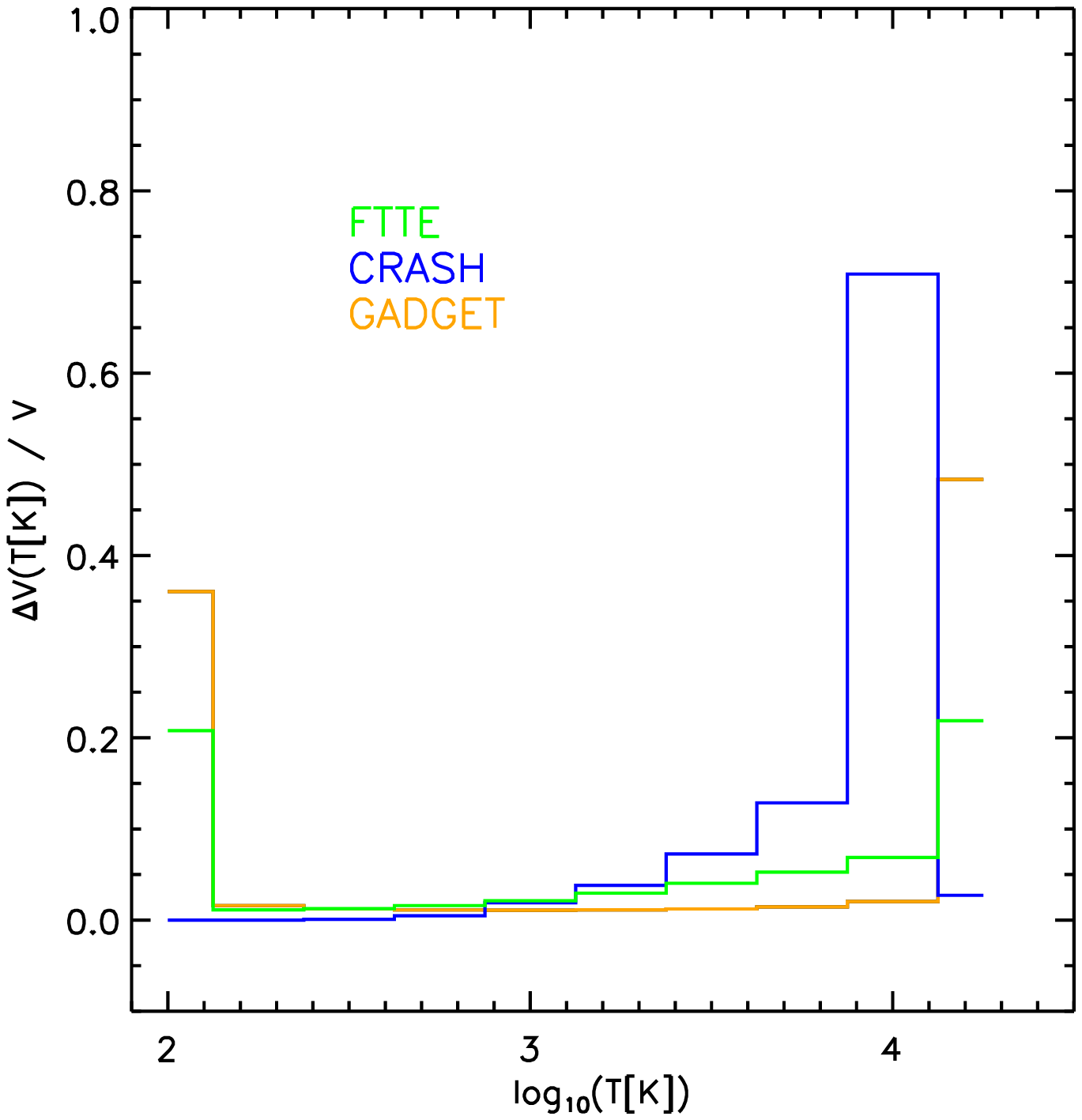}
\hspace{1cm}
\includegraphics[width=0.25\textwidth, clip]{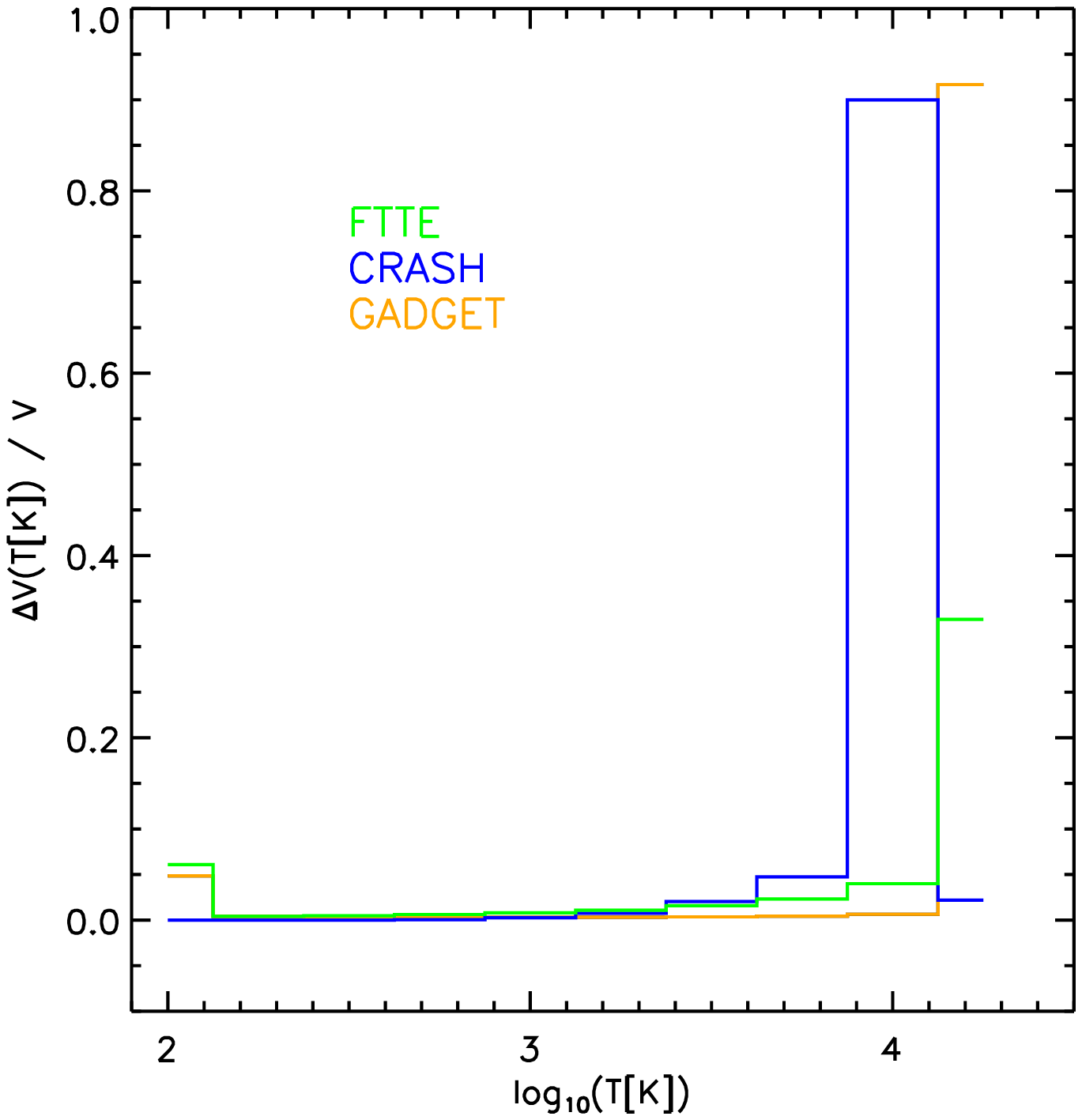}
\caption{Volume fraction of the temperature at three different times
  $t=0.05$, $0.2$ and $0.4 \, \rm Myr$ (left to right). Results from
  {\small GADGET} are compared with results from {\small CRASH} and
  {\small FTTE} from \citet{Iliev2006b}. All codes produce different
  histograms.  {\small FTTE} and {\small GADGET} agree better with
  each other at later evolution times. {\small CRASH} produces higher
  temperatures due to its use of a different spectral distribution.
 \label{fig:histtempcosmo}}
\end{figure*}

We further compare the volume faction of the temperature and the
ionised fraction from {\small GADGET}, {\small CRASH} and {\small
  FTTE} in Figures \ref{fig:histioncosmo} and \ref{fig:histtempcosmo},
at three different times $t=0.05$, $0.2,$ and $0.4 \, \rm Myr$. The
temperature volume fractions do not match particularly well due to the
different photoheating mechanisms that {\small GADGET}, {\small CRASH}
and {\small FTTE} use, but they find a similar maximum temperature.
The volume fractions of the ionised fraction for {\small GADGET},
{\small CRASH} and {\small FTTE} have similar shapes. {\small GADGET},
in contrast to the other codes, produces less intermediately ionised
gas. However, overall the histograms are in a reasonably good
agreement with each other, suggesting that our moment-based scheme is
quite capable in describing the reionisation process and produces
results of similar accuracy as other established radiative transfer
codes.

\section{Summary and Conclusions} \label{SecConclusions}

We have presented a novel method for solving the radiative transfer
equations within SPH, which is based on moments of the radiative
transfer equation that are closed with a variable Eddington
tensor. The radiation transport effectively becomes an anisotropic
diffusion problem in this formulation. We have developed a new
discretization scheme for anisotropic diffusion in SPH together with
an implicit time integration method which for the first time allows a
calculation of such anisotropic diffusion in SPH. Together with a
scheme to estimate Eddington tensors based on the optically thin
approximation, this yields a very fast approximate treatment of
radiative transfer that can be used in dynamical SPH calculations.

We have implemented our method into the cosmological simulation code
{\small GADGET-3} and presented several test problems where we varied
the initial conditions and different numerical parameters to
investigate the accuracy of the method. Our test results agree in
general very well with analytical predictions and data from other
simulations, except that the long-term evolution of sharp geometric
shadows is clearly not followed accurately. While this clearly limits
the range of applicability of the method, we expect that the method
can still provide reasonably accurate results for problems where
shadowing is comparatively unimportant, such as cosmological
reionisation, where the SPH-based variable Eddington tensor approach
can be competitive with other techniques.  However, our method has two
important strengths not shared by most other techniques: It can easily
cope with an arbitrary number of sources since its speed is
essentially independent of the number of sources, and furthermore, it
is fast enough to be included into a cosmological simulation code where
radiative transfer is calculated on-the-fly together with the ordinary
dynamics. This is especially promising for future calculations of
galaxy formation and reionisation that we want to carry out with our
new code.

The epoch of reionisation is an important and extremely interesting
process in the evolution of the Universe. There is already very
important observational evidence that constrains reionisation, but few
precise statements about the onset, duration, and end of reionisation
can be made at this point. The situation will likely change in the
near future through upcoming observations, ranging from new CMB
observations with PLANCK, to 21cm tomography at high redshift with
radio telescopes such as LOFAR. In the meantime, we heavily rely on
cosmological simulations to advance our understanding of the
reionisation process.

However, the simulations run so far leave many questions still
unanswered, and their lack of self-consistency with the actual
dynamics and the rapidly evolving source population may have
introduced sizable inaccuracies. It is therefore highly desirable to
have new numerical techniques that do not rely on solving the
radiative transfer equation in a post-processing mode by treating only
individual snapshot files from simulations.  Rather, accurate
calculations should account for radiative transfer `on-the-fly', so
that the gas properties are affected simultaneously by gravity and
radiation. Our new method promises to be an important step in this
direction. In future work, one of our most important goals is
therefore to carry out high-resolution simulations of structure
formation where the build-up of stellar and quasar sources in galaxies
is followed self-consistently with the radiation field, allowing us to
make more accurate predictions for the temporal evolution of
reionisation, and the topology of reionised regions as a function of
time.

All tests problems we have presented in this thesis agree well with
theoretical predictions or results obtained with other radiative
transfer codes. We should be able to obtain a realistic and accurate
temperature evolution of the Universe during reionisation, which is
important for setting the `cosmic equation of state' that regulates
the absorption seen as Lyman-$\alpha$ forest in the spectra of distant
quasars. Finally, we plan to include the photoionisation of other
elements besides hydrogen, most importantly of helium. Helium
reionisation probably happened sometime at redshift $z\sim 2-4$, where
it may have left a sizable imprint in the temperature evolution of the
intergalactic medium. Surprisingly, recent observations suggest that
the temperature-density relation of the IGM may be inverted
\citep{Bolton2007}, which could be caused by radiative transfer
effects related to helium reionisation. Whether this is indeed
possible can only be clarified with simulations. Studying this
question with our new methods would therefore be particularly timely.

\section*{Acknowledgements}
This research was supported by the DFG cluster of excellence ‘Origin and
Structure of the Universe’ (www.universe-cluster.de).

\appendix

\section{Cooling rates}\label{AppendixA}
Here we present all cooling rates that we have used in our
simulations with temperature evolution (Section
\ref{nonisothermal}). The rates have been obtained from
\citet{Cen1992} and are given in $\rm erg \, cm^{-3} \, s^{-1}$.

\begin{enumerate}
\item
Recombination cooling rate
{\small 
\be \Lambda_{r} = 8.7 \times 10^{-27}\sqrt{T}
\left(\frac{T}{10^3}\right)^{-0.2}\left/ \left[1+\left(\frac{T}{10^6}
\right)^{0.7} \right]n_e n_{\rm HI} \right. , \ee }

\item
Collisional ionisation cooling rate
{\small 
\be \Lambda_{ci} = 1.27 \times 10^{-21} \sqrt{T} \left(1+
\sqrt{\frac{T}{10^5}}\right) {\rm exp}\left
(\frac{-157809.1}{T}\right)n_e n_{\rm HI} , \ee }

\item
Collisional excitation cooling rate
{\small 
\be \Lambda_{ce} = 7.5 \times
10^{-19}\left(1+\sqrt{\frac{T}{10^5}}\right)^{-1} {\rm
exp} \left(\frac{-118348}{T}\right)n_e n_{\rm HI} , \ee }

\item
Bremsstrahlung cooling rate
{\small 
\be \Lambda_{\rm B} = 1.42\times 10^{-27} g_{\rm ff} \sqrt{T} n_e n_{\rm
HI} , \ee }

\end{enumerate}

where $g_{\rm ff} = 1.3$ is the Gaunt factor.

\section{Method of Steepest Descent}\label{AppendixB}
In this Appendix, we give a derivation of the CG
technique for solving linear problems, which we employ in our implicit
time integration scheme of the anisotropic diffusion problem. As the
CG scheme is closely related to the method of steepest
decent, we start with an explanation of this more general technique,
and then specialize to the CG method.

The method of steepest descent is a scheme to solve
a linear system of equations given by  $A
\vec{x} = \vec{b}$. The idea is to obtain the solution
as the minimum of the quadratic form
\be
f(\vec{x}) = \frac{1}{2}\vec{x}^TA\vec{x}-\vec{b}^T\vec{x},
\ee
such that
\be
f'(\vec{x}) = \frac{1}{2}A^T\vec{x} + \frac{1}{2}A\vec{x} -\vec{b}.
\ee
This equation reduces to
\be
f'(\vec{x}) = A\vec{x} -\vec{b}
\ee
if $\vec{A}$ is symmetric and positive definite, i.e.~if
$\vec{x}^TA\vec{x} >0$
for all $\vec{x}\ne 0$.

We now consider an iteration scheme that tries to find the solution
$\vec{x}$.  As we take steps, we choose the direction of the next step
in the direction in which the quadratic form $f$ decreases most
rapidly, which is in the direction opposite to the gradient
$f'(\vec{x})$.  Therefore, the next step should be proportional to
$-f'(\vec{x}_{(i)})=\vec{b}-A\vec{x}_{(i)}$. Here the index $i$
denotes the number of the step we take towards the correct value of
$\vec{x}$. Let us denote the difference between the numerical and the
exact solution as $\vec{e}_{(i)} = \vec{x}_{(i)} - \vec{x}$, and the
residual as $\vec{r}_{(i)} = \vec{b} - A\vec{x}_{(i)} =
-f'(\vec{x}_{(i)})$. Therefore, the next step taken is given by
$\vec{x}_{(1)} = \vec{x}_{(0)} + \alpha\vec{r}_{(0)}$. The optimum
value of $\alpha$ is chosen such that the directional derivative
$\frac{{\rm d}}{{\rm d} \alpha}f(\vec{x}_{(1)})$ equals 0, i.e. the
vectors $f'(\vec{x}_{(1)})$ and $\vec{r}_{(0)}$ should be chosen
orthogonal:
\be \frac{{\rm d}}{{\rm d} \alpha}f(\vec{x}_{(1)}) =
f'(\vec{x}_{(1)})^T\frac{{\rm d}}{{\rm d} \alpha}\vec{x}_{(1)} =
f'(\vec{x}_{(1)})^T\vec{r}_{(0)}=0 \ee

\noindent We further notice that $f'(\vec{x}_{(1)})=-\vec{r}_{(1)}$ and
therefore
\begin{eqnarray*}
\vec{r}_{(1)}^T\vec{r}_{(0)} &=& 0\\
(\vec{b}-\vec{A}\vec{x}_{(1)})^T\vec{r}_{(0)} &=& 0\\
(\vec{b}-\vec{A}(\vec{x}_{(0)}+\alpha\vec{r}_{(0)}))^T\vec{r}_{(0)} &=& 0\\
(\vec{b}-\vec{A}\vec{x}_{(0)})^T\vec{r}_{(0)} - \alpha
(\vec{A}\vec{r}_{(0)})^T\vec{r}_{(0)} &=& 0\\
(\vec{b}-\vec{A}\vec{x}_{(0)})^T\vec{r}_{(0)} &=& \alpha
(\vec{A}\vec{r}_{(0)})^T\vec{r}_{(0)}\\
\vec{r}_{(0)}^T\vec{r}_{(0)} &=& \alpha
\vec{r}_{(0)}^T(\vec{A}\vec{r}_{(0)}) \\
\alpha &=& \frac{\vec{r}_{(0)}^T\vec{r}_{(0)}}{\vec{r}_{(0)}^T\vec{A}\vec{r}_{(0)}}
\end{eqnarray*}
Finally, putting it all together, the method of steepest descent is as
follows
\bea
\vec{r}_{(i)} &=& \vec{b} - \vec{A}\vec{x}_{(i)}\\
\alpha_{(i)} &=&
\frac{\vec{r}_{(i)}^T\vec{r}_{(i)}}{\vec{r}_{(i)}^T\vec{A}\vec{r}_{(i)}}\\
\vec{x}_{(i+1)} &=& \vec{x}_{(i)} + \alpha_{(i)}\vec{r}_{(i)}\\
\vec{r}_{(i+1)} &=& \vec{r}_{(i)} - \alpha_{(i)}\vec{A}\vec{r}_{(i)}
\eea

\section{Conjugate Gradient Method}\label{AppendixC}

In the method of steepest descent the value of $\vec{x}$ is determined via
successively adding the search directions $\vec{r}_{(i)}$. Let us define the set of
directions \{$\vec{d}_{(i)}$\} as the search directions for the CG
method, such that $\vec{x}_{(i+1)} = \vec{x}_{(i)} +
\alpha_{(i)}\vec{d}_{(i)}$ and $\vec{r}_{(i+1)} = \vec{r}_{(i)} -
\alpha_{(i)}A\vec{d}_{(i)}$. We further require that the vectors
\{$\vec{d}_{(i)}$\} are $\vec{A}$-conjugate, i.e. $\vec{d}_{(i)}\vec{A}\vec{d}_{(j)}
= 0$, which means
 \be \vec{d}_{(i+1)} = \vec{r}_{(i+1)} + \sum_{k=0}^{i} \beta_{ik}
\vec{d}_{(k)}. \ee

Using Gram-Schmidt orthogonalization, the coefficients $\beta_{ik}$
are found to be \be \beta_{ik} = \left\{
\begin{array}{cl}
\frac{1}{\alpha_{(i-1)}}
\frac{\vec{r}_{(i)}^T\vec{r}_{(i)}}{\vec{d}_{(i-1)}^T\vec{A}\vec{d}_{(i-1)}}
& \mbox{for $i=k+1$}\\
\\
0 & \mbox{for $i>k+1$}
\end{array}
\right.
\ee
Thus
\be
\beta_{(i)} \equiv \beta_{i, i-i} = \frac{\vec{r}_{(i)}^T\vec{r}_{(i)}}{\vec{r}_{(i-1)}^T\vec{r}_{(i-1)}}.
\ee
Therefore, the CG method can be summarized as follows:
\bea
\vec{r}_{(0)} &=& \vec{b} - \vec{A}\vec{x}_{(0)}\\
\vec{d}_{(0)} &=& \vec{b} - \vec{A}\vec{x}_{(0)}\\
\alpha_{(i)} &=&
\frac{\vec{r}_{(i)}^T\vec{r}_{(i)}}{\vec{d}_{(i)}^T\vec{A}\vec{d}_{(i)}}\\
\vec{x}_{(i+1)} &=& \vec{x}_{(i)} + \alpha_{(i)}\vec{d}_{(i)}\\
\vec{r}_{(i+1)} &=& \vec{r}_{(i)} - \alpha_{(i)}\vec{A}\vec{d}_{(i)}\\
\beta_{(i+1)} &=&
\frac{\vec{r}_{(i+1)}^T\vec{r}_{(i+1)}}{\vec{r}_{(i)}^T\vec{r}_{(i)}}\\
\vec{d}_{(i+1)} &=& \vec{r}_{(i+1)} - \beta_{(i+1)}\vec{A}\vec{d}_{(i)}
\eea

The interesting feature of the CG method is that each subsequent
correction to the solution vector is orthogonal to all previous ones,
while at the same time it points into the direction where the error in
the solution decreases most quickly. This normally produces a
comparatively rapid convergence of the scheme.

\bibliographystyle{mn2e}
\bibliography{paper}

\label{lastpage}

\bsp

\end{document}